\documentclass[12pt,oneside]{article}  
\pdfoutput=1 	% use "amsart" instead of "article" for AMSLaTeX format
\usepackage{geometry}                		% See geometry.pdf to learn the layout options. There are lots.
\usepackage{graphicx}	
\usepackage{titling}			% Use pdf, png, jpg, or eps§ with pdflatex; use eps in DVI mode
\usepackage{amssymb}
\usepackage[dvipsnames]{xcolor}

\def\hhref#1{\href{http://arxiv.org/abs/#1}{arXiv:#1}} 

\usepackage{amsmath,amssymb,amsthm}

\theoremstyle{thmstylethree}

 \usepackage[T1]{fontenc}
 \usepackage{hyperref}

\hypersetup{colorlinks,linkcolor=blue,urlcolor=Blue}

\newtheorem{exercise}{Exercise}[section]

\title{Introductory Lectures on Resurgence\\~\\
{\large CERN Summer School:  {\it Continuum Foundations of Lattice Gauge Theories\\
July 2024}}}
\author{Gerald V. Dunne\\
Physics Department,
University of Connecticut, CT 06269, USA}
\date{}

\begin{document}

\maketitle

\begin{abstract}A set of four introductory  lectures on Resurgent Asymptotics for Physics (``resurgence")  at the CERN Summer School: {\it Continuum Foundations of Lattice Gauge Theories},  July 2024. Lecture 1: The Airy function and the Stokes phenomenon. Lecture 2: The nonlinear Stokes phenomenon. Lecture 3: Resurgence in QFT: the Heisenberg-Euler effective action. Lecture 4: Resurgent continuation and summation. The emphasis of these lectures  is on physically motivated examples. The lectures include many exercises designed to illustrate some of the key ideas of resurgence.
 \end{abstract}

\tableofcontents

\section*{Introduction and Motivation}
\label{sec:intro}
\numberwithin{equation}{section}

The goal of these lectures is to give a beginner's introduction to some of the main ideas of resurgent asymptotics (``resurgence''), by means of illustrative physical  examples.  The primary physical motivation is to gain a deeper appreciation of the  connections between perturbative physics and nonperturbative physics. A key underlying theme is to understand how to compute physical quantities using the quantum path integral, written symbolically as a sum over all configurations $\phi$  of the Feynman phase factor involving the classical action $S[\phi]$:
\begin{eqnarray}
\int \mathcal D \phi\, \exp\left(\frac{i}{\hbar}\, S[\phi]\right)\quad .
\label{eq:path-integral}
\end{eqnarray}
The physical meaning of this expression is clear, as a sum of quantum amplitudes incorporating quantum evolution and interference effects, but of course the actual {\it computation} of the path integral remains challenging in many physically important situations:

\noindent$\bullet$ Minkowski versus Euclidean path integrals.
\\
$\bullet$ 
Strong coupling versus weak coupling.
\\
$\bullet$ Non-equilibrium and real time processes.
\\
$\bullet$ Finite density.
\\
$\bullet$  High intensity background fields.

We note the {\it dual role} of the path integral:

\noindent$\bullet$ The path integral is the generator of Feynman perturbation theory.
\\
$\bullet$ The path integral is the generator of a saddle point expansion.

These two roles of the path integral produce different types of expansions and involve very different methods, and yet they must ultimately be equivalent. The basic premise  is that resurgence provides a general framework to understand quantitatively how these different-looking methods fit together. 

Another underlying theme of resurgence is the observation that perturbation theory generically produces expansions which are {\bf asymptotic} expansions, with coefficients that grow factorially fast. So these expansions have {\it zero} radius of convergence. Therefore we need a formalism that brings such perturbative expansions under mathematical  control, so that we can extract physical predictions. A powerful method is Borel summation, which is significantly extended in its scope using resurgence, leading to what is known as Borel-\'Ecalle summation \cite{ecalle,ecalle2,costin-book,marcos-lectures,sauzin,dorigoni,annrev,aniceto}. Illustrative physical examples will be discussed here. 

An important idea is the modern development of an old observation of Hardy 
\begin{quote}
{\it ``The only scales of infinity that are of any practical importance in analysis are those which may be constructed by means of the logarithmic and exponential functions.''}\\
\hskip 2 cm 
G. H. Hardy, {\it Orders of Infinity}  \cite{hardy}
\end{quote}
This is a precursor to the concept of a {\it ``transseries''}, generated by powers (and iterations) of  the following three {\it ``transmonomial''} building blocks in terms of a small parameter $\hbar$:
\begin{eqnarray}
\hbar\qquad; \qquad e^{-1/\hbar}\qquad ; \qquad \ln(\hbar)
\label{eq:monomials}
\end{eqnarray}
Formal asymptotic series in powers of $\hbar$ do not uniquely specify a function, while a transseries is designed to encode all the analytic information about the function. This is achieved by treating powers (and iterated compositions) of $\hbar$, $e^{-1/\hbar}$ and $\ln(\hbar)$ in a unified manner, within a single object: a {\it transseries}. 
Under certain circumstances one can show that such transseries are closed under all relevant operations of analysis \cite{ecalle,ecalle2,costin-book,sauzin}, leading to a conjecture that transseries are widely applicable in physical problems based on systems of equations (``natural problems''). Operationally, a transseries encodes a systematic implementation of the Stokes phenomenon (discussed in these lectures). This generality is an extremely deep idea, whose scope is still being explored, but there are several illustrative physical examples which convey the basic ideas and goals. For certain mathematical systems, such as finite dimensional exponential integrals, and systems of nonlinear differential, difference or integral equations, there exist rigorous results. In quantum field theory, which is inherently infinite dimensional, there are some beautiful nontrivial examples and the general case is an active field of current research.

\section{Airy and Stokes}
\label{sec:airy-stokes}
 Some of the key ideas of resurgence can be traced back to fundamental work by Airy and Stokes, studying the physics of rainbows. Remarkably, already in the 1830s there were laboratory studies of rainbows that generated up to 30 ``spurious'' (now called ``supernumerary'') rainbows. Airy wrote an important paper about the physics of rainbows \cite{airy}, based on the propagation of light rays through droplets of water, in which he derived an integral representation of a function, now known as the Airy function, whose square describes the intensity pattern of such rainbows:
\begin{eqnarray}
{\rm Ai}(x)=\frac{1}{2\pi} \int_{-\infty}^\infty dt\, e^{i(t^3/3+x\, t)}
\label{eq:airy}
\end{eqnarray}
Here $x$ plays the role of the transverse distance from the main rainbow.
Airy was unable to evaluate the integral in \eqref{eq:airy} beyond the second peak of the intensity pattern, due to the rapid oscillation of the integrand (as a function of the integration variable $t$) when $x$ becomes large. This can be viewed as {\it ``the original sign problem''}, a  zero-dimensional version of the sign problem \cite{alexandru} for the Feynman path integral \eqref{eq:path-integral}.

Stokes later wrote two breakthrough papers on this topic \cite{stokes1,stokes2}. The first paper \cite{stokes1} used the (modern at the time) methods of Cauchy to evaluate the Airy function as an asymptotic large $x$ expansion using saddle point (or steepest descent) methods \cite{dingle,carl-book}, leading to the well-known leading asymptotic expansions on the real axis:
\begin{eqnarray}
 {\rm Ai}\left( x\right)
 \sim 
\begin{cases}
\frac{e^{-\frac{2}{3}\, x^{3/2}}}{2\sqrt{\pi}\, x^{1/4}} \qquad\qquad , \quad x\to +\infty\cr\cr
\frac{\cos\left(\frac{2}{3}\, (-x)^{3/2}-\frac{\pi}{4}\right)}{\sqrt{\pi}\,(- x)^{1/4}}
\quad, \quad x\to -\infty
\end{cases}
\label{eq:airy-cases}
\end{eqnarray}
The second expression completely solved Airy's problem, producing intensity patterns which matched beautifully the multi-band observations.
%\begin{quote}
%{\color{Red} {\it "Stokes, by mathematical supersubtlety,
%transformed Airy's integral into a form by which the
%light at any point of any of those thirty bands, and any
%desired greater number of them, could be calculated
%with but little labour ..."}}\\~ \hskip 5cm Lord Kelvin in Stokes's Obituary, 1903
%\end{quote}
This example is  now standard  in complex analysis and mathematical physics texts. It also illustrates the fact that asymptotic series are often more useful than convergent series.\footnote{For example, compare the ten-term Taylor expansion of ${\rm Ai}(x)$ about $x=0$ (an expansion with infinite radius of convergence) with the leading term of the asymptotic expansion of ${\rm Ai}(x)$ about $x=-\infty$ (an expansion with zero radius of convergence).} 

Stokes's second paper \cite{stokes2} was even more profound. Stokes stressed that the behavior of the (real) function ${\rm Ai}(x)$ is governed by {\it generically complex saddle points} in the complex $t$ plane. As the {\it phase} of the external parameter $x$ varies (e.g. $x$ from positive to negative), these saddle points move around in the complex $t$ plane, and the associated steepest descent contours are also correspondingly deformed. There exist special phases of $x$ at which the {\it topology} of the contributing steepest descent contours ``jump'', leading to a saddle point becoming either relevant or irrelevant for the quantity being computed. This is now known as the {\bf Stokes phenomenon}, and it plays a crucial role in resurgence. In Lecture 2 we encounter a nonlinear generalization, in Lecture 3 we see some physical examples, and in Lecture 4 some numerical analytic continuation aspects.

\subsection{``Perturbation Theory'' for the Airy Function}
\label{sec:airy}

We first study the asymptotics of the Airy function as a ``perturbative expansion'', as $x\to +\infty$. In Section \ref{sec:np-airy} we use a nonperturbative saddle expansion. The Airy equation, 
\begin{eqnarray}
y^{\prime\prime}(x)=x\, y(x)
\label{eq:airy-eq}
\end{eqnarray}
can be analyzed by the  Liouville-Green (``WKB'') substitution, $y(x)=e^{-S(x)}/\sqrt{S^\prime(x)}$, which leads  to $S(x)\sim \pm \frac{2}{3} x^{3/2}$, as $x\to +\infty$. This motivates an ansatz
\begin{eqnarray}
y_\pm(x)\sim \#\,  \frac{e^{{\color{Red} \mp\frac{2}{3} x^{\frac{3}{2}}}}}{{\color{blue} x^{\frac{1}{4}}}} \sum_{n=0}^\infty \frac{c_n^{(\pm)}}{\left({\color{Red} x^{3/2}}\right)^n}
\quad, \quad x\to +\infty
\label{eq:airy-eq2}
\end{eqnarray}
where $\#$ is an arbitrary constant, fixed by convention.
The resulting recursion relation for the coefficients $c_n^{(\pm)}$ can be solved:
\begin{eqnarray}
c_n^{(\pm)}=\left(\mp1\right)^n \frac{\Gamma\left(n+\frac{1}{6}\right) \Gamma\left(n+\frac{5}{6}\right)}{{\color{blue} 2\pi}\, n!\, \left({\color{Red} \frac{4}{3}}\right)^n}= \left\{{\color{blue} 1,\mp \frac{5}{48},\frac{385}{4608}, \mp \frac{85085}{663552},} \dots\right\}
\label{eq:airy-cn}
\end{eqnarray}
This {\it perturbative} approach generates two independent solutions as $x\to +\infty$: 
 \begin{eqnarray}
y_\pm (x)= \left\{ \begin{matrix} {\rm Ai}(x) \cr  \frac{1}{2}\, {\rm Bi}(x) \end{matrix}  \right\} \sim \frac{e^{\mp \frac{2}{3} x^{\frac{3}{2}}} }{2\, \pi^{\frac{1}{2}}\, x^{\frac{1}{4}} } \sum_{n=0}^\infty \left(\mp1\right)^n \frac{ \Gamma\left(n+\frac{1}{6}\right) \Gamma\left(n+\frac{5}{6}\right)}{(2\pi)\, n!\, \left(\frac{4}{3}\, x^{3/2}\right)^n}
% \quad, \quad x\to +\infty
 \label{eq:airy-yplus-yminus}
 \end{eqnarray}
 The 1/2 factor with ${\rm Bi}$ is a convention. Notice that the coefficients $c_n^{(\pm)}$ in \eqref{eq:airy-cn} grow factorially in magnitude, so the solutions in \eqref{eq:airy-eq2}-\eqref{eq:airy-yplus-yminus}  are {\it formal} asymptotic series.

These perturbative solutions illustrate a generic feature of resurgent functions: they have {\it large-order/low-order resurgence relations},  a duality between the large $n$ behavior of the expansion coefficients $c_n^{(\mp)}$, and the large $x$ expansion of the functions $y_\pm(x)$. To see this for the Airy function, we note from \eqref{eq:airy-cn} that at large order $(n\to\infty)$:
\begin{eqnarray}
c_n^{(-)}\sim {\color{blue} \frac{1}{2\pi}} \frac{(n-1)!}{\left({\color{Red} \frac{4}{3}}\right)^n}\left(
{\color{blue} 1}- {\color{blue}\frac{5}{36}} \frac{1}{n}+
{\color{blue} \frac{25}{2592}}\frac{1}{n^2}-\dots\right) 
\qquad, \qquad n\to +\infty
\end{eqnarray}
This can be re-written in terms of decreasing factorials (see Exercise \ref{ex:1.1}):
\begin{eqnarray}
c_n^{(-)}\sim \frac{(n-1)!}{{\color{blue} (2\pi)}\left({\color{Red} \frac{4}{3}}\right)^n}\left(
{\color{blue} 1}- \left({\color{Red} \frac{4}{3}}\right){\color{blue}\frac{5}{48}}\frac{1}{ (n-1)}+
\left({\color{Red} \frac{4}{3}}\right)^2{\color{blue} \frac{385}{4608}} \frac{1}{(n-1)(n-2)}-\dots\right)
\label{eq:cn-resurgent}
\end{eqnarray}
When written in this form, the coefficients of the subleading terms of the  large $n$ behavior match the coefficients of the subleading terms of the large $x$ behavior of the other saddle solution $y_+(x)$. The factor $\frac{4}{3}$ corresponds to the difference between the ``actions'', $\pm \frac{2}{3}$, of the two saddle points. The expansion about one saddle point is directly related to the expansion about other saddle points. This is generic.  See Exercise \ref{ex:1.2} for an example where the coefficients are not just numerical coefficients, but are functions of another parameter, different from the expansion variable. 
{\color{Blue}
\begin{exercise}
\label{ex:1.1}
 {\bf Resurgent form of large order growth:}\\
  In resurgence it is  convenient to re-write a factorial large order growth expression, with power-law corrections, as an expansion in  "diminishing" factorials. Namely,
$$b_n\sim \Gamma(n+a) \sum_{m=0}^\infty \frac{d_m}{n^m} \qquad,\quad  n\to +\infty
$$
can be re-expressed as 
$$
b_n\sim \sum_{k=0}^\infty c_k\, \Gamma(n+a-k) \qquad, \quad n\to +\infty
$$
\begin{enumerate}
\item
Show that the coefficients $c_k$ are expressed in terms of the $d_m$ via the Stirling numbers of the first kind (hint: \href{https://dlmf.nist.gov/26.8.ii}{[dlmf.26.8.ii}])\footnote{In these lectures we will often refer to mathematical identities from the NIST Digital Library of Mathematical Functions at \href{https://dlmf.nist.gov}{https://dlmf.nist.gov}.}
$$
c_k=\sum_{l=0}^k S^{(1)}(k, l) \sum_{j=0}^l(-a)^l {{j-l}\choose{j}} d_{l-j} 
$$
\item
Verify this with some examples.
\end{enumerate}
\end{exercise}
}

{\color{Blue} 
\begin{exercise}
\label{ex:1.2}
{\bf Large Order/Low Order Resurgence Relations for
Bessel Functions:} 
The modified Bessel function $I_\nu(x) $ has the large $x$ asymptotic expansion (\href{https://dlmf.nist.gov/10.40.E5}{[dlmf.10.40.E5]}):
$$
\hskip -.7cm I_\nu(x) \sim \frac{e^{x}}{\sqrt{2 \pi x}} \sum_{n=0}^\infty(-1)^n \frac{\alpha_n(\nu)}{x^n} \pm i e^{i \nu \pi} \frac{e^{-x}}{\sqrt{2 \pi x}} \sum_{n=0}^\infty \frac{\alpha_n(\nu)}{x^n}, \quad \left|{\rm arg}(x) - \frac{\pi}{2}\right| <\pi
$$
The coefficients depend on the Bessel index parameter $\nu$
$$
\alpha_n(\nu)=(-1)^n\, \frac{{\color{Red}\cos (\pi  \nu)}}{\pi}  \frac{  \Gamma \left(n+\frac{1}{2}-\nu\right) \Gamma
   \left(n+\frac{1}{2}+\nu\right)}{ 2^n\,  \Gamma (n+1)}
$$
\vskip -.3cm 
\begin{enumerate}
\item
Show that the large-order growth ($n\to \infty$) is
\begin{eqnarray}
\hskip -1.5cm \alpha_n(\nu)\sim \frac{1}{\pi} 
\frac{(-1)^n (n-1)!}{2^n} \left({\color{Red} \alpha_0(\nu)}
- \frac{2\, {\color{Red} \alpha_1(\nu)}}{(n-1)}
+ \frac{2^2\, {\color{Red}\alpha_2(\nu)}}{(n-1)(n-2)}
-\dots\right)
\nonumber
\label{eq:self-resurgence}
\end{eqnarray}

\item 
Verify this large-order behavior numerically.

\item
What is the significance of the ${\color{Red}\cos (\pi  \nu)}$ prefactor?
\end{enumerate}
\end{exercise}
}

One might be tempted by the asymptotic expansions in \eqref{eq:airy-yplus-yminus} to identify the two functions ${\rm Ai}(x)$ and ${\rm Bi}(x)$ under the rotation $x\to e^{\mp 2\pi i/3}\, x$, ($x^{3/2}\to -x^{3/2}$):
\begin{eqnarray}
{\rm Ai}\left(e^{\mp 2\pi i/3}\, x\right) \overset{?}{\to} \frac{1}{2} e^{\pm \pi i/6}\, {\rm Bi}\left( x\right) 
 \qquad, \quad x\to +\infty
\label{eq:ai-bi1}
\end{eqnarray}
However, this is not correct. The correct {\it ``connection formula''} is (\href{https://dlmf.nist.gov/9.2.E11}{[dlmf.9.2.E11]})
\begin{eqnarray}
{\rm Ai}\left(e^{\mp 2\pi i/3}\, x\right) =\frac{1}{2} e^{\pm \pi i/6}\, {\rm Bi}\left( x\right) 
{\color{Red} + \frac{1}{2} e^{\mp \pi i/3}\, {\rm Ai}\left( x\right) }
 \qquad, \quad x\to +\infty
\label{eq:ai-bi2}
\end{eqnarray}
There is an ``extra'' nonperturbative piece, $\frac{1}{2} e^{\mp \pi i/3}\, {\rm Ai}\left( x\right) $, on the RHS of \eqref{eq:ai-bi2}, which is exponentially smaller than the first term as $x\to +\infty$, and so is neglected in the formal asymptotic series \eqref{eq:airy-eq2}. Eq.\eqref{eq:ai-bi2} is an example of a transseries with two different exponential terms, $e^{\pm \frac{2}{3}x^{\frac{3}{2}}}$, and in which the two fluctuation series are quantitatively related.

This Airy function example  illustrates a general feature of  formal asymptotic perturbative expansions: treated naively, they can miss the existence (and evaluation) of nonperturbative contributions that are beyond all perturbative orders. Fortunately there is a method for detecting and evaluating such nonperturbative terms: Borel summation.

{\color{Blue} 
\begin{exercise}
\label{ex:1.3}
{\bf Connection Formula for the Airy Function}
\begin{enumerate}

\item
Plot the real part of ${\rm Ai}\left(e^{\mp 2\pi i/3}\, x\right)$ and the real part of $\frac{1}{2} e^{\pm \pi i/6}\, {\rm Bi}\left( x\right)$ on the same figure, for $x>0$,  and notice how close they are, even at small $x$.

\item Plot the real part of the {\it difference}  ${\rm Ai}\left(e^{\mp 2\pi i/3}\, x\right)-\frac{1}{2} e^{\pm \pi i/6}\, {\rm Bi}\left( x\right)$, and compare it with the real part of $ \frac{1}{2} e^{\mp \pi i/3}\, {\rm Ai}\left( x\right)$. Note that it is exponentially small at large $x$.

\item Plot the real part of ${\rm Ai}\left(x\right)$ in the complex $x$ plane, and identify the different {\it Stokes lines} (lines along which the behavior is  exponentially growing or decaying),  {\it anti-Stokes lines} (lines along which the behavior is exponentially oscillatory),  and the associated {\it Stokes sectors} bounded by these lines.

\end{enumerate}
\end{exercise}
}

\subsection{Borel Summation: the Basic Idea}
\label{sec:borel}

Borel summation (and its extension, Borel-\'Ecalle summation \cite{ecalle,ecalle2,costin-book,sauzin}) is a method of converting formal asymptotic expansions to an integral representation that can, in principle, encode the full analytic structure of the function under consideration. It is a rich subject, with many intricacies. The conditions for Borel summation are well described in the literature  \cite{ecalle,ecalle2,costin-book,sauzin}, but here we  illustrate the subtleties through representative physical examples. Associated numerical considerations are discussed in Lecture 4.

The basic idea can be illustrated with the classic example studied by Euler \cite{euler} 
\begin{eqnarray}
\sum_{n=0}^\infty (-1)^n \frac{n!}{x^{n+1}}=\quad ???
\label{eq:euler1}
\end{eqnarray}
which has the generic factorial growth of the perturbative (large $x$) expansion coefficients.
Write $n!=\int_0^\infty dt\, e^{-t}\, t^n$, and interchange summation and integration (therein lies the interesting mathematics) to obtain 
\begin{eqnarray}
\sum_{n=0}^\infty (-1)^n \frac{n!}{x^{n+1}} \rightarrow \int_0^\infty dt\, e^{-t}\, \frac{1}{x}\sum_{n=0}^\infty \left(\frac{-t}{x}\right)^n 
=\int_0^\infty dt\, e^{-x t} \frac{1}{1+t}
\label{eq:euler2}
\end{eqnarray}
The final integral is a Laplace integral, with {\it ``Borel transform"} ${\mathcal B}(t)=\frac{1}{1+t}$, and  the integral is finite and convergent for all $x>0$. So if we read the equation backwards we recognize the original formal series as the asymptotic $x\to +\infty$ expansion of the convergent integral. We  then {\it define} the integral as the {\it Borel sum} of the formal series. Borel summation provides a {\it regularization} of the formal asymptotic series, bringing it under mathematical control.

This raises the  important questions of uniqueness, and of analytic continuation. For example, consider expanding as $x\to -\infty$, instead of as $x\to +\infty$. This amounts to considering {\it sign-non-alternating} factorial coefficients. The same formal manipulations result in a Laplace integral
\begin{eqnarray}
\sum_{n=0}^\infty  \frac{n!}{x^{n+1}} \rightarrow \int_0^\infty dt\, e^{-t}\, \frac{1}{x}\sum_{n=0}^\infty (t/x)^n 
=\int_0^\infty dt\, e^{-x t} \frac{1}{1-t}
\label{eq:euler3}
\end{eqnarray}
with a singularity (here, a simple pole) {\it on the contour of integration}. We must therefore provide a prescription for dealing with this pole. This is often fixed by the physical context, and/or a global condition. If we begin with the integral representation, then rotating $x\to -x$ can be compensated by rotating $t\to -t$. But there are two ways to rotate, $t\to e^{\mp i \pi}\, t$, and the difference between these two yields a residue term, $\pm i\pi e^{-x}$, which is nonperturbatively small as $x\to +\infty$.

In the example \eqref{eq:euler1} we recognize the Borel-Laplace integral \eqref{eq:euler2} as the integral representation of an incomplete gamma function (\href{https://dlmf.nist.gov/8.6.E5}{[dlmf.8.6.E5]}):
\begin{eqnarray}
e^x\, \Gamma(0,x)=\int_0^\infty dt\, e^{-x t} \frac{1}{1+t}
\label{eq:inc-gamma1}
\end{eqnarray}
The nonperturbative connection formula (\href{https://dlmf.nist.gov/8.2.E10}{[dlmf.8.2.E10]}) under $x\to -x$ is:
\begin{eqnarray}
e^{-x}\, \Gamma(0,e^{i\pi} \, x) - e^{-x} \, \Gamma(0,e^{-i\pi} \, x)=-2\pi i \, e^{-x}
\qquad, \qquad x>0
\label{eq:inc-gamma2}
\end{eqnarray}
This is strange. It means that for $x>0$, the sign-non-alternating  series in \eqref{eq:euler3}, in which every term is real,  has both a real and imaginary part:
\begin{eqnarray}
{\rm Re} \left[e^{-x} \,\Gamma(0,- x)\right]&=&-{\mathcal P} \int_0^\infty dt\, e^{-x t} \frac{1}{1-t}
\label{eq:inc-gamma-real}
\\
{\rm Im} \left[e^{-x}\, \Gamma(0,e^{\pm i\pi} \, x)\right]&=&\mp\, \pi \, e^{-x}
\label{eq:inc-gamma-imag}
\end{eqnarray}
Here ${\mathcal P}$ denotes the principal value integral. 

Similar behavior can be seen when the Borel transform has an infinite number of poles:
{\color{Blue}
\begin{exercise}
\label{ex:1.4}
{\bf Borel Singularities and "nonperturbative Completion''}
\\
Consider the asymptotic expansion of the "trigamma" function $\psi^{(1)}(z):=\frac{d^2}{dz^2}\ln \Gamma(z)$  \begin{eqnarray}
\psi^{(1)}\left(\frac{1+x}{2}\right) \sim \sum_{n=0}^\infty \frac{2^{n+1} {B}_n\left(\frac{1}{2}\right)}{x^{n+1}}\qquad, \quad x\to +\infty
\nonumber
\label{eq:borel-ex1b}
\end{eqnarray}
(${B}_n\left(q\right)$ is the Bernoulli polynomial)
\vskip -.3cm 
\begin{enumerate}
\item
Use Borel summation to show that
\begin{eqnarray}
 \psi^{(1)}\left(\frac{1+x}{2}\right)
= 2 \int_0^\infty dt\, e^{-x\, t}\frac{t}{\sinh(t)}
\nonumber
\label{eq:borel-ex1a}
\end{eqnarray}
Hint: \href{http://dlmf.nist.gov/24.4.E27}{[dlmf.24.4.E27]} \& \href{http://dlmf.nist.gov/24.7.E2}{[dlmf.24.4.E2]}. There is an {\it infinite number} of Borel poles.
\item
Show that the real part of this function, for $x$ along the imaginary axis, has an infinite series of exponential terms
\begin{eqnarray}
{\rm Re}\left[\psi^{(1)}\left(\frac{1+i\, x}{2}\right)\right] \sim 0\,\, {\color{Red} -2\pi^2  \sum_{k=1}^\infty (-1)^k k\,  e^{-k\, \pi\, x}}\quad, \quad x\to+\infty 
\nonumber
\end{eqnarray}
Also show that these are precisely the terms that are required for consistency with  \href{https://dlmf.nist.gov/5.15.E6}{[dlmf.5.15.E6]}, the nonperturbative reflection formula. 
\end{enumerate}
\end{exercise}
}
The Euler example \eqref{eq:euler1}-\eqref{eq:euler2} generalizes to a Borel transform with a branch point rather than a pole:
{\color{Blue} 
\begin{exercise}
\label{ex:1.5}
{\bf Borel Singularities and Nonperturbative Terms}
\\
 Consider the Borel transform $B(t)=\frac{1}{(1+t)^\beta}$, which has a {\it branch point} at $t=-1$, with exponent $0<\beta<1$, and a branch cut along the negative axis: $t\in (-\infty, -1]$. See \href{https://dlmf.nist.gov/8.6.E5}{[dlmf.8.6.E5]}:
\begin{eqnarray}
x^{\beta-1} \, e^{x} \, \Gamma\left(1-\beta, x\right) = \int_0^\infty dt\, e^{-x\, t}\, \frac{1}{(1+t)^\beta}
\nonumber
\label{eq:borel-ex2a}
\end{eqnarray}
\begin{enumerate}
\item
Generate an expression for the $x\to+\infty$ asymptotic expansion of the function $x^{\beta-1} \, e^{x} \, \Gamma\left(1-\beta, x\right)$. 
\item
Using the discontinuity across the cut of the Borel transform function $B(t)=\frac{1}{(t+1)^\beta}$, derive the general  connection formula \href{https://dlmf.nist.gov/8.2.E10}{[dlmf.8.2.E10]} for the incomplete gamma function:
\begin{eqnarray}
e^{\pi\, i\, \beta}  \Gamma\left(1-\beta , x\, e^{\pi i}\right)-e^{-\pi\, i\, \beta}  \Gamma\left(1-\beta, x\, e^{-\pi i}%
\right)=\frac{2\pi i}{\Gamma\left(\beta\right)}
\nonumber
\label{eq:incomplete-gamma-connection}
\end{eqnarray}
\end{enumerate}
\end{exercise}
}

These  examples illustrate several important aspects of Borel summation:
\begin{itemize}
\item
Borel summation maps a formal series to a convergent integral. This can be viewed as a {\it regularization} of the formal asymptotic series.

\item
Uniqueness of Borel summation is interesting and important.

\item
The Borel transform is an inverse Laplace transform. Nonperturbative contributions arise from the singularities of the Borel transform as the contour of the Borel integral is deformed.
\item
A formal series in which all terms are real can produce nonperturbative imaginary contributions, whose sign should be fixed by global and/or physical considerations.
\item
A formal series with all real and positive coefficients can produce a real part that changes sign. This means that positivity and convexity arguments based on formal series must be treated with care.

\end{itemize}

In general, the Borel transform can be defined as a map:
\begin{eqnarray}
f(x)\sim \sum_{n=0}^\infty \frac{c_n}{x^{n+1}} \quad \longrightarrow \quad 
{\mathcal B}[f](t):=\sum_{n=0}^\infty \frac{c_n}{n!} \, t^n
\label{eq:borel1}
\end{eqnarray}
If $c_n\sim n!$ as $n\to\infty$, then dividing out the factorial growth of the $c_n$ coefficients produces a Borel transform function ${\mathcal B}[f](t)$ with a {\it finite} radius of convergence, and hence with at least one  singularity in the (finite) complex $t$ plane. The directional Borel sum of the formal series is defined by the integral representation
\begin{eqnarray}
{\mathcal S}_\theta f(x):= \int_0^{\infty\, e^{i\, \theta} } dt\, e^{-x t}\, {\mathcal B}[f](t)
\label{eq:borel2}
\end{eqnarray}
As $\theta$ rotates, the integration contour may cross singularities of the Borel transform. If these are isolated poles or branch points we generically pick up exponential terms of the form $\sim e^{-x\, t_k}$ for the $k^{\rm th}$ Borel singularity $t_k$. 

The moral is:
\begin{quote}
\centerline{Borel singularities $\quad \longleftrightarrow \quad $ nonperturbative physics}
\end{quote}
We next illustrate this correspondence by
using Borel summation  to recover the "missing" nonperturbative term in the Airy connection formula \eqref{eq:ai-bi2}

\subsection{Borel Summation of the Airy Function}
\label{sec:borel-airy}

Recall that the formal series in \eqref{eq:airy-yplus-yminus} both satisfy the Airy  equation, $y''=x\,y$, but {\it do not satisfy} the global connection formula \eqref{eq:ai-bi2}. The Borel transform \eqref{eq:borel1} for  ${\rm Ai}(x)$   is
\begin{eqnarray}
\sum_{n=0}^\infty \left(-1\right)^n \frac{ \Gamma\left(n+\frac{1}{6}\right) \Gamma\left(n+\frac{5}{6}\right)}{(2\pi)\, n!} {\color{blue}\frac{t^n}{n!}}= ~_2 F_1\left(\frac{1}{6}, \frac{5}{6}, 1; -t\right)
\label{eq:airy-borel1}
\end{eqnarray}
Therefore ${\rm Ai}(x)$ is expressed as a Borel integral:
\begin{eqnarray}
{\rm Ai}(x)=\frac{e^{- \frac{2}{3} x^{\frac{3}{2}}} }{2\sqrt{\pi}\, x^{\frac{1}{4}} }\left( \frac{4}{3} x^{\frac{3}{2}} \right)
\int_0^\infty dt \, e^{-\frac{4}{3} x^{\frac{3}{2}} t} ~_2 F_1\left(\frac{1}{6}, \frac{5}{6}, 1; -t\right)
\label{eq:airy-borel2}
\end{eqnarray}
The Borel transform in \eqref{eq:airy-borel1}-\eqref{eq:airy-borel2} is a hypergeometric function with argument $-t$, so it has a cut along the negative $t$ axis: $t\in (-\infty, -1]$. 
As we rotate the phase of $x$ by $\pm 2\pi/3$, we counter-rotate $t$ by $\mp \pi$. The difference between these is given by the jump across the cut of the Borel transform (\href{https://dlmf.nist.gov/15.2.E3}{[dlmf.15.2.E3]}):
\begin{eqnarray}
\hskip -.5cm {\color{Red}~_2F_1\left(\frac{1}{6}, \frac{5}{6}, 1; t+i\, \epsilon\right)
-
~_2F_1\left(\frac{1}{6}, \frac{5}{6}, 1; t-i\, \epsilon\right)}
={\color{blue} i\,~_2F_1\left(\frac{1}{6}, \frac{5}{6}, 1; 1-t \right)}
\, ,\, 
t\geq 1
\label{eq:airy-jump}
\end{eqnarray}
Note that the jump across the cut is expressed in terms of the very same hypergeometric function, but with argument $(1-t)$. This jump across the cut produces the ``missing'' nonperturbative term in the connection formula \eqref{eq:ai-bi2}. See Exercise \ref{ex:1.6}.

The lesson from this example is that the singularity of the Borel transform  encodes the connection formula. Therefore the Borel integral, with suitable analytic continuation of the Borel transform function and associated contour deformation, is capable of encoding the correct analytic continuation properties of the original function, even though the formal series does not manifestly contain this information. This is one of the simplest examples of the origin of a ``transseries'': the formal asymptotic series in \eqref{eq:airy-yplus-yminus} should be ``completed'' by the addition of a nonperturbative term. 

{\color{Blue} 
\begin{exercise}
\label{ex:1.6}
{\bf nonperturbative Bessel Connection Formula.} The modified Bessel function $K_\nu(x)$ has a Borel representation (Airy is associated with $\nu=\frac{1}{3}$) for $x>0$
\begin{eqnarray}
K_\nu(x)= \sqrt{2\pi x}\, e^{-x}\int_0^\infty dt\, e^{-2\, x\, t}  ~_2F_1\left(\frac{1}{2}-\nu, \frac{1}{2}+\nu, 1; -t\right) 
\nonumber
\label{eq:knu1}
\end{eqnarray}
\vskip -.4cm 
\begin{enumerate}
\item
Derive the asymptotic expansion (\href{https://dlmf.nist.gov/10.40.E2}{[dlmf.10.40.E2]})
\begin{eqnarray}
K_{\nu}\left(x\right)\sim\left(\frac{\pi}{2x}\right)^{\frac{1}{2}}e^{-x}\sum_{k=0}^{\infty}\frac{a_{k}(\nu)}{x^{k}}
\quad, \quad x\to\infty, \quad |{\rm arg}(x)| \leq \frac{3\pi}{2} -\delta
\nonumber
\label{eq:knu2}
\end{eqnarray}
\begin{eqnarray}
a_{k}(\nu)
=\frac{ \cos (\pi  \, \nu)}{\pi} \left(-\frac{1}{2}\right)^k \frac{  \Gamma \left(k+\frac{1}{2}-\nu\right) \Gamma
   \left(k+\frac{1}{2}+\nu\right)}{  \Gamma (k+1)}
\nonumber
\label{eq:ak-nu}
\end{eqnarray}

\item
Use the discontinuity of the hypergeometric function \href{https://dlmf.nist.gov/15.2.E3}{[dlmf.15.2.E3]} to derive the nonperturbative connection formula of $K_\nu(x)$ \href{https://dlmf.nist.gov/10.34.E2}{[dlmf.10.34.E2]}:
$$K_{\nu}\left(ze^{m\pi i}\right)=e^{-m\nu\pi i}K_{\nu}\left(z\right)-\pi i\sin
\left(m\nu\pi\right)\csc\left(\nu\pi\right)I_{\nu}\left(z\right)$$

\end{enumerate}
\end{exercise}
}

\subsection{``Nonperturbative'' Expansion of the Airy Function}
\label{sec:np-airy}

Having shown how we can generate the transseries for the Airy function starting from its formal asymptotic expansion (an analogue of "perturbation theory"), we now approach the problem from the other direction, generating the transseries from the original integral representation (an analogue of a ``nonperturbative'' saddle expansion of a path integral). The two approaches begin by looking very different, but of course must be the same.

\subsubsection{Leading Saddle Point Expansion}
\label{sec:airy-saddle}

Recall the integral representation of the Airy function \cite{airy}:
\begin{eqnarray}
{\rm Ai}(x)=\frac{1}{2\pi}\int_{-\infty}^{+\infty} dt\, e^{i\left(x\, t+\frac{1}{3} t^3\right)}
\label{eq:airy1}
\end{eqnarray}
Consider (\ref{eq:airy1})  as a ``zero-dimensional'' Minkowski space path integral in the variable $t$, in which the integrand is an oscillatory phase, depending on another parameter, $x$. Due to the rapid oscillations of the integrand there is no practical numerical method to evaluate \eqref{eq:airy1} as an integral along the real $t$ axis. 
We are interested in the behavior of the Airy function ${\rm Ai}(x)$ for large $|x|$, for which these oscillations become even more intense. This is the zero-dimensional {\it ``sign problem''}.  Stokes solved this problem by contour deformation and saddle point analysis  \cite{stokes1,dingle}. One physical motivation for studying resurgence is to see if this approach can be generalized in a computationally meaningful way for infinite dimensional path integrals \cite{witten}.

Write $x$ as an amplitude and a phase, $x=r\, e^{i\,\theta}$, so  the asymptotic limit is $r\to\infty$, with $r$ real and positive, and change the integration variable from $t$ to $z$ by writing as $t=-i\, \sqrt{r}\, z$. The large parameter $r$ (actually $r^{3/2}$) now multiplies the entire exponent function:
\begin{eqnarray}
{\rm Ai}(x)=\frac{\sqrt{r}}{2\pi i}\int_{i\, \mathbb R} dz\, e^{r^{3/2}\left(e^{i \theta}\, z-\frac{1}{3} z^3\right)} 
\label{eq:airy2}
\end{eqnarray}
and the integration contour $\gamma$ is now  the imaginary axis.
This is an exponential integral
\begin{eqnarray}
\int_\gamma dz\, e^{\frac{1}{\hbar} \, S(z)}
\qquad, \quad S(z; \theta) =e^{i \theta}\, z-\frac{1}{3} z^3
\label{eq:exp-int}
\end{eqnarray}
integrated over a contour $\gamma$, with $\hbar$ identified with $\frac{1}{r^{3/2}}$. The Airy ``action function'' is given by $S(z; \theta)$,
which depends parametrically on $\theta$, the phase of $x$ (the argument of the Airy function in (\ref{eq:airy1})).

For the integral in (\ref{eq:exp-int}) to be well defined the contour $\gamma$ must tend asymptotically (as $|z|\to \infty$) to  directions in which $z^3>0$: 
\begin{eqnarray}
{\rm arg}(z) \to 0\quad {\rm or}  \quad \frac{2\pi}{3}\quad {\rm or}  \quad \frac{4\pi}{3}\quad, \qquad {\rm as} \quad |z| \to \infty
\label{eq:airy-asymptotes}
\end{eqnarray}
\begin{figure}[htb]
\centering
\includegraphics[scale=.25]{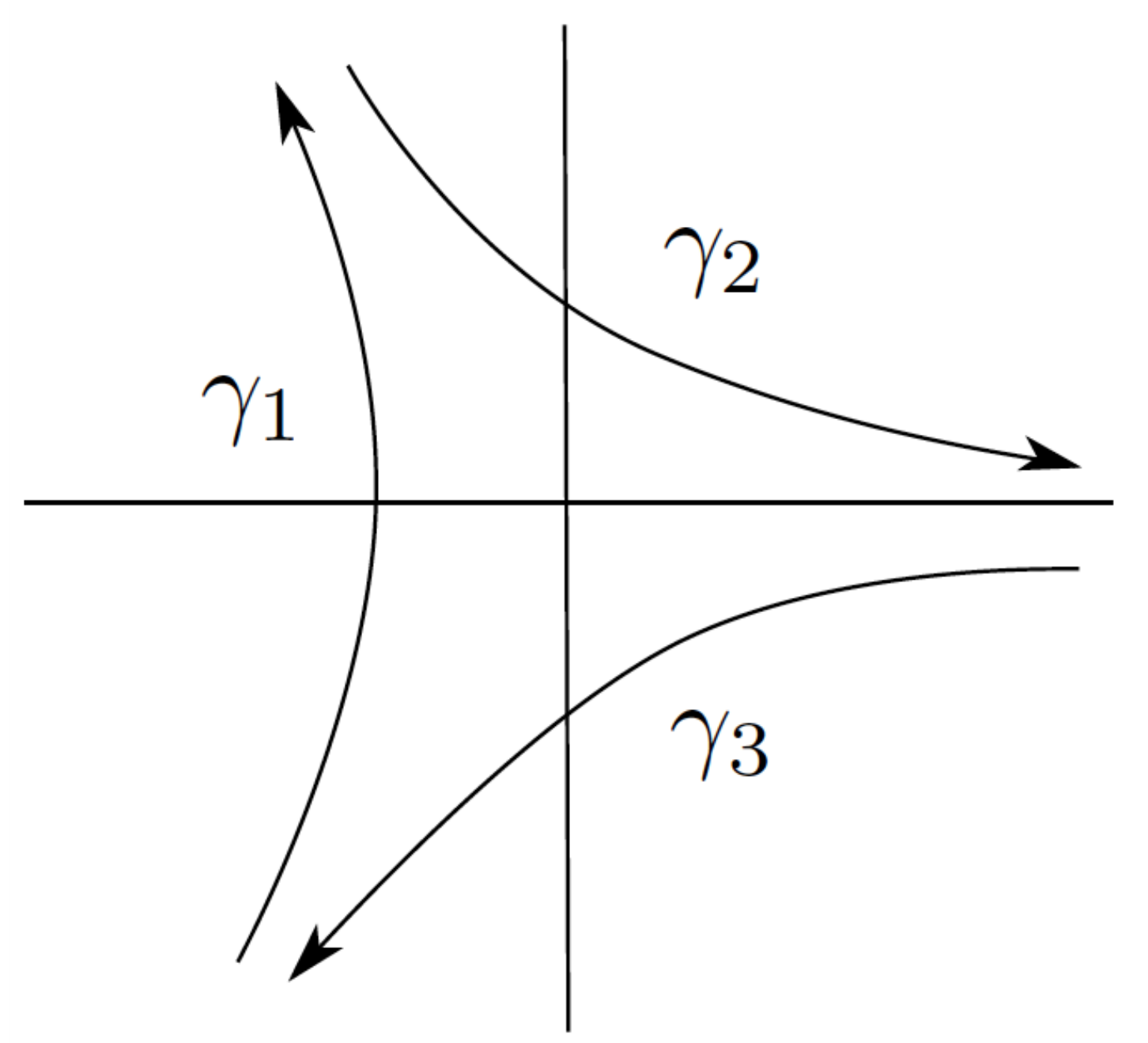}
\caption{The three basis contours for the Airy function integral in (\ref{eq:exp-int}). These curves tend asymptotically to the directions ${\rm arg}(z)=0, \frac{2\pi}{3}, \frac{4\pi}{3}$. }
\label{fig:airy-three-curves}
\end{figure}
There are 3 such possible contours, up to simple deformations (there are no singularities of the integrand in the finite $z$ plane). See Figure \ref{fig:airy-three-curves}. Thus, the integral in (\ref{eq:exp-int}) should be deformed into some linear combination of these contours.  Only two of these contours are independent since $\gamma_1+\gamma_2+\gamma_3\equiv 0$. If these are chosen to be steepest descent contours, as below, then the integral becomes well behaved and can be evaluated straightforwardly in a number of ways. The important observation is that this linear combination of contours changes as $\theta$ changes. This is the origin of the {\it ``Stokes phenomenon''}.

We first identify the saddle points, the stationary points of the action function $S(z; \theta)$. Since $S(z; \theta)$ is cubic in $z$, there are just 2 saddle points:
\begin{eqnarray}
z_\pm =\pm \, e^{i \theta/2}
\label{eq:zpm}
\end{eqnarray}
The action evaluated at each saddle point is 
\begin{eqnarray}
S_\pm\equiv S(z_\pm)=\pm\, \frac{2}{3}\, e^{3i\,\theta/2}
\label{eq:spm}
\end{eqnarray}
The saddle point evaluation of the Airy integral involves deforming the original integration contour $\gamma$ into a linear combination of the well-defined basis contours $\gamma_1, \gamma_2, \gamma_3$, such that the contours pass through the saddle points along directions of steepest descent, on which the integral becomes a well-defined  integral. In the large $r$ limit the dominant contribution comes from the behavior in the vicinity of the saddle points. However, we can also go beyond this Gaussian approximation.

We begin by considering the difference between $x>0$ (i.e, $\theta=0$) and $x<0$ (i.e., $\theta=\pi$), in order to understand the saddle point explanation of the difference between the exponentially decaying form of the Airy function when $x>0$, and the oscillatory form of the Airy function when $x<0$. 
Then we comment on the general $\theta$ situation.

\subsubsection{Leading exponential behavior when $\theta=0$:} For $x>0$ on the real axis, $\theta=0$, the saddle points are at $z_\pm =\pm 1$. Deform the contour $\gamma$ to pass through $z_-=-1$. In the limit $r\to\infty$ (i.e., $x\to +\infty$) we expand near $z_-$ as:
\begin{eqnarray}
z=-1+ e^{i\,\phi}\, \xi \quad\Rightarrow\quad S(z)=z-\frac{1}{3} z^3 \approx -\frac{2}{3} +e^{2i\, \phi}\, \xi^2+O(\xi^3)
\label{eq:zm-expansion}
\end{eqnarray}
truncated at quadratic order in the leading asymptotic approximation. 
For a well-defined Gaussian integral, $\int d\xi \, e^{r^{3/2} S(\xi)}$, the coefficient of $\xi^2$ in $S(\xi)$ must be negative, so we choose $\phi=\frac{\pi}{2}$. The contour passes ``vertically" through $z_-$, giving  leading behavior as $x\to +\infty$:
\begin{eqnarray}
{\rm Ai}(x) \sim \frac{\sqrt{r}}{2\pi i} e^{-\frac{2}{3}\, r^{3/2}} \int_{-\infty}^\infty i\, d\xi\, e^{-r^{3/2}\, \xi^2} 
=\frac{1}{2\sqrt{\pi}\, x^{1/4}}\, e^{-\frac{2}{3}\, x^{3/2}}
\label{eq:x-pos}
\end{eqnarray}
as in \eqref{eq:airy-cases}. We have effectively deformed the original contour of integration into $\gamma_1$, passing through the saddle at $z_-=-1$, in the notation of Figure \ref{fig:airy-three-curves}.
Had we deformed the contour through the saddle point $z_+$, we would have obtained an asymptotically growing solution, corresponding to the other Airy function, ${\rm Bi}(x)$.

\subsubsection{ Leading oscillatory behavior when $\theta=\pi$:} For $x<0$ on the real axis, $\theta=\pi$, the saddle points are at $z_\pm =\pm i$. Deform the contour $\gamma$ to pass through $z_-=-i$. In the limit $r\to\infty$ (i.e., $x\to -\infty$) we expand in the vicinity of $z_-$ as (note that since $\theta=\pi$, the action function is now $S=-z-\frac{1}{3} z^3$):
\begin{eqnarray}
z=-i+ e^{i\,\phi}\, \xi \quad\Rightarrow\quad S(z)=-z-\frac{1}{3} z^3 \approx \frac{2 i}{3} + i e^{2i\, \phi}\, \xi^2+O(\xi^3)
\label{eq:zpp-expansion2}
\end{eqnarray}
To have a well-defined Gaussian integral, $\int d\xi \, e^{r^{3/2} S(\xi)}$, 
we choose $\phi=\frac{\pi}{4}$. In other words, the contour passes ``diagonally" through $z_-$,  at angle $\frac{\pi}{4}$, giving the leading behavior as $x\to -\infty$:
\begin{eqnarray}
 \frac{\sqrt{r}}{2\pi i} e^{\frac{2i}{3}\, r^{3/2}} \int_{-\infty}^\infty e^{i \frac{\pi}{4}} \, d\xi\, e^{-r^{3/2}\, \xi^2} 
= e^{-i \frac{\pi}{4}} \,  \frac{1}{2\sqrt{\pi}\, (-x)^{1/4}}\, e^{\frac{2i}{3}\, (-x)^{3/2}}
\label{eq:x-neg2}
\end{eqnarray}
We have effectively deformed part of the original contour of integration into $-\gamma_3$. In order to match the original contour along the imaginary axis we also need to consider the contour $-\gamma_2$, passing through $z_+=+i$.
Expand near $z_+$ to quadratic order
\begin{eqnarray}
z=+i+ e^{i\,\phi}\, \xi \quad\Rightarrow\quad S(z)=-z-\frac{1}{3} z^3 \approx -\frac{2 i}{3} -i e^{2i\, \phi}\, \xi^2+O(\xi^3)
\label{eq:zpp-expansion1}
\end{eqnarray}
 for the leading asymptotic approximation. 
We choose $\phi=\frac{3\pi}{4}$,
so the contour passes  ``diagonally'' through $z_+$,  at angle $\frac{3\pi}{4}$, giving the leading behavior as $x\to -\infty$:
\begin{eqnarray}
\frac{\sqrt{r}}{2\pi i} e^{-\frac{2i}{3}\, r^{3/2}} \int_{-\infty}^\infty e^{i \frac{3\pi}{4}} \, d\xi\, e^{-r^{3/2}\, \xi^2} 
=e^{i \frac{\pi}{4}} \,  \frac{1}{2\sqrt{\pi}\, (-x)^{1/4}}\, e^{-\frac{2i}{3}\, (-x)^{3/2}}
\label{eq:x-neg1}
\end{eqnarray}
The contribution from each of the saddle points, $z_\mp$, has the same magnitude, so we sum them, leading to the familiar result as $x\to -\infty$:
\begin{eqnarray}
{\rm Ai}(x)
\sim \frac{1}{\sqrt{\pi} (-x)^{1/4}}\, \cos\left(\frac{2}{3} (-x)^{3/2}-\frac{\pi}{4}\right)
\label{eq:x-neg3}
\end{eqnarray} 
We have effectively deformed the original contour of integration, along the imaginary $z$ axis, into the linear combination $-\gamma_3-\gamma_2$, passing through the saddles at $z_\pm=\pm i$, in the notation of Figure \ref{fig:airy-three-curves}.

\subsubsection{Beyond the Gaussian approximation:}
\label{sec:beyond-gaussian}

The standard analysis of the previous section only probed the {\it leading} large $|x|$ (i.e., $r\to\infty$) behavior, based on a {\it Gaussian approximation} to the integral in the vicinity of the saddle points. We can do better than this. Keeping {\it all terms} in the expansion about the saddle point, along the steepest descent contour, produces a well-defined integral that can be evaluated in several different ways, analytically or numerically, and which is exact.

\begin{figure}[htb]
\centering
\includegraphics[scale=.65]{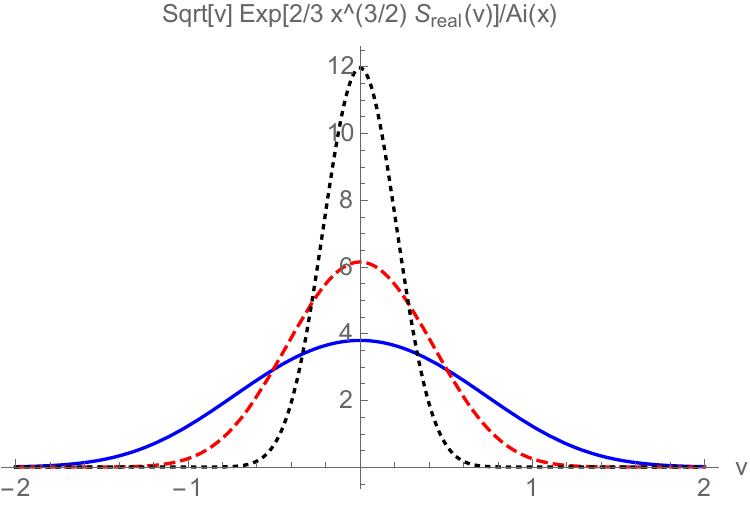}
\caption{The \underline{integrand} function, $e^{x^{3/2}\,  S_{\rm real}(v)}=e^{-\frac{2}{3}\, x^{3/2}\, \left(1+\frac{4}{3}v^2\right) \sqrt{1+\frac{1}{3} v^2}}$, from \eqref{eq:res} and  (\ref{eq:airy-pos}), plotted along the steepest descent contour \eqref{eq:sd-0},  with $x=1$ (solid blue), $x=2$ (dashed red) and $x=5$ (dotted black). Each curve has been normalized by dividing by ${\rm Ai}(x)$. The integrand becomes more localized along the steepest descent contour as $x\to\infty$.} 
\label{fig:bell-shaped-plot}
\end{figure}

\noindent\underline{\bf {$\theta=0$:}} First consider $x$ real and positive, so that $\theta=0$ and $r\to\infty$ in the asymptotic limit. We have already seen that the contour $\gamma$ should be deformed into one that passes through the negative saddle point $z_-=-1$,  parallel to the imaginary axis in the vicinity of the saddle point. The full ``steepest descent'' contour through $z_-$ is defined by the condition that the imaginary part of the action function $S(z)$ remains constant along the contour. Writing $z=u+i\, v$, when $\theta=0$ we have
\begin{eqnarray}
S(z)=
\left(u+u v^2-\frac{1}{3}u^3\right) +i\left(v-u^2 v+\frac{1}{3}v^3\right)
\label{eq:s-re-im}
\end{eqnarray}
Since $S(z_-)=-\frac{2}{3}$ is purely real, the constant value of the imaginary part of $S(z)$ along the steepest descent contour $\gamma_s$ passing through $z_-$ must  be zero. Therefore the steepest descent contour through $z_-=-1$ has $u=-\sqrt{1+\frac{1}{3} v^2}$, with the minus sign chosen so that $u=-1=z_-$ when $v=0$. This steepest descent contour can be parametrized as
\begin{eqnarray}
z= -\sqrt{1+\frac{1}{3} v^2}+i\, v\qquad, \quad v\in (-\infty, \infty)
\label{eq:sd-0}
\end{eqnarray}
Deforming the contour $\gamma$ into the steepest descent contour $\gamma_s$ yields a Jacobian factor:
\begin{eqnarray}
dz=\frac{dz}{dv}\, dv= -\left(\frac{v}{3\sqrt{1+\frac{1}{3} v^2}}-i\right)dv
\label{eq:jac1}
\end{eqnarray}
By construction, on $\gamma_s$  the {\it imaginary part} of $S(z)$ is zero, and the {\it real part} of $S(z)$ has a simple form as a function of the parametric variable $v$
\begin{eqnarray}
S_{\rm real}(v)=-\frac{2}{3}\left(1+\frac{4}{3}v^2\right) \sqrt{1+\frac{1}{3} v^2}
\label{eq:res}
\end{eqnarray}
$S_{\rm real}(v)<0$ on the entire contour, and  $e^{r^{3/2} S(z)}$ has a nice ``bell-shaped'' structure along the the entire contour, localized around the saddle point at $v=0$. See Fig. \ref{fig:bell-shaped-plot}.

{\it Here we have made  no approximation}. We have simply deformed the contour $\gamma$ in (\ref{eq:exp-int}), the imaginary axis, into the steepest descent contour passing through $z_-=-1$, along which the integral is simple to evaluate. So we have the exact integral expression (for $x>0$):
\begin{eqnarray}
{\rm Ai}(x)
&=& \frac{\sqrt{r}}{2\pi i} \int_{-\infty}^\infty dv\, \left(i-\frac{v}{3\sqrt{1+\frac{1}{3} v^2}}\right)  
e^{-\frac{2}{3} \, r^{3/2}  \left(1+\frac{4}{3}v^2\right) \sqrt{1+\frac{1}{3} v^2}} \\
&=& {\color{blue}  \frac{\sqrt{x}}{2\pi} \, e^{-\frac{2}{3} \, x^{3/2}} \int_{-\infty}^\infty dv\, e^{- x^{3/2}\, v^2}}
{\color{Red} e^{ x^{3/2} \left(\frac{2}{3}+v^2-\frac{2}{3} \left(1+\frac{4}{3}v^2\right) \sqrt{1+\frac{1}{3} v^2}\, \right)}}
\label{eq:airy-pos}
\end{eqnarray}
The factor in blue in \eqref{eq:airy-pos} is the Gaussian approximation that produces the familiar leading $x\to+\infty$ behavior  (\ref{eq:x-pos}). The factor in red in \eqref{eq:airy-pos} contains all the information beyond the Gaussian approximation. At this point we have several options. The simplest is to evaluate the final integral in (\ref{eq:airy-pos}) numerically. The integrand is well-behaved and localized near $v=0$, becoming more and more localized as $x=r \to +\infty$. See Fig. \ref{fig:bell-shaped-plot}. It can be integrated to very high precision in various ways, for example using a Monte Carlo integration algorithm.  Contrast this with the original Airy function integral \eqref{eq:airy}. Alternatively we can expand the factor in red in  \eqref{eq:airy-pos} as a series in $v^2$, in which case each term is a Gaussian moment integral of the form 
\begin{eqnarray}
\int_{-\infty}^\infty dv\, e^{-x^{3/2}\, v^2} v^{2n}= \frac{1}{x^{3/4}}\, \frac{\Gamma\left(n+\frac{1}{2}\right)}{x^{3n/2}}
\label{eq:moments}
\end{eqnarray}
This produces the formal $x\to+\infty$ asymptotic series for ${\rm Ai}(x)$ in \eqref{eq:airy-eq2}-\eqref{eq:airy-cn}.

{\color{Blue}
\begin{exercise}
\label{ex:1.7}
\noindent{\bf All-orders saddle analysis of the Airy integral for $x\geq 0$:}
\begin{enumerate}
\item
Verify \eqref{eq:airy-pos}  for $x\geq 0$ by evaluating the integral numerically.
\item
Verify that if one expands the red factor in (\ref{eq:airy-pos}), one obtains the following all-orders expansion as $x\to +\infty$:
\begin{eqnarray}
 {\rm Ai}(x)
\sim \frac{e^{-\frac{2}{3} \, x^{3/2}} }{2 \sqrt{\pi} \, x^{1/4}} 
\left(1 {\color{Red} -\frac{5}{48\, x^{3/2}}+\frac{385}{4608\,  x^3}-\frac{85085}{663552 \, x^{9/2}}+
\dots} \right)
\label{eq:airy-full}
\end{eqnarray}
The terms in red in (\ref{eq:airy-full}) represent the fluctuation corrections beyond the leading Gaussian approximation. 

\end{enumerate}
\end{exercise}
}

\noindent\underline{\bf {$\theta=\pi$:}}  Now consider the all-orders steepest descent derivation of the asymptotic expansion of the Airy function integral as $x\to -\infty$. 
Recall that $x\equiv r\, e^{i\theta}$. Writing $z=u+i\, v$, when $\theta=\pi$ we have
\begin{eqnarray}
S(z; \pi)=
\left(-u+u v^2-\frac{1}{3}u^3\right) +i\left(-v-u^2 v+\frac{1}{3}v^3\right)
\label{eq:s-re-im-pi}
\end{eqnarray}
In this case the saddle points lie on the imaginary axis: $z_\pm=\pm i$, and the contour is deformed into a linear combination of two steepest descent curves. 
Since ${\rm Im}\left[S(z_-)\right]=\frac{2}{3}$, the constant value of the imaginary part of $S(z)$ along the steepest descent contour $\gamma_s$ passing through $z_-$ must  be $+\frac{2}{3}$. In this case, the steepest descent contour through $z_-=-i$ can be parametrized in terms of $u$, $u\in (-\infty, \infty)$, the real part of $z$:
\begin{eqnarray}
z=u+i\left(\frac{-i(1+u^2)}{\left(i+u\sqrt{3+3u^2+u^4}\right)^{1/3}}+i \left(i+u\sqrt{3+3u^2+u^4}\right)^{1/3}\right)
\label{eq:vu-minus}
\end{eqnarray}
Similarly the  steepest descent contour through $z_+=+i$ can be parametrized as:
\begin{eqnarray}
z=u+i\left(\frac{i(1+u^2)}{\left(i+u\sqrt{3+3u^2+u^4}\right)^{1/3}}-i \left(i+u\sqrt{3+3u^2+u^4}\right)^{1/3}\right)
\label{eq:vu-plus}
\end{eqnarray}
{\color{Blue}
\begin{exercise}
\label{ex:1.8}
\noindent{\bf All-orders saddle analysis of the Airy integral for $x\leq 0$:}
\begin{enumerate}
\item
Plot the \underline{integrands} of the Airy integral on each of the two steepest descent contours, in \eqref{eq:vu-minus}-\eqref{eq:vu-plus}, when $x\to -\infty$, and confirm that the integrands are localized.
\item
Hence write an integral representation along each steepest descent contour (for $x\leq 0$), and verify numerically the agreement with ${\rm Ai}(x)$ for $x\leq 0$. 
\end{enumerate}
\end{exercise}
}

\noindent\underline{General $\theta$:}  Recall that the saddle points are at $z_\pm=\pm e^{i \theta/2}$. As $\theta$ varies, the saddle points move, and the steepest descent contours through these saddles are deformed. To determine the steepest descent contours passing through the saddle points $z_\pm$, we decompose the action function $S(z; \theta)$ into its real and imaginary parts:
\begin{equation}
S(z; \theta)
= \left(u\, \cos\theta -v\, \sin\theta +u v^2-\frac{1}{3} u^3\right) +i\left(u\, \sin\theta +v\cos \theta -u^2 v+\frac{1}{3}v^3\right)
\label{eq:stheta-re-im}
\end{equation}
The action at each of the saddle points is $S_\pm =\pm \frac{2}{3} e^{3i \theta/2}$, so the condition for the steepest descent contour is
\begin{eqnarray}
u\, \sin\theta +v\cos \theta -u^2 v+\frac{1}{3}v^3= \pm \frac{2}{3}\, \sin\left(\frac{3\theta}{2}\right)
\label{eq:theta-condition}
\end{eqnarray}
Equation \eqref{eq:theta-condition} is a quadratic equation in $u$ (the real part of $z$) and a cubic equation in $v$ (the imaginary part of $z$). When $0\leq \theta <\frac{2\pi}{3}$, the steepest descent curve is most easily parameterized (for $v\in (-\infty, \infty)$) as
\begin{eqnarray}
z=\left(\frac{\sin (\theta )}{2 v}-\frac{\left(\frac{\sin \left(\frac{\theta
   }{2}\right)}{v}+1\right) \sqrt{3 \cos ^2\left(\frac{\theta }{2}\right)+v^2-2 v \sin
   \left(\frac{\theta }{2}\right)}}{\sqrt{3}}\right)+i\, v
   \label{eq:zv-theta}
\end{eqnarray}
which generalizes \eqref{eq:sd-0} to  $0\leq \theta <\frac{2\pi}{3}$. When $\theta=\frac{2\pi}{3}$ there is a Stokes jump and now we need to deform the original contour into the union of {\it two} steepest descent contours, which are best parametrized in terms of $u$. This involves  solving the cubic equation \eqref{eq:theta-condition} for $v(u)$, so the corresponding expressions are awkward and not particularly illuminating.
This exercise makes it clear that for more complicated action functions this approach is  impractical, and likewise it is unfeasible for higher dimensional integrals, let alone infinite dimensional QM or QFT path integrals. This motivates a different approach, that of generating Lefschetz thimbles by gradient flow methods.

\subsubsection{Steepest Descent Contours and Gradient Flow}
\label{sec:gradient}

\begin{figure}[h!]
\centerline{\includegraphics[scale=0.6]{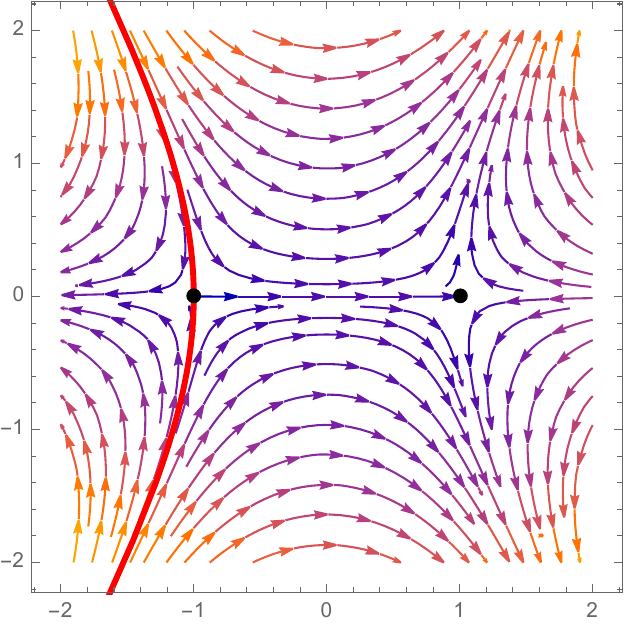}\quad
\includegraphics[scale=0.6]{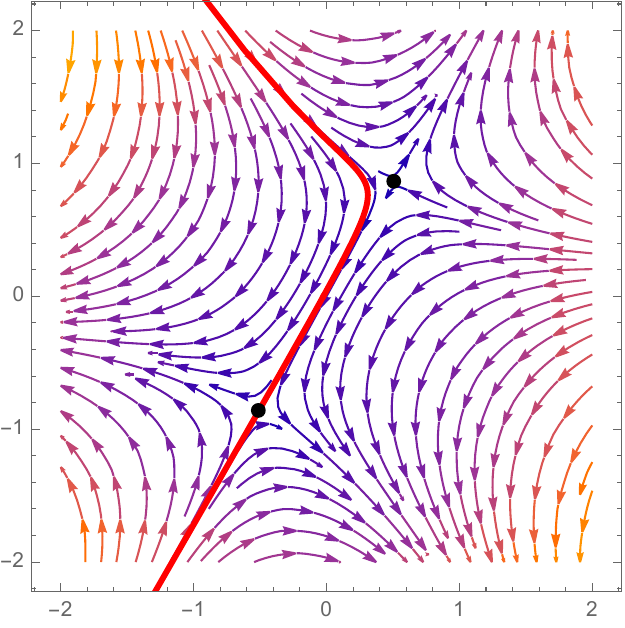}}
\centerline{\includegraphics[scale=0.6]{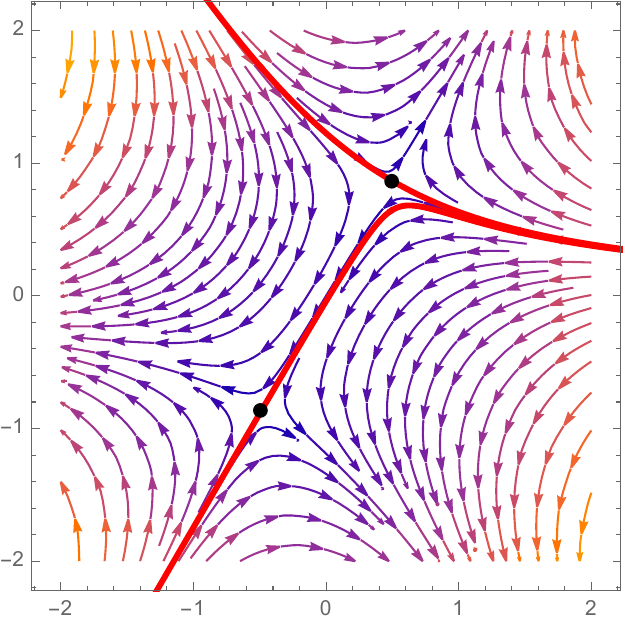}\quad
\includegraphics[scale=0.6]{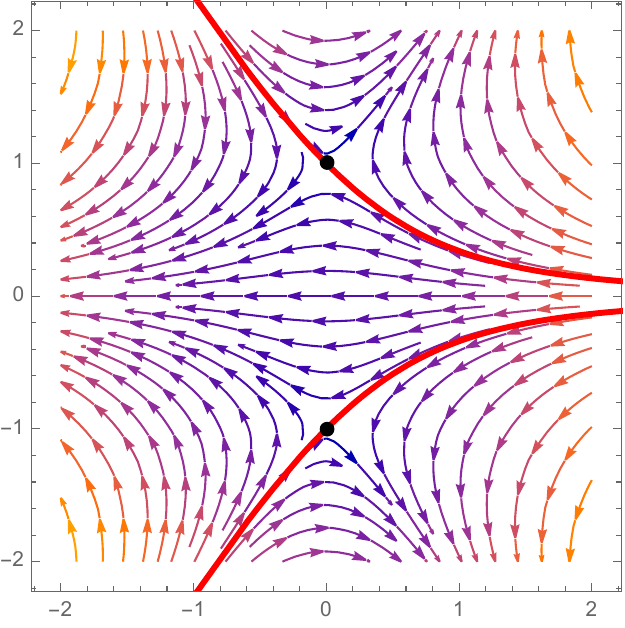}}
\caption{The Airy thimble contours (red solid lines) through the saddle points (black dots), and the stream plot of the vector field $\overline{\vec{\nabla} S}$. The plots are shown for $\theta=0, \frac{99}{100}\frac{2\pi}{3} , \frac{101}{100}\frac{2\pi}{3} , \pi$, from top left to bottom right, where $\theta={\rm arg}(x)$. As $\theta$ changes the saddles move and the thimble contours deform. There is a Stokes jump at $\theta=\frac{2\pi}{3}$, between the 2nd and 3rd plots.}
\label{fig:airy-thimbles}
\end{figure}
To overcome the branch limitations mentioned in the last section, we reformulate the problem in terms of gradient flow. Introduce a ``flow time'' $\tau$, 
and solve the following complex gradient-flow problem
\begin{eqnarray}
\frac{dz}{d\tau}= -\overline{\left(\frac{\partial S}{\partial z}\right)}
\label{eq:th1}
\end{eqnarray}
where the overbar denotes the complex conjugate. These evolution equations imply that the imaginary part of $S(z(\tau))$ is preserved by the evolution with $\tau$, and the real part of $S(z(\tau))$ changes monotonically:
\begin{eqnarray}
\frac{d}{d\tau} \left(S-\overline{S}\right) = \frac{\partial S}{\partial z}\, \dot{z}-\overline{\left(\frac{\partial S}{\partial z}\right)}\, \dot{\bar{z}}
=0
\label{eq:s-imag}
\end{eqnarray}
\begin{eqnarray}
\frac{d}{d\tau} \left(S+\overline{S}\right) = \frac{\partial S}{\partial z}\, \dot{z}+\overline{\left(\frac{\partial S}{\partial z}\right)}\, \dot{\bar{z}}
= -2\left | \frac{\partial S}{\partial z}\right |^2 
<0
\label{eq:s-real}
\end{eqnarray}
For each saddle, these flow equations define a ``Lefschetz thimble'', $J_{\rm saddle}$,  and a ``dual thimble'', $K_{\rm saddle}$:
\begin{eqnarray}
J_{\rm saddle} &=&\left\{ z(\tau=0)\in \mathbb C\,\, \Big | \lim_{\tau\to -\infty}z(\tau)=z_{\rm saddle} \right\}
\label{eq:thimble}
\\
K_{\rm saddle} &=&\left\{ z(\tau=0)\in \mathbb C\,\, \Big | \lim_{\tau\to +\infty}z(\tau)=z_{\rm saddle} \right\}
\label{eq:dual-thimble}
\end{eqnarray}
For example, for the Airy function we write $z=u+i\, v$ as before, and recalling that $S(z; \theta)=e^{i\,\theta}\, z-\frac{1}{3}\, z^3$, we obtain two coupled ODEs for $u(\tau)$ and $v(\tau)$:
\begin{eqnarray}
\dot{u}&=&\cos\theta-u^2+v^2
\label{eq:udot}
\\
\dot{v}&=& -\sin\theta+2\,u\, v
\label{eq:vdot}
\end{eqnarray}
Figure \ref{fig:airy-thimbles} shows the $(u,v)$ plane and the steepest descent contours (red solid lines) through the saddle points $z_\pm=\pm e^{i\theta/2}$ (black dots), and the stream plot of the vector field $\overline{\vec{\nabla} S}$. The plots are shown for $\theta=0, \frac{99}{100}\frac{2\pi}{3} , \frac{101}{100}\frac{2\pi}{3} , \pi$, from top left to bottom right. As $\theta$ changes the saddles move and the thimble contours deform. There is a Stokes jump at $\theta=\frac{2\pi}{3}$, as the original contour is deformed into two steepest descent contours.

This gradient flow approach generalizes to multi-dimensions and under certain circumstances to infinite dimensions \cite{witten,alexandru}.

\section{Resurgence and the Nonlinear Stokes Phenomenon}
\label{sec:nonlinear-stokes}

In quantum mechanics (QM) and quantum field theory (QFT), we typically confront problems with an {\it infinite} number of saddle points; for example, an infinite order ``multi-instanton expansion". This leads to the {\it nonlinear Stokes phenomenon}.  In a nonlinear problem, when an exponentially suppressed nonperturbative term appears in a transseries expansion, nonlinearities will generically produce higher powers of it, leading to an infinite-order transseries. To introduce the idea, we consider a nonlinear generalization of the Airy equation, a special case of the Painlev\'e II equation:\footnote{Note that integrability is not important for resurgence. However, many examples of resurgence in the literature have been studied using integrable ODEs because integrable ODEs occur in many physical contexts, and also have rich mathematical structure.}
\begin{eqnarray}
y^{\prime\prime}=x\, y(x) {\color{Red} + 2\, y^3(x)}
\label{eq:p2}
\end{eqnarray}
The Painlev\'e equations play an important role in the solution of matrix models \cite{marcos-lectures}, which in some sense bridge the gap between QM and QFT.

The Painlev\'e transcendents can be viewed as {\it "nonlinear special functions"} \cite{clarkson}, and they are as prevalent in nonlinear problems as the familiar {\it "linear special functions"} (Bessel, Parabolic Cylinder, etc...) are in linear problems. Indeed, there is a hierarchy that matches the hierarchy of linear special functions (\href{https://dlmf.nist.gov/32.2.vi}{dlmf/32.2.vi}). 
Painlev\'e and collaborators classified all second order nonlinear ODEs with the property that 
the only moveable singularities (i.e., those associated with boundary conditions) are poles. This classification leads to a  set of six Painlev\'e equations, with solutions known as  Painlev\'e transcendents. Painlev\'e II is particularly interesting and illustrative, as it is the integrable nonlinear generalization of the Airy equation, which describes universal behavior near a linear turning point. Painlev\'e II is {\it universal} in both physics and mathematics \cite{clarkson,tracy-widom} in the sense that it describes the behavior near phase transitions in many nonlinear problems: double-scaling limit in unitary matrix models, 2d Yang-Mills and 2d supergravity; non-intersecting brownian motions; correlators in polynuclear growth; directed polymers; Tracy-Widom law for statistics of maximum eigenvalue for Gaussian random matrices; longest increasing subsequence in random permutations; ... . 

{\color{Blue} 
\begin{exercise} 
\label{ex:2.1}
 {\bf Meromorphic Expansion of Painlev\'e II Solution.} 
 
 \begin{enumerate}
 
 \item
 Show that the general Painlev\'e II solution to \eqref{eq:p2} has a meromorphic expansion with {\color{Blue} only poles for moveable singularities} (those associated with boundary conditions):
 \begin{eqnarray}
 \hskip -.5cm y(x)=\frac{1}{x-{\color{Red}x_0}}-\frac{x_0}{6}(x-x_0)-\frac{1}{4}(x-x_0)^2+{\color{Red}h_0}(x-x_0)^3+\frac{x_0}{72}(x-x_0)^4+\dots
\label{eq:p2poles}
\end{eqnarray}
where all coefficients are expressed in terms of the pair {\color{Red} $(x_0, h_0)$}, for any pole $x_0$.

\item
Change the nonlinearity of the equation from $y^3(x)$ to $y^4(x)$ and show that the Painlev\'e integrability condition fails (comment: nevertheless, despite being nonintegrable,  resurgent transseries analysis still holds \cite{cleri}).
\end{enumerate}
\end{exercise}
}

In numerical analysis of the PII equation, it was observed that there is a special separatrix solution known as the Hastings-McLeod solution \cite{hm}, which is exponentially sensitive to the boundary conditions. This solution arises frequently in physical applications. We discuss one such physical example in section \ref{sec:gww}.

\begin{figure}[h!]
 \centerline{
\includegraphics[scale=1]{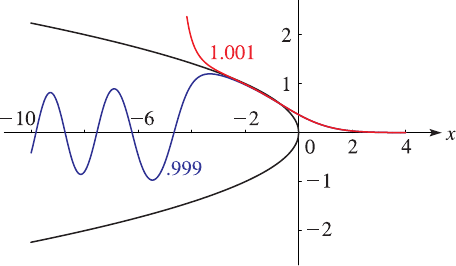}}
 \caption{Plots of the solution to the Painlev\'e II equation $y''=x\, y+2y^3$. The red curve shows the solution satisfying $y(x)\sim 1.001 {\rm Ai}(x)$ as $x\to +\infty$, and the blue curve shows the solution satisfying $y(x)\sim 0.999 {\rm Ai}(x)$ as $x\to +\infty$. The black curve is the separatrix $\sqrt{-x/2}$. The Hastings-McLeod solution hugs the separatrix as $x\to -\infty$, and  $y_{\rm HM}(x)\sim  {\rm Ai}(x)$ as $x\to +\infty$. From \url{https://dlmf.nist.gov/32.3.ii}.}
 \label{fig:hm}
 \end{figure}

Since ${\rm Ai}(x)$ is exponentially small as $x\to +\infty$, it is natural to seek a solution that behaves like the Airy solution for large $x$, because the $y(x)^3$ term in \eqref{eq:p2} is negligible. However, as $x\to -\infty$ we can seek a smooth solution in which the second derivative is small, by balancing the two terms on the RHS of \eqref{eq:p2}, leading to $y(x)\sim \sqrt{-x/2}$. The Hastings-McLeod solution is the unique real solution on $\mathbb R$ that matches ${\rm Ai}(x)$ asymptotics as $x\to+\infty$ with $\sqrt{-\frac{x}{2}}$ asymptotics as $x\to -\infty$. 
See Fig. \ref{fig:hm}. 
 
 The Hastings-McLeod solution also has the interesting property that the poles of the solution $y_{\rm HM}(x)$  are restricted to two $\pi/3$ wedges, centered on the positive and negative imaginary axis. This will be important below: see section \ref{sec:gww-stokes}.

\subsection{Painlev\'e II Transseries as $x\to+\infty$.}
\label{sec:p2-pos}

To study the $x\to +\infty$ regime, we can write \eqref{eq:p2} as an exact integral equation:
\begin{eqnarray}
y(x)={\color{blue} \sigma_+}\, {\rm Ai}(x)+2\pi \int_x^\infty dz\, y^3(z)\left[{\rm Ai}(x) \, {\rm Bi}(z)-{\rm Ai}(z) \, {\rm Bi}(x)\right]
\label{eq:p2-integral}
\end{eqnarray}
Here $\sigma_+$  arises as the coefficient of the homogeneous linearized equation (the Airy equation). 
Iterating this integral equation leads immediately to a transseries solution
\begin{eqnarray}
y_+(x)\sim \sum_{n=0}^\infty \sigma_+^{2n+1} \, Y_{[2n+1]}(x)\qquad, \quad x\to+\infty
\label{eq:p2-transseries}
\end{eqnarray}
where $\sigma_+$ plays the role of the {\it "transseries parameter"}.
The first few terms are:
\begin{eqnarray}
Y_{[1]}(x)&=& {\rm Ai}(x)
\label{eq:int-sols1} \\
Y_{[3]}(x)&=& 2\pi \left({\rm Ai}(x) \int_x^\infty {\rm Ai}^3(z) {\rm Bi}(z)\, dz -{\rm Bi}(x) \int_x^\infty {\rm Ai}^4(z) \, dz\right)
\label{eq:int-sols2}\\
Y_{[5]}(x)&=& 6\pi \left({\rm Ai}(x) \int_x^\infty Y_{[3]}(z) \left(Y_{[1]}(z)\right)^{2} {\rm Bi}(z)\, dz \right.
\nonumber\\
&&\left. \qquad 
-{\rm Bi}(x) \int_x^\infty Y_{[3]}(z)  \left(Y_{[1]}(z)\right)^{2} {\rm Ai}(z) \, dz\right)\label{eq:int-sols}
\label{eq:int-sols3}
\end{eqnarray}
Counting the $x\to +\infty$ exponential factors associated with the factors of Ai and Bi, we see that $Y_{[2k+1]}(x)\sim \left(\frac{e^{-\frac{2}{3}\, x^{3/2}}}{2\,\sqrt{\pi}\, x^{1/4}}\right)^{2k+1}$, so the transseries expansion \eqref{eq:p2-transseries} takes the form
\begin{eqnarray}
y_+(x)\sim \sum_{k=0}^\infty \left({\color{Red} \frac{\sigma_+ \, e^{-\frac{2}{3}\, x^{3/2}}}{2\,\sqrt{\pi}\, x^{1/4}}}\right)^{2k+1} {\mathcal F}_{[2k+1]}(x)
 \qquad, \quad x\to +\infty
\label{eq:p2plus}
\end{eqnarray}
This is an expansion in odd powers of an ``Airy instanton factor'', 
with each ``instanton order'' $(2k+1)$ multiplied by a fluctuation term $ {\mathcal F}_{[2k+1]}(x)$, with coefficients $c_n^{[2k+1]}$:
\begin{eqnarray}
 {\mathcal F}_{[2k+1]}(x) \sim \sum_{n=0}^\infty \frac{c_n^{(2k+1)}}{x^{3n/2}} 
 \qquad, \quad x\to +\infty
 \label{eq:p2plus-fluc}
 \end{eqnarray}
 The transseries parameter $\sigma_+$ will be fixed by matching to a prescribed behavior in the $x\to -\infty$ regime.
 
\subsection{Painlev\'e II Transseries as $x\to -\infty$.}
\label{sec:p2-neg}

Now consider the opposite direction: $x\to -\infty$. Here we seek a
formal series solution
\begin{eqnarray}
y_-(x)\sim {\color{blue} \sqrt{\frac{-x}{2}}} \left(1
-\frac{1}{8(-x)^{\color{Red}3}}-\frac{73}{128 (-x)^{\color{Red}6}}-\frac{10567}{1024 (-x)^{\color{Red}9}} - \dots\right)
\label{eq:p2minus}
\end{eqnarray}
The numerical coefficients in \eqref{eq:p2minus} are generated by a  recursion formula. They grow factorially in magnitude, and all beyond the first term are negative. See Exercise \ref{ex:2.2}.  

There is no free parameter in \eqref{eq:p2minus}. This implies that \eqref{eq:p2minus} cannot be the full solution. Something is missing!  These are  the nonperturbative corrections.
To find these missing terms, we use the following simple method \cite{costin-book}. Write the solution as 
\begin{eqnarray}
y(x)=y_{\rm pert}(x)+ \epsilon \, y_{\rm nonpert}(x)
\label{eq:ynp}
\end{eqnarray}
where $y_{\rm nonpert}(x)$ is a nonperturbative term, beyond all perturbative orders, and $\epsilon$ is a small parameter that {\it grades} this nonperturbative order. Linearizing the PII equation with this ansatz, the $O(\epsilon^0)$ term produces the  perturbative expansion in \eqref{eq:p2minus}, $y_{\rm pert}=y_-$, while the $O(\epsilon^1)$ term yields:
\begin{eqnarray}
y_{\rm nonpert}^{\prime\prime} &=& \left(6\, y_{\rm pert}^{2}+x\right)y_{\rm nonpert}
\\
& \sim&  \left(-2\, x-\frac{3}{4x^2} +\dots \right)y_{\rm nonpert}
\qquad, \qquad x\to -\infty
\label{eq:p2np}
\end{eqnarray}
This is a {\it homogeneous linear} equation, which is solved by an exponential ansatz: 
$y_{\rm nonpert}(x)\sim (-x)^\beta e^{-\gamma (-x)^{\frac{3}{2}}} \left(1+\dots\right)$.
Matching terms we find 
\begin{equation}
y_{\rm nonpert}(x)\sim \frac{{\color{blue}\sigma_-}}{(-x)^{1/4}} e^{-{\color{Red} \sqrt{2}}\, \frac{2}{3} (-x)^{3/2}}\left(1-\frac{{\color{blue} \frac{17}{72}}}{{\color{Red}\sqrt{2}}\, \frac{2}{3} (-x)^{3/2}}
+\frac{{\color{blue} \frac{1513}{10368}}}{({\color{Red}\sqrt{2}}\, \frac{2}{3} (-x)^{3/2})^2}
-\dots \right)
\label{eq:p2npsol}
\end{equation}
The ${\color{Red}\sqrt{2}}$ factor in the exponent  is not a misprint! It arises because the leading term in \eqref{eq:p2np} is an Airy equation with a factor of -2: $y_{\rm nonpert}^{\prime\prime} \sim -2\, x\, y_{\rm nonpert}$. We comment further on this $\sqrt{2}$ factor later.

Just like for the {\it perturbative} series solution, $y_{-}(x)$ in \eqref{eq:p2minus}, for the nonperturbative solution $y_{\rm nonpert}(x)$ there is no closed formula for the fluctuation coefficients  (in blue in \eqref{eq:p2npsol}), but they are generated by a simple recursion formula, up to the overall coefficient $\sigma_-$. 

Returning to the full nonlinear Painlev\'e II equation, we see that as $x\to -\infty$ this solution generates a transseries solution of the form
\begin{eqnarray}
y_-(x)\sim \sqrt{\frac{-x}{2}} \sum_{k=0}^\infty \left(\frac{\sigma_-\, e^{-\frac{2{\color{Red} \sqrt{2}}}{3} (-x)^{3/2}}}{2\sqrt{\pi}\, (-x)^{1/4}}\right)^{k} {\mathcal Y}_{[k]}(x)
\qquad, \qquad x\to -\infty
\label{eq:p2neg}
\end{eqnarray}
Remarkably, we find a low order/large order resurgence relation connecting the perturbative expansion \eqref{eq:p2minus} and the fluctuation factor in \eqref{eq:p2npsol}. The large order behavior of the perturbative expansion coefficients in \eqref{eq:p2minus}  can be written as decreasing factorials
\begin{eqnarray}
 c_n^{(0,-)}  \sim  -\frac{1}{\pi}\sqrt{\frac{2}{3\pi}}
\frac{\Gamma\left(2n-\frac{1}{2}\right)}{ \left(\frac{2{\color{Red} \sqrt{2}}}{3}\right)^{2n}}
\left(1-\frac{{\color{blue} \frac{17}{72}}}{(2n-\frac{3}{2})} +\frac{{\color{blue} \frac{1513}{10368}}}{(2n-\frac{3}{2})(2n-\frac{5}{2})}- \dots\right)
\label{eq:cl3}
\end{eqnarray}
Compare with the fluctuations around the first nonperturbative exponential in \eqref{eq:p2npsol}:
\begin{itemize}
\item
The $2\sqrt{2}/3$ factor in the nonperturbative exponent in \eqref{eq:p2npsol} appears in the large order growth \eqref{eq:cl3} of the perturbative coefficients from \eqref{eq:p2minus}.
\item
The subleading  fluctuation coefficients in \eqref{eq:p2npsol} agree with the subleading  coefficients in the large-order growth of the perturbative coefficients in \eqref{eq:p2minus}-\eqref{eq:cl3}.
\end{itemize}
{\color{Blue} 
\begin{exercise}
\label{ex:2.2}
{\bf Resurgence in Nonlinear ODEs -- Painlev\'e II.}
\begin{enumerate}

\item
Generate many terms, and verify the leading large order behavior of the coefficients of the formal $x\to -\infty$ series, $y_{-}(x)$,  for the Painlev\'e II Hastings-McLeod solution in \eqref{eq:p2minus}.

 \item
Numerically identify the Stokes constant in  \eqref{eq:cl3} to high precision
$$0.1466323...= \frac{1}{\pi}\sqrt{\frac{2}{3\pi}}$$

\item
Confirm the large-order/low-order resurgence relation between \eqref{eq:p2npsol} and  \eqref{eq:cl3}.

\end{enumerate}
\end{exercise}
}

\subsection{Nonlinear Stokes Phenomenon for Painlev\'e II}.
\label{sec:nls-p2}

Our remaining task is to match the $x\to +\infty$ solution in \eqref{eq:p2plus} to the $x\to -\infty$ solution in \eqref{eq:p2neg}. 
Each of the two solutions, $y_\pm(x)$, has the form of a transseries, but note that they have a very different structure, including a different exponential factor (``instanton action''). Furthermore, each involves a single {\it ``transseries parameter''}, $\sigma_\pm$. The result of this connection formula is that the two solutions match smoothly iff
\begin{eqnarray}
{\rm Hastings-McLeod\,\, solution} \quad : \qquad \sigma_+=\sigma_-=1
\label{eq:p2hm}
\end{eqnarray}
Physically, as $x\to 0^\pm$ we need all orders of the instanton expansion of \eqref{eq:p2plus} and all orders of the instanton expansion of \eqref{eq:p2minus} in order to match near $x\approx 0$. Moreover, we require $\sigma_+=\sigma_-=1$. This is an explicit example of the phenomenon of instanton condensation, or {\it transmutation of a transseries} \cite{gww}: the single instanton solution on either side breaks down at small $|x|$, where the instanton exponentials, $e^{-\frac{2}{3}\, x^{3/2}}$ and $e^{-\frac{2\sqrt{2}}{3}\, (-x)^{3/2}}$, are no longer small. Approaching $x=0$ from either side all the instanton terms resum into a different instanton expansion on the other side. 

In fact, as in the Airy equation, we should really view this transmutation of the transseries {\it in the complex $x$ plane}: this is the method of {\it transasymptotics} \cite{costin-book}. At large $|x|$, as the phase approaches $\frac{\pi}{3}$, the single instanton approximation breaks down and when we cross this anti-Stokes line the transseries turns into the meromorphic  pole expansion in \eqref{eq:p2poles}. Rotating further, at phase  $\frac{2\pi}{3}$ we exit the pole region and the solution transmutes again into a transseries that is different from the one in the region $|{\rm arg}(x)|\leq \frac{\pi}{3}$.
This is the nonlinear Stokes phenomenon: a singularity in the Borel plane is repeated in all integer multiples, so when rotating in the complex plane we pick up an infinite number of nonperturbative terms, which must be resummed for consistent global matching. Yet another way to understand this connection formula is that the condition $\sigma_\pm=1$ pushes all the poles of the solution to infinity except for a $\pi/3$ wedge centered on each of the positive and negative imaginary axes.  

In the next section we will see a physical example of a phase transition associated with the nonlinear Stokes phenomenon for this Hastings-McLeod solution.

\subsection{The Gross-Witten-Wadia Unitary Matrix Model}
\label{sec:gww}

The Gross-Witten-Wadia (GWW) unitary matrix model is equivalent to  2  dimensional  $U(N)$ lattice gauge theory, which can be reduced to $N^2$ copies of a single plaquette \cite{gw,wadia}. This model illustrates the basic features of many matrix models \cite{marcos-lectures}: large $N$ expansions, phase transitions, double-scaling limits, ... . 

The GWW partition function is an integral over   $N\times N$ unitary matrices $U$:
\begin{eqnarray}
Z(t, N)=\int_{U(N)} DU \, \exp\left[\frac{N}{t} {\rm tr}
\left(U+U^\dagger\right)\right]
\label{eq:gww1}
\end{eqnarray}
Here $t=N g^2$ is the 't Hooft coupling. In the limit $g^2\to 0$ and $N\to \infty$, with $t$ fixed, the system has a 3rd order phase transition, at the critical 't Hooft coupling, $t_c=1$, developing a kink in the specific heat (per degree of freedom) \cite{gw,wadia}. 

The partition function \eqref{eq:gww1} can be expressed as an $N\times N$ determinant with entries being modified Bessel functions:
\begin{eqnarray}
Z(t, N)=\det\left[I_{j-k}\left(\frac{N}{t}\right)\right]_{j, k=1 ...N}
\label{eq:gww2}
\end{eqnarray}
Large determinants are complicated, so an alternative approach is to use an {\it ``order parameter''}, $\Delta(t, N) \equiv \langle \det U \rangle$, which satisfies a nonlinear ODE of Painlev\'e type \cite{rossi,gww}:
\begin{eqnarray}
t^2 \Delta^{\prime\prime} +t \Delta^\prime + \frac{{\color{Red} N^2} \Delta}{t^2}\left(1- \Delta^2 \right)  = \frac{\Delta}{1-\Delta^2}\left({\color{Red} N^2} -t^2 \left(\Delta^\prime\right)^2\right)
\label{eq:rossi}
\end{eqnarray}
This is a Painlev\'e III equation (in Okamoto form) \cite{gww}. We can therefore use transseries methods similar  to those for PII in the previous section. Notice that $N^2$ is merely a parameter in this equation \eqref{eq:rossi}, and it plays a role similar to $1/\hbar^2$ in the Schr\"odinger equation. Physically, $N^2\to \infty$ is like a thermodynamic limit, or a semiclassical limit. Hence we expect that a formal large $N$ series, in powers of $1/N^2$, would receive a nonperturbative completion resulting in the transseries form
\begin{eqnarray}
\Delta(t, N)=\sum_n \frac{c_n^{(0)}(t)}{N^{2n}} {\color{Red} + e^{-N\, S(t)} \sum_n \frac{c_n^{(1)}(t)}{N^n} 
+ e^{-2N\, S(t)} \sum_n \frac{c_n^{(2)}(t)}{N^n}  +\dots }
\label{eq:gww-delta}
\end{eqnarray}
Given a solution $\Delta(t, N)$, there is a direct mapping to the partition function \cite{rossi,gww}, and hence to other physical thermodynamic quantities such as the free energy and specific heat. These therefore inherit the same basic transseries structure as in \eqref{eq:gww-delta}.

\subsubsection{Strong 't Hooft Coupling Regime of the GWW Model}
\label{sec:gww-strong}

In the strong coupling regime, $t>1$, one finds that $\Delta(t, N)$ is exponentially small, so we linearize the Rossi equation \eqref{eq:rossi}:
\begin{eqnarray}
t^2 \Delta^{\prime\prime} +t \Delta^\prime + \frac{N^2}{t^2}\left(1-t^2\right)  \Delta \approx 0
\label{eq:rossi-strong}
\end{eqnarray}
This is a homogeneous linear ODE, a Bessel equation, with solution $J_N\left(\frac{N}{t}\right)$ decreasing at large $t$.\footnote{This is an analog of the linearization of PII to the Airy equation as $x\to +\infty$, discussed in Section \ref{sec:p2-pos}. This will be important for the GWW double-scaling limit discussed in Section \ref{sec:gww-stokes}.}
The large $N$ Debye asymptotics (\href{https://dlmf.nist.gov/10.19.ii}{dlmf/10.19.ii}) yields 
\begin{eqnarray}
\Delta(t, N)&\approx& {\color{blue} \sigma_{\rm strong}}\, J_N\left(\frac{N}{t}\right)
\\
 &\sim& {\color{blue} \sigma_{\rm strong}}\, \frac{\sqrt{t}\, e^{-N S_{\rm strong}(t)}}{\sqrt{2\pi N}\, (t^2-1)^{1/4}}  \sum_{n=0}^{\infty}\frac{U_n\left(t\right)}{N^{n}}+\dots 
\nonumber
\end{eqnarray}
with the strong-coupling large $N$ instanton action
\begin{eqnarray}
S_{\rm strong}(t) = {\rm arccosh(t)} - \sqrt{1-\frac{1}{t^2}}
\label{eq:gww-strong}
\end{eqnarray}
The fluctuation factors are formal series in $1/N$ with coefficients $U_n(t)$ that are functions of the 't Hooft coupling $t$,  generated recursively. Iterating with the nonlinear terms in \eqref{eq:rossi}, we generate an infinite-order transseries with all odd powers of the instanton factor
\begin{eqnarray}
\left( \sigma_{\rm strong} \frac{e^{-N S_{\rm strong}(t)}}{\sqrt{t\, S_{\rm strong}^\prime (t)}}\right)
 \label{eq:gww-strong-instanton}
 \end{eqnarray}
 each multiplied by a fluctuation series in powers of $1/N$, with coefficients that are functions of $t$.

{\color{Blue}
\begin{exercise}
\label{ex:2.3}
{\bf Transseries for $\Delta(t, N)$:}\\
 Consider the  GWW model Rossi equation \eqref{eq:rossi} (Painlev\'e III in Okamoto form):
\begin{eqnarray}
t^2 \Delta^{\prime\prime} +t \Delta^\prime + \frac{{\color{Red} N^2} \Delta}{t^2}\left(1- \Delta^2 \right)  = \frac{\Delta}{1-\Delta^2}\left({\color{Red} N^2} -t^2 \left(\Delta^\prime\right)^2\right)
\nonumber
\end{eqnarray}
\begin{enumerate}
\item
Show that in the $t>1$ region this equation linearizes to
\begin{eqnarray}
t^2 \Delta^{\prime\prime} +t \Delta^\prime + \frac{N^2}{t^2}\left(1-t^2\right)  \Delta \approx 0
\nonumber
\end{eqnarray}
solved by the Bessel functions $J_N\left(\frac{N}{t}\right)$ and $Y_N\left(\frac{N}{t}\right)$.

\item
Derive the leading  large $N$ Debye asymptotics by
solving this Bessel equation using the Liouville-Green substitution: $\Delta= \frac{e^{-N\, S(t)}}{\sqrt{t\, S^\prime(t)}}$.
\end{enumerate}
\end{exercise}
}

\subsubsection{Weak 't Hooft Coupling Regime of the GWW Model}
\label{sec:gww-weak}

At  weak coupling, $t<1$, the large $N$ limit produces a different kind of transseries. From \eqref{eq:rossi}, balancing the $N^2$ terms produces an {\it algebraic equation} with $\Delta(t, N)\sim \sqrt{1-t}$, so we generate a formal  perturbative expansion
\begin{eqnarray}
 \Delta(t, N) \sim 
	\sqrt{1-t}\sum_{n = 0}^\infty \frac{d^{(0)}_n(t)}{N^{2n}} 
	\qquad, \quad N\to\infty \quad (t<1)
	\label{eq:gww-weak-pert}
\end{eqnarray}
The coefficients $d_n^{(0)}(t)$ are functions of $t$, determined recursively in the index $n$.
As for the PII Hastings-McLeod solution in \eqref{eq:p2minus} on the $x<0$ axis, this  solution is a formal asymptotic expansion  with no free parameter, indicating that something is missing. This leads to a nonperturbative completion via a transseries expansion
\begin{eqnarray}
 \Delta(t, N) \sim 
	\sqrt{1-t}\sum_{n = 0}^\infty \frac{d^{(0)}_n(t)}{N^{2n}} 
	{\color{Red} -\frac{\sigma_{\rm weak} }{2\sqrt{2\pi N}}\,  \frac{t\, e^{- N S_\text{weak}(t)}}{(1-t)^{1/4}} \sum_{n = 0}^\infty \frac{d^{(1)}_n(t)}{N^n} +\dots}
	\label{eq:gww-weak-trans}
\end{eqnarray}
Inserting this ansatz into the Rossi equation \eqref{eq:rossi} determines the weak-coupling large $N$ instanton action
\begin{eqnarray}
S_\text{weak}(t) = \frac{2 \sqrt{1-t}}{t}-2\, {\rm arctanh}\left(\sqrt{1-t}\right)
\label{eq:gww-weak}
\end{eqnarray}
Iterating with the nonlinear terms in \eqref{eq:rossi}, we generate an infinite order transseries containing powers of the instanton factor
\begin{eqnarray}
\left( \sigma_{\rm weak} \frac{e^{-N S_{\rm weak}(t)}}{\sqrt{t\, S_{\rm weak}^\prime (t)}}\right)
 \label{eq:gww-weak-instanton}
 \end{eqnarray}
 each multiplied by a fluctuation series in powers of $1/N$. Note that all the fluctuation coefficients $d_n^{(k)}(t)$ are nontrivial functions of the 't Hooft coupling $t$.
 
 The transseries structure in \eqref{eq:gww-weak-trans} exhibits the generic resurgent large-order/low-order duality introduced in Section \ref{sec:airy}. To see this, one can verify that the large order growth of the perturbative coefficients $d_n^{(0)}(t)$ is
 \begin{equation}
 d^{(0)}_n(t) \sim  \frac{-1}{\sqrt{2} (1-t)^{3/4} \pi^{3/2}}    \frac{\Gamma(2n-\frac{5}{2})}
		{{\color{blue} (S_{\rm weak}(t))^{2n-\frac{5}{2}}}}
		\left[1+{\color{Red}\frac{(3 t^2-12 t-8)}{96 (1-t)^{3/2}}} \frac{{\color{blue} S_{\rm weak}(t)}}{(2n-\frac{7}{2})}+\dots\right]
\label{eq:gww-weak-large}
\end{equation}
This should be compared with the low orders of the expansion about the leading instanton term in \eqref{eq:gww-weak-trans}, which is generated directly by the Rossi equation \eqref{eq:rossi}:
\begin{eqnarray}
\sum_{n = 0}^\infty \frac{d^{(1)}_n(t)}{N^n} = 1 +{\color{Red} \frac{(3 t^2-12 t-8)}{96 (1-t)^{3/2}} } \frac{1}{N} 
+\dots 
\label{eq:gww-weak-fluc}
\end{eqnarray}
Notice the large-order/low-order relation between \eqref{eq:gww-weak-large} and \eqref{eq:gww-weak-fluc}. In particular, observe that this resurgent relation is a relation not just between numbers, but between {\it nontrivial functions} of the 't Hooft parameter $t$ (for all $t<1$). This is known as {\it parametric resurgence}, as it arises from an expansion in terms of the {\it parameter} $N$ appearing in the ODE \eqref{eq:rossi}, rather than in terms of the {\it variable} $t$ of the ODE.
{\color{Blue}
\begin{exercise}
\label{ex:2.4}
{\bf Resurgence relation in the weak coupling GWW Model} 
\begin{enumerate}

\item Generate many coefficients $d_n^{(0)}(t)$ of the formal large $N$ solution for $\Delta(t, N)$ in the small $t$ regime
$$
\Delta(t, N)  \sim \sqrt{1-t}\sum_{n=0}^\infty \frac{d_n^{(0)}(t)}{N^{2n}}
$$

\item
Derive the form of the first nonperturbative correction to this formal large $N$ expansion, including the first few fluctuation corrections:
$$
\Delta_{NP}(t, N)\sim f(t)\,  e^{-N \, S_{\rm weak}(t)} \sum_{n=0}^\infty \frac{d_n^{(1)}(t)}{N^{n}} +\dots
$$
\item
Show that the subleading  large $n$ growth corrections of the coefficient functions $d_n^{(0)}(t)$ are associated with the expansion terms $d_n^{(1)}(t)$ of the first nonperturbative correction to the formal large $N$ expansion in \eqref{eq:gww-weak-large}-\eqref{eq:gww-weak-fluc}.

\end{enumerate}
\end{exercise}
}

\subsubsection{The GWW Phase Transition as a Nonlinear Stokes Transition}
\label{sec:gww-stokes}

\begin{figure}[h!]
\centerline{\includegraphics[scale=0.2]{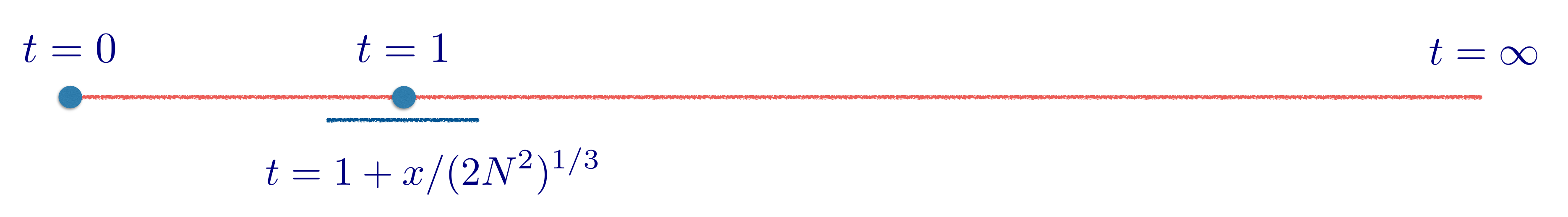}}
\caption{The vicinity of the GWW phase transition is probed by a double-scaling limit.}
\label{fig:gww-ds}
\end{figure}

We have shown that in the thermodynamic limit ($N\to\infty$) the order parameter $\Delta(t, N)$ is expanded in a large $N$ transseries, which is {\it different} at weak coupling ($t<1$) and strong coupling ($t>1$). The transition at $t=1$ can be understood in terms of a nonlinear Stokes phenomenon, where the transseries is governed by different saddle points on either side of the transition. The vicinity of the phase transition can be probed more finely by considering a {\it double-scaling} limit, in which we express the behavior near $t=1$ as (see Figure \ref{fig:gww-ds})
\begin{eqnarray}
t\sim 1+\frac{x}{(2N^2)^{1/3}} \qquad;\qquad \Delta(t, N)=\left(\frac{2t}{N}\right)^{1/3}\,  y(x)
\label{eq:gww-ds}
\end{eqnarray}
The scaling factors are determined by the Rossi equation \eqref{eq:rossi}, and are consistent with the coalescence of PIII to PII (\href{https://dlmf.nist.gov/32.2.vi}{dlmf/32.2.vi}). It also matches  the uniform scaling reduction of Bessel functions to the Airy function:
\begin{eqnarray}
\lim_{N\to\infty} J_N(N-N^{1/3} \, x) = \left(\frac{2}{N}\right)^{1/3} {\rm Ai}\left(2^{1/3} \, x \right)
\label{eq:gww-ds2}
\end{eqnarray}
In the double-scaling limit \eqref{eq:gww-ds} the Rossi equation \eqref{eq:rossi} reduces to the PII eqn 
$ y'' =x y+2 y^3$.
Furthermore, approaching the phase transition from above, $t\to 1^+$,
\begin{eqnarray}
N\, S_{\rm strong}(t) \sim N\, \frac{2\sqrt{2}}{3} (t-1)^{3/2} \sim N\, \frac{2\sqrt{2}}{3} \frac{x^{3/2}}{\sqrt{2}\, N} = \frac{2}{3}\, x^{3/2}
\label{eq:gww-ds-strong}
\end{eqnarray}
On the other hand,  approaching the phase transition from below,  $t\to 1^-$,
\begin{eqnarray}
N\, S_{\rm weak}(t) \sim N\,\frac{4}{3} (1-t)^{3/2}  \sim N\, \frac{4}{3} \frac{(-x)^{3/2}}{\sqrt{2}\, N} = \frac{2\sqrt{2}}{3} \,(-x)^{3/2}
\label{eq:gww-ds-weak}
\end{eqnarray}
This double-scaling region analysis gives another complementary perspective on  the extra factor of $\sqrt{2}$ in the exponent of the transseries saddle action on the negative $x$ axis of the PII equation, as discussed in Section \ref{sec:p2-neg}.

Physically, this means that the immediate vicinity of the phase transition region of the GWW model is described by the Hastings-McLeod PII solution. The PII equation is {\it universal} in the sense that it describes the local phase transition regions in many different matrix models. It is the (integrable) nonlinear Airy equation,  consistent with the fact that the Airy equation universally describes the behavior near a linear turning point.
\begin{figure}[h!]
\centerline{\includegraphics[scale=0.2]{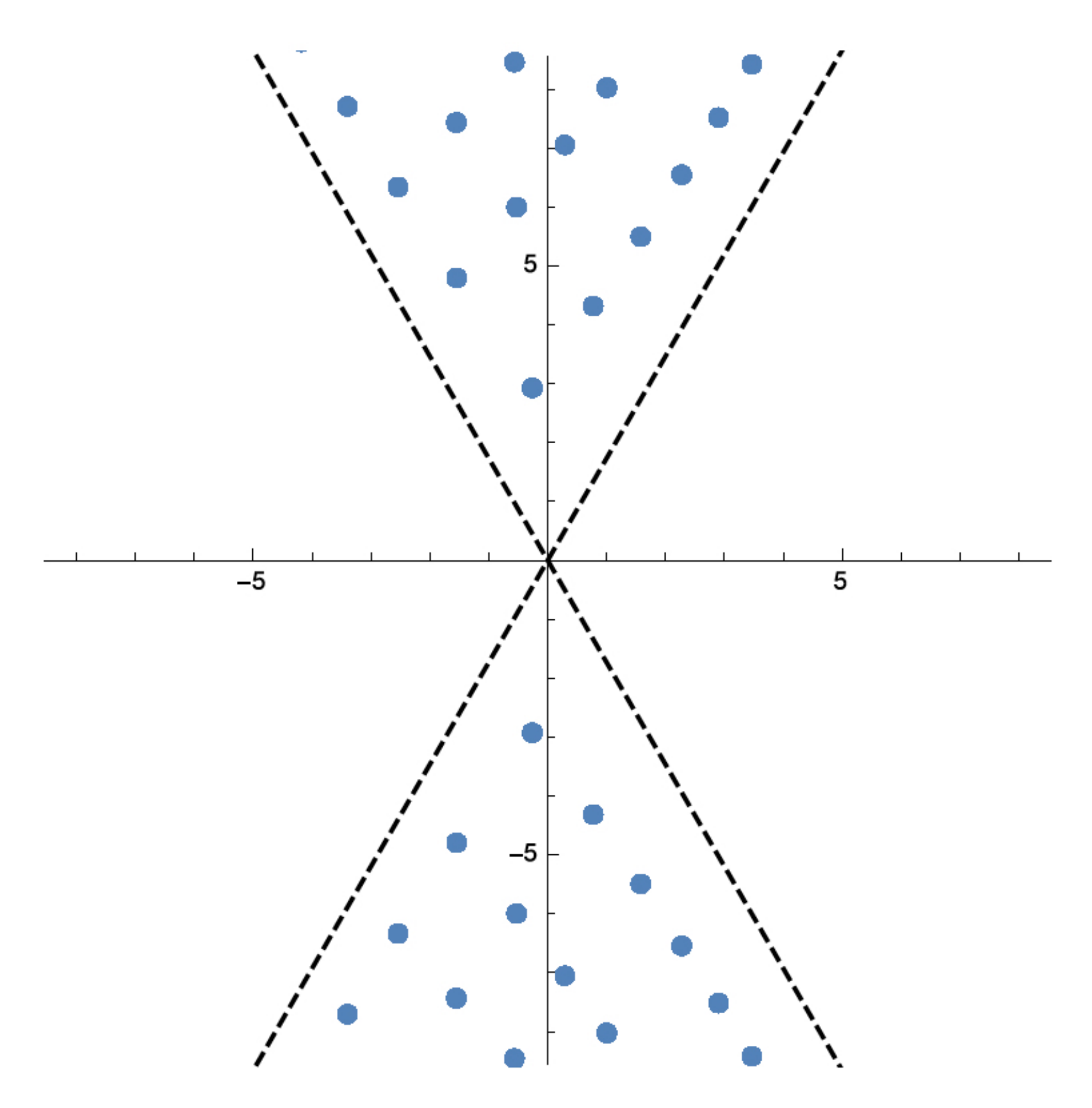}
}
\caption{ Poles of the Painlev\'e II Hastings-McLeod solution, which are confined to two $\pi/3$ wedges \cite{huang}. The wedges meet at the origin in the complex $x$ plane, and as $N\to\infty$ they pinch the axis at $x=0$, (i.e. $t=1$; recall \eqref{eq:gww-ds}), the GWW phase transition \cite{gww}. }
\label{fig:gww-lee-yang}
\end{figure}
One final comment is that the Gross-Witten-Wadia phase transition can be understood in terms of  Lee-Yang zeros \cite{lee-yang}. Lee and Yang proposed that in certain statistical systems a phase transition can be seen as the complex zeros of the finite $N$ partition function (e.g.,  $N$ characterizes the size of the lattice) pinching the real axis at the phase transition point.  See Figure \ref{fig:gww-lee-yang}, which shows the complex zeros of the Hastings McLeod PII solution, which are confined to two wedges of width $\pi/3$ about the positive and negative imaginary axes \cite{huang}. Since the $x$ variable here is the deviation from $t=1$, with the $1/N^{2/3}$ scaling in \eqref{eq:gww-ds}, this means that the $x$ zeros nearest the origin approach the physical phase transition at $t_c=1$, also with the correct scaling behavior \cite{damgaard}. The left-right asymmetry of the zeros in Fig \ref{fig:gww-lee-yang} can be traced to the fact that the transseries in the pole-free region on the right has a different structure from the transseries in the pole-free region on the left, and hence the transasymptotic matchings at the left and right boundaries are different \cite{costin-book}.

\section{Heisenberg-Euler QED Effective Action}
\label{sec:he}

One of the clearest examples of resurgence in QFT is the 1935 computation by Heisenberg and Euler of the 1-loop QED effective action \cite{he,dr-qed,kogan,schwartz}. The paper \cite{he} is one of the first QFT computations, and it is certainly the first {\it nonperturbative} QFT computation. Ref. \cite{he}, and Euler's PhD thesis, were also the first to discuss {\it effective field theory}. In modern language, Heisenberg and Euler integrated out the heavy QED fields (the electrons and positrons), yielding the nonlinear effective action for the light QED fields (the photon). This computation showed how both dispersive and absorptive effects arise from quantum vacuum fluctuations: photon-photon scattering, vacuum birefringence, electron-positron pair production (when an external electric field ``ionizes'' the Dirac sea) \cite{gies,dipiazza,karbstein}. It also contained the 
genesis of charge renormalization, identifying logarithmic corrections to the effective charge \cite{he,viki}.

\begin{figure}[h!]
\centerline{\includegraphics[scale=0.75]{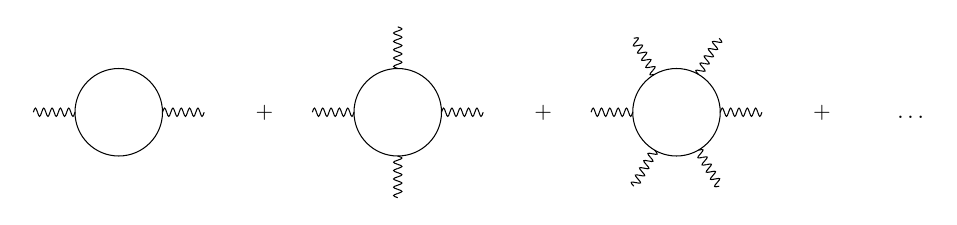}}
\caption{The diagrammatic perturbative expansion of the one loop effective action (\protect{\ref{eq:he-action}}).}
\label{digexp}
\end{figure}

The one-loop QED effective action for electrons in the presence of a classical background electromagnetic field is \cite{schwinger,kogan,schwartz}
\begin{eqnarray}
\mathcal S=-i \ln \det (i D \hskip -7pt \slash-m)=-\frac{i}{2}\ln \det(D \hskip -6pt \slash^{\hskip 2pt 2}+m^2)
\label{eq:he-action}
\end{eqnarray}
where the Dirac operator is $D \hskip -7pt \slash =\gamma^\nu\left(\partial_\nu+ie A_\nu\right)$,  $A_\nu$ is a
fixed classical gauge potential with field strength tensor $F_{\mu\nu}=\partial_\mu A_\nu- \partial_\nu A_\mu$, and $m$ is the electron mass. The one-loop effective action has a natural perturbative expansion in powers of the external photon field $A_\mu$, as illustrated in Fig. \ref{digexp}. By Furry's theorem (charge conjugation symmetry of QED), the expansion is in terms of {\it even} numbers of external photon lines.
Heisenberg and Euler showed that in the low energy limit for the external photon lines, taking the background field strength $F_{\mu\nu}$ to be constant, it is possible to compute a relatively simple integral representation of the effective action, which generates {\sl all orders} of the perturbative one-loop diagrams in Fig. \ref{digexp}. 
\begin{equation}
{\mathcal L}(B,E)
=
-\frac{e^2}{8\pi^2} \int_0^{\infty}\frac{d \xi }{\xi}
\,e^{- m^2 \xi}
\left\{
\frac{ E\, B}{\tanh(e B \xi)\tan(e E \xi)} -\frac{1}{e^2\xi^2}
-\frac{1}{3} (B^2-E^2)
\right\}
\label{eq:he1}
\end{equation}
For definiteness we consider the case where $\vec E\cdot \vec{B}\neq 0$, so we can choose a frame in which $\vec{B}$ and $\vec{E}$ are parallel, with magnitudes $B$ and $E$, respectively. A constant $F_{\mu\nu}$ can be represented by a gauge field $A_\mu=-\frac{1}{2}F_{\mu\nu}x^\nu$, which is linear in $x$. Thus, in an appropriate basis, the Dirac operator factorizes into two independent Landau level problems, of ``cyclotron'' frequencies $\frac{e B}{mc}$ and $\frac{i e E}{mc}$, with Landau degeneracy factors $\frac{e B}{2\pi}$ and $\frac{e E}{2\pi}$, respectively. Therefore the effective action can be evaluated via zeta function regularization of the Dirac determinant \cite{he,dr-qed,kogan,schwartz}. See Exercises \ref{ex:3.1}-\ref{ex:3.2}.

The result \eqref{eq:he1} has the form of a Borel integral -- the summation of the perturbative expansion in Fig. \ref{digexp}. So this is a natural QFT system in which to study resurgence and the connection between perturbative and nonperturbative physics. Eq. \eqref{eq:he1} is the fully resummed, and renormalized, effective action. The overall constant contains one power of the fine structure constant $\alpha=\frac{e^2}{4\pi}$; \eqref{eq:he1} is the {\it one-loop} effective action. The first subtraction term inside the integral is the zero-field subtraction, and the second subtraction term is the logarithmic charge renormalization term, proportional to the Maxwell Lagrangian, ${\mathcal L}_{\rm Maxwell}=\frac{1}{2}(\vec{E}^2-\vec{B}^2)$.

\subsection{Physical Consequences of the Heisenberg-Euler Effective Action}
\label{sec:physical-he}

\subsubsection{ The QED Vacuum as a Medium: Nonlinear Interactions.} 
\label{sec:nl}

The Euler-Heisenberg effective Lagrangian (\ref{eq:he1}) is {\it nonlinear} in the electromagnetic fields. The weak field expansion of \eqref{eq:he1} generates the perturbative diagrammatic expansion in Figure  \ref{digexp}.
The first two perturbative orders are:
\begin{eqnarray}
{\mathcal L}(B, E)
 \sim \frac{e^4\left[(E^2-B^2)^2 + 7(E B)^2\right]}{360 m^4\pi^2 }
  +\frac{e^6 (E^2-B^2)\left[2(E^2-B^2)^2+13 (E B)^2\right]}{1260  m^8 \pi^2}
 +\dots
  \label{eq:he-weak-terms}
\end{eqnarray}
These terms represent new nonlinear interactions. The first of these new interactions is light-light scattering,
which does not occur in the classical Maxwell theory.
 The nonlinearities can be viewed as dielectric effects, with the quantum vacuum behaving as a polarizable medium. In Weisskopf's words \cite{viki}:
{\color{Blue}
\begin{quote}
``{\sl When passing through electromagnetic fields, light will behave as if the vacuum, under the action of the fields, were to acquire a dielectric constant different from unity.}"
\end{quote}
}
\subsubsection{Low-energy effective field theory} 
\label{sec:eft}

The Heisenberg-Euler result (\ref{eq:he1}) is the paradigm of a {\it "low energy effective field theory"} \cite{manohar}, describing the physics of the light degrees of freedom (the photon field) at energies much lower than the scale above which one has integrated out the heavy degrees of freedom (the electron/positron field). The effective Lagrangian is expanded in terms of gauge and Lorentz invariant operators $O^{(n)}$ for the light fields, respecting the relevant symmetries:
\begin{equation}
{\mathcal L}_{\rm eff}=m^4\, \sum_n a_n\, \frac{O^{(n)}}{ m^n}\ .
\label{effectivelag}
\end{equation}
By power counting, the operators $O^{(n)}$ are balanced by appropriate powers of the heavy mass scale $m$. 
Euler's thesis also discussed the general structure when $F_{\mu\nu}(x)$ is spacetime dependent. The resulting {\it derivative expansion} is discussed below, in Section \ref{sec:beyond}.
Correspondingly, the polarization tensor has an operator-product-expansion (OPE)\cite{novikov}
\begin{eqnarray}
\Pi_{\mu\nu}=(q_\mu q_\nu -q^2
g_{\mu\nu})\sum_n c_n(Q^2) \,\langle O^{(n)}\rangle
\label{ope}
\end{eqnarray}
Different polarizations propagate at different speeds:  QED vacuum birefringence is a tiny effect, not yet directly measured, and is the subject of intensive experimental and theoretical efforts \cite{gies,dipiazza,karbstein}.

\subsubsection{Pair Production from Vacuum in an Electric Field} 
\label{pairprod}

\begin{figure}[h!]
\centerline{\includegraphics[scale=0.2]{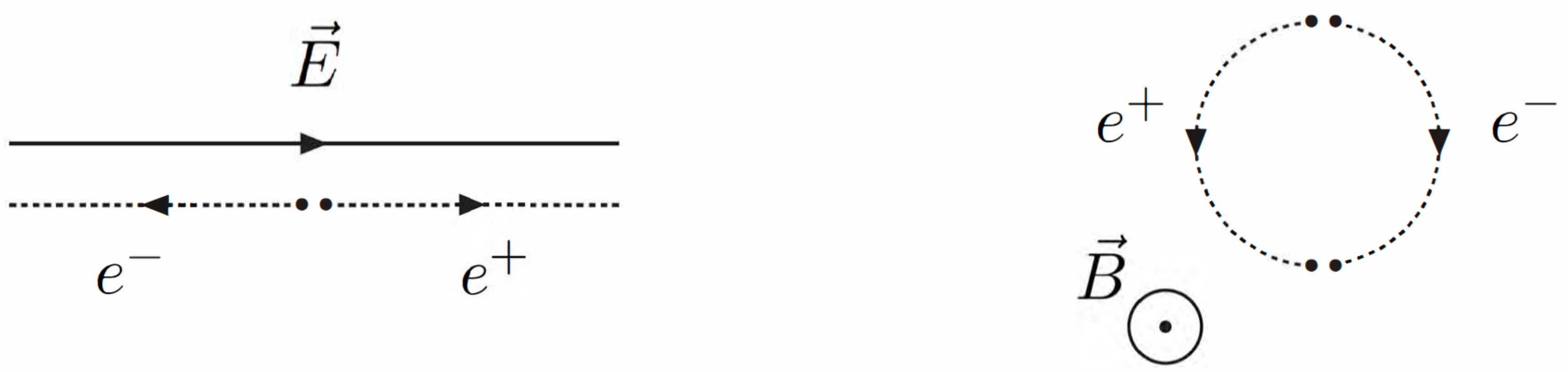}}
\caption{A static electric field can tear apart a virtual $e^+ e^-$ pair from the vacuum, producing an asymptotic electron and positron. A static magnetic field does not break apart a virtual dipole.}
\label{evsb}
\end{figure}

A background electric field destabilizes the QED vacuum,  accelerating apart the virtual vacuum dipole pairs, leading to $e^+e^-$ particle production: Fig. \ref{evsb}. Heisenberg and Euler \cite{he} computed the leading pair production rate in a weak electric field:
\begin{eqnarray}
\Gamma \sim\, \frac{e^2 E^2}{4\pi^3}
\exp\left[-\frac{m^2\pi }{eE}\right].
\label{imag}
\end{eqnarray}
This rate is deduced from the imaginary part of the effective Lagrangian (\ref{eq:he1}) when the background is purely electric.
In modern language, the imaginary part gives the rate of vacuum non-persistence due to pair production \cite{schwinger}. The rate is extremely small for typical electric field strengths, becoming more appreciable when the $E$ field approaches a critical value  (the {\it Schwinger critical field})
\begin{eqnarray}
{\mathcal E}_c=\frac{m^2 c^3}{e\hbar} \approx 10^{16}\,{\rm Vcm}^{-1}
\label{eq:critical}
\end{eqnarray}
where the work done accelerating a virtual pair apart by a Compton wavelength is of the order of the rest mass energy for the pair. Such electric field strengths are well beyond current technological capabilities, even in the most intense lasers. The effective critical field can be enhanced by interacting high-intensity lasers with high energy electron beams \cite{dipiazza,karbstein}. And even though the condition of a constant electric field is unrealistic, the Heisenberg and Euler effective Lagrangian \eqref{eq:he1} provides the starting point for more detailed analyses which incorporate time-dependent electric fields, as discussed in Section \ref{sec:beyond}.

\subsubsection{Charge Renormalization, $\beta$-functions and the Strong Field Limit.}
\label{charge}

The Heisenberg and Euler result (\ref{eq:he1}) also incorporates charge renormalization.  
Weisskopf \cite{viki} emphasized the characteristic strong-field logarithmic behavior of the Heisenberg-Euler result (\ref{eq:he1}), for example in a strong magnetic background:
\begin{eqnarray}
\frac{{\mathcal L}_{\rm spinor}}{{\mathcal L}_{\rm Maxwell}}\sim \left(\frac{e^2}{4\pi}\right)\, \left(-\frac{1}{3\pi}\right) \log\left(\frac{2e  B}{m^2}\right)
\qquad , \qquad B\to\infty\ .
\label{eq:he-strong}
\end{eqnarray}
In these units, the fine structure constant is $\alpha=\frac{e^2}{4\pi}$.
The coefficient, $\frac{-1}{3\pi}$, of the logarithmic dependence in \eqref{eq:he-strong} is the one-loop QED $\beta$-function coefficient \cite{schwartz}.

Exercises 3.1 - 3.2 work through the derivation of the Heisenberg-Euler effective Lagrangian using zeta function regularization \cite{dr-qed,kogan}:
\begin{eqnarray}
\ln \det ({\rm operator}) := -\zeta^\prime (0) \qquad {\rm where } \qquad \zeta(s):=\sum_{{\rm spectrum}\, \lambda} \frac{1}{\lambda^s}
\label{eq:zeta}
\end{eqnarray}
where the zeta function is defined as a trace over the spectrum of the operator.

{\color{Blue} 
\begin{exercise}
\label{ex:3.1}
{\bf Analytic Continuation of the Hurwitz Zeta Function}
\begin{enumerate}
\item
Analytically continue the integral representation of the Hurwitz zeta function
\begin{eqnarray}
\zeta_H(s, z)=\sum_{n=0}^\infty \frac{1}{(n+z)^s}
= \frac{1}{\Gamma(s)} \int_0^\infty dt\, e^{-z\, t}\frac{t^{s-1}}{1-e^{-t}} 
\quad , \quad {\rm Re}(s) >1\, , \, {\rm Re}(z)>0
\nonumber
\end{eqnarray}
(see \href{https://dlmf.nist.gov/25.11.vii}{dlmf.25.11.vii}) into the region ${\rm Re}(s)>-2$ to obtain
\begin{eqnarray}
\zeta_H(s, z)=
\frac{z^{1-s}}{s-1}+\frac{z^{-s}}{2}+\frac{s\, z^{-1-s}}{12}
+ \frac{2^{s-1}}{\Gamma(s)} \int_0^\infty \frac{dt}{t^{1-s}}\, e^{-2z\, t}\left(\coth(t)-\frac{1}{t}-\frac{t}{3}\right)
\nonumber
\end{eqnarray}

\item
Hence show that $\zeta_H(-1, z)=-\frac{1}{12}+\frac{z}{2}-\frac{z^2}{2}$, and 
\begin{eqnarray}
\zeta_H^\prime(-1, z)=
\frac{1}{12}-\frac{z^2}{4}-\zeta_H(-1, z)\, \ln z 
- \frac{1}{4}\int_0^\infty \frac{dt}{t^2}\, e^{-2z\, t}\left(\coth(t)-\frac{1}{t}-\frac{t}{3}\right)
\nonumber
\end{eqnarray}

\end{enumerate}
\end{exercise}
}

{\color{Blue} 
\begin{exercise}
\label{ex:3.2}
{\bf Heisenberg-Euler via the Zeta Function:} \\
The eigenvalues of the operator $(D \hskip -6pt \slash^{\hskip 2pt 2}+m^2)$ in a constant magnetic field $B$ are  Landau levels
$$
\lambda_n^\pm =m^2+p_\perp^2+e B(2n+1\pm 1)\qquad\quad, \qquad n=0, 1, 2, ...
$$
with Landau degeneracy factor $\frac{eB}{2\pi}$. 
\begin{enumerate}
\item Derive the following expression for the Heisenberg-Euler effective action:
\begin{eqnarray}
\mathcal L(B)=
 \frac{e^2 B^2}{2\pi^2} \left\{ \zeta_H^\prime\left(-1, \frac{m^2}{2eB}\right) + \zeta_H\left(-1, \frac{m^2}{2eB}\right)\, \ln\left(\frac{m^2}{2eB}\right)
-\frac{1}{12} +\frac{1}{4} \left(\frac{m^2}{2eB}\right)^2\right\}
\nonumber
\label{eq:he-hurwitz}
\end{eqnarray}
\item
Use the result of Exercise \ref{ex:3.1} to derive the expression \eqref{1lspmag} for the Heisenberg-Euler effective lagrangian in a magnetic background.
\end{enumerate}
\end{exercise}
}

\subsection{Weak Field Expansions of Heisenberg-Euler}
\label{sec:he-weak}

\underline{Purely magnetic background} : If the background is purely magnetic, of strength $B>0$,  the integral representation (\ref{eq:he1}) reduces to
\begin{equation}
{\mathcal L}(B)=-\frac{e^2 B^2}{8\pi ^{2}} \int_{0}^{\infty} \frac{ds}{s^{2}}\; 
\left(\coth s-\frac{1}{s}-\frac{s}{3}\right)\,e^{-m^2s/(eB)}
\label{1lspmag}
\end{equation}

\noindent\underline{Purely electric background} : If the background is purely electric, of strength $E>0$, the integral representation (\ref{eq:he1}) reduces formally to
\begin{equation}
{\mathcal L}(E)=-\frac{e^2 E^2}{8\pi ^{2}} \int_{0}^{\infty} \frac{ds}{s^{2}}\; 
\left(\cot s-\frac{1}{s}+\frac{s}{3}\right)\,e^{-m^2s/(eE)}\ .
\label{1lspelec}
\end{equation}
The weak field expansion is generated by expanding the Borel transform functions:
\begin{eqnarray}
\coth(s)= \sum_{n=0}^\infty \frac{2^{2n} \mathcal B_{2n}}{(2n)!} s^{2n-1}
\qquad; \qquad 
\cot(s) = \sum_{n=0}^\infty (-1)^n \frac{2^{2n} \mathcal B_{2n}}{(2n)!} s^{2n-1}
\label{eq:he-coth}
\end{eqnarray}
($\mathcal B_{2n}$ are the Bernoulli numbers \href{https://dlmf.nist.gov/24.2.i}{dlmf/24.2.i}).
We find the weak-field expansions
\begin{eqnarray}
{\mathcal L}(B) &\sim& -\frac{m^4}{8\pi^2} \sum_{n=0}^\infty 
\frac{{\mathcal B}_{2n+4}} {(2n+4)(2n+3)(2n+2)}\left(\frac{2eB}{m^2}\right)^{2n+4}
\label{eq:he-borel-B}
\\
{\mathcal L}(E) &\sim& \frac{m^4}{8\pi^2} \sum_{n=0}^\infty  
\frac{(-1)^n {\mathcal B}_{2n+4}} {(2n+4)(2n+3)(2n+2)}\left(\frac{2eE}{m^2}\right)^{2n+4}
\label{eq:he-borel-E}
\end{eqnarray}
The ${\mathcal B}_{2n}$ alternate in sign and grow factorially  in magnitude \cite{ww}:
\begin{eqnarray}
{\mathcal B}_{2n}=(-1)^{n+1}\,2 \,\frac{(2n)!}{(2\pi)^{2n}}\, \zeta(2n)
\label{bernoulli}
\end{eqnarray}
Here $\zeta(n)$ denotes the Riemann zeta function:
$
\zeta(n)=\sum_{k=1}^\infty \frac{1}{k^n}
$, 
which is exponentially close to 1 for large integers  $n$. Therefore, the weak-field expansions \eqref{eq:he-borel-B}-\eqref{eq:he-borel-E} are formal asymptotic series. To recover the nonperturbative information we apply Borel summation, as in section \ref{sec:borel}. For e.g., using \eqref{bernoulli} it is a straightforward exercise to Borel sum the factorially divergent weak-field series \eqref{eq:he-borel-B}. The $k$ sum of the zeta function factor leads to the Borel transform as an infinite sum over pole terms:
\begin{eqnarray}
\sum_{k=1}^\infty\, \frac{-2s}{k^2\pi^2(s^2+k^2\pi^2)}=\frac{1}{s^2}\left({\rm coth} s-\frac{1}{s}-\frac{s}{3}\right)
\label{eq:cothsum}
\end{eqnarray}
The Borel transform is {\it meromorphic}  (only pole singularities). Recall Ex. \ref{ex:1.4}.

Alternatively, resumming all the Borel poles leads to a simpler modified Borel integral representation:
\begin{equation}
\mathcal{L}\left(B\right)=
\frac{B^2}{4\pi^4}\int_0^\infty ds\,  {\rm Li}_2\left(e^{-\pi \, m^2\, s/B}\right) \left[\frac{s}{1+s^2}\right]
\label{eq:l1b-dilog}
\end{equation}
Here ${\rm Li}_2(x)$ is the classical dilogarithm function:
\begin{eqnarray}
    {\rm Li}_2(x)=\sum_{k=1}^\infty \frac{x^k}{k^2}
    \label{eq:li2}
\end{eqnarray}
Analytic continuation to a constant {\it electric} field ($B^2\to -E^2$) yields both a real and imaginary part:
\begin{eqnarray}
\mathcal{L}\left(E\right)&=&
 \frac{m^4}{8\pi}\left(\frac{E}{\pi m^2}\right)^2 {\mathcal P} \int_0^\infty ds \,{\rm Li}_2\left(e^{-\pi \, m^2\, s/E}\right)\left[ \frac{s}{1-s^2} \right]
 \nonumber\\
 &&  + i\,  \frac{m^4}{8\pi}\left(\frac{E}{\pi m^2}\right)^2\, {\rm Li}_2\left(e^{-\pi \, m^2/E}\right)
\label{eq:l1e-dilog}
\end{eqnarray}
The modified Borel transform resums all instanton orders.

For a magnetic background field, the formal weak-field expansion \eqref{eq:he-borel-B} is {\it alternating in sign} and factorially divergent, so the Borel integral \eqref{1lspmag} is real. On the other hand,  for an electric background field, the formal weak-field expansion \eqref{eq:he-borel-E} is {\it non-alternating in sign} and factorially divergent, so the Borel integral \eqref{1lspelec} has a real part and a nonperturbatively small imaginary part, which gives the vacuum pair production rate. This is analogous to the Zeeman (magnetic field) and Stark (electric field) effects in atomic physics, in which the perturbative weak field expansions of the electronic energy levels (in the atom or molecule) also grow factorially in magnitude and are alternating or non-alternating in sign for the Zeeman or Stark effect, respectively \cite{silverstone}. In the Zeeman effect there are real Zeeman splittings of the energy levels, while in the Stark effect there are real Stark shifts as well as imaginary energy shifts associated with ionization.

The Heisenberg-Euler example is analogous to Dyson's beautiful physical argument (not a proof!) for the divergence of QED perturbation theory \cite{dyson}. Dyson argued that a perturbative QED computation yields an expansion in the fine structure constant, hence an expansion in $e^2$. Therefore, if such an expansion had a nonzero finite radius of convergence, this would effectively say that the QED vacuum with $e^2>0$ could be smoothly connected to the QED vacuum with $e^2<0$. Dyson suggested that such a situation would not be stable. Therefore the radius of convergence should be zero, with a cut along the negative $e^2$ axis.  Similarly, one can say that the phenomenon of ionization implies that the perturbative Stark effect expansion must be factorially divergent; otherwise the sum of real terms could not produce an imaginary part to describe ionization. Analogously, in the Heisenberg-Euler case, the expected physical instability of the QED vacuum due to tunneling from the Dirac sea in the presence of an electric field implies that the weak field expansion must be factorially divergent, once again in order for a sum of real terms to produce an imaginary nonperturbative term. These examples emphasize the physical significance of the Borel singularities.

\subsection{Strong-field expansions of Heisenberg-Euler}
\label{sec:he-strong}

Strong-field expansions of the Heisenberg-Euler effective Lagrangians (\ref{eq:he1}) follow from another integral representation  of the Hurwitz zeta function terms in \eqref{eq:he-hurwitz} \cite{ww}:
\begin{eqnarray}
\zeta_H^\prime(-1,z)= \zeta^\prime(-1)-\frac{z}{2}\ln (2\pi) -\frac{z}{2}(1-z)+\int_0^z \ln \Gamma(x) dx\ .
\label{zetalog}
\end{eqnarray}
The Taylor expansion \cite{ww} of $\ln\Gamma(x)$:
\begin{eqnarray}
\ln\Gamma(x)=-\ln x-\gamma x +\sum_{n=2}^\infty \frac{(-1)^n}{n}\zeta(n)\, x^n
\label{loggamma}
\end{eqnarray}
therefore leads to the strong-field expansion:
\begin{eqnarray}
{\mathcal L}(B)&=&\frac{(eB)^2}{2\pi^2}\left\{ -\frac{1}{12}+\zeta^\prime(-1)-\frac{m^2}{4eB}+\frac{3}{4}\left(\frac{m^2}{2eB}\right)^2 -\frac{m^2}{4 eB}\, \ln (2\pi)\right.
\nonumber\\
&&\left. +\left[-\frac{1}{12}+\frac{m^2}{4eB}-\frac{1}{2}\left(\frac{m^2}{2eB}\right)^2\right] \ln \left(\frac{m^2}{2eB}\right) -\frac{\gamma}{2}\left(\frac{m^2}{2eB}\right)^2 \right.
\nonumber\\
&&\left.+\frac{m^2}{2eB}\left(1-\ln\left(\frac{m^2}{2eB}\right)\right) +\sum_{n=2}^\infty \frac{(-1)^n \zeta(n)}{n(n+1)} \left(\frac{m^2}{2eB}\right)^{n+1}\right\}
\label{spmagstrong}
\end{eqnarray}
The  expansion is {\it convergent}, but also involves logarithms.
We deduce the {\it leading strong field} limit \eqref{eq:he-strong},
and  the first beta function coefficient of QED: $-\frac{1}{3\pi}$.

{\color{Blue} 
\begin{exercise}
\label{ex:3.3}
{\bf Heisenberg-Euler for Scalar QED:}
In scalar QED the  Klein-Gordon operator spectrum in a constant $B$ field has no spin projection term:
$$
\lambda_n =m^2+p_\perp^2+e B(2n+1)\qquad\quad, \qquad n=0, 1, 2, ...
$$

\begin{enumerate}
\item
Hence use the zeta function method to show that the Heisenberg-Euler effective action for scalar QED has the integral representation \cite{viki}
$$
\mathcal L_{\rm scalar}(B) =\frac{e^2 B^2}{16\pi^2} \int_0^\infty ds\, \frac{1}{s^2}\left(\frac{1}{\sinh(s)} -\frac{1}{s}+\frac{s}{6}\right)
$$
\item
Generate the  weak field expansion of the scalar QED effective action and compare the large-order behavior of the expansion coefficients  with the spinor QED case.

\item
Compute the leading strong-field behavior by inspection of the Borel integral representation, and relate this to the scalar QED beta function.

\end{enumerate}
\end{exercise}
}

\subsection{Beyond the Constant Background Field Limit}
\label{sec:beyond}

The nonlinear QED effects due to vacuum fluctuations are tiny effects, so they are difficult to probe experimentally. To reach the necessary field intensities  requires ultra-intense laser fields, for which the electric and magnetic fields are very far from being constant. This is also a feature of atomic and molecular physics in intense laser fields, albeit at lower intensity scales (set by the atomic and molecular energy scales rather than the electron rest mass scale). There have been dramatic advances, both theoretical and experimental, which are  being systematically explored in the strong field regime of QED \cite{gies,dipiazza,karbstein}. 

\subsubsection{Semiclassical Effective Actions}
\label{sec:wkb}

An important result is due to Keldysh  \cite{keldysh}, shortly after the development of the laser, who used semiclassical methods to study atomic ionization not in a constant electric field but in a monochromatic linearly polarized time-dependent electric field:
\begin{eqnarray}
E(t)={\mathcal E}\, \cos(\omega t)
\label{eq:et}
\end{eqnarray}
Keldysh's result was adapted to  QED vacuum pair production by Br\'ezin \& Itzykson \cite{brezin}, and  Popov \& Marinov \cite{popov}. The time dependence introduces a new physical scale $\omega$.  Keldysh defined a dimensionless {\it inhomogeneity parameter}, $\gamma$, distinguishing two  different physical regimes, which we can now interpret as being separated by a Stokes transition. For QED:
\begin{eqnarray} 
\gamma\equiv \frac{m \omega}{eE} 
\label{eq:gamma}
\end{eqnarray}
 For clarity, consider scalar QED (spinor QED adds some technical details concerning the spin components).
Expand the field operator in Klein-Gordon solutions \cite{brezin}
\begin{eqnarray}
-\ddot{\phi}-(p_z-eA_z)^2\phi=(m^2+p_\perp^2)\phi
\label{kgeq}
\end{eqnarray}
with scattering boundary conditions
\begin{eqnarray}
\phi&\sim & e^{-it\sqrt{m^2+p^2}}+b_{\vec{p}}\,
e^{it\sqrt{m^2+p^2}}\quad , \quad t\to-\infty
\nonumber\\[1mm]
&\sim & a_{\vec{p}}\, e^{-it\sqrt{m^2+p^2}}\quad , \hskip 2.5cm
t\to+\infty
\label{scattbc}
\end{eqnarray}
Particles (antiparticles) are viewed as propagating forward (backward) in time
\cite{feynman,nambu}. The pair creation probability is expressed via the reflection coefficient $b_{\vec{p}}$
\begin{eqnarray}
P\approx \int \frac{d^3 p}{(2\pi)^3}\, |b_{\vec{p}}|^2\ .
\label{scattprob}
\end{eqnarray}
This is  ``over-the-barrier'' scattering, with  leading semiclassical expression \cite{landau}
\begin{eqnarray}
|b_{\vec{p}}|^2\approx \exp\left[-2\,{\rm Im} \oint\sqrt{m^2+p_\perp^2+[p_z-e A_z(t)]^2}\,\,dt\right]
\label{bpwkb}
\end{eqnarray}
integrating around the turning points. The dominant exponential factor is
\begin{eqnarray}
P\sim \exp\left[-\pi\, \frac{m^2\,c^3}{e\, \hbar\, \mathcal E}\, g(\gamma)\right]
\label{wkbprob}
\end{eqnarray}
For the monochromatic field, $E(t)=\mathcal E\, \cos(\omega t)$, we obtain 
\cite{brezin,popov}:
\begin{eqnarray}
 g(\gamma)=\frac{2}{\pi} \int_{-1}^1\sqrt{\frac{1-u^2}{1+\gamma^2 u^2}}\,du
&=&\frac{4}{\pi}\frac{\sqrt{1+\gamma^2}}{\gamma^2}
\left[{\bf K}\left(\frac{\gamma^2}{1+\gamma^2}\right)-
{\bf E}\left(\frac{\gamma^2}{1+\gamma^2}\right)\right]
\label{eq:mono}
\\
&\sim& \begin{cases}1-\frac{1}{8}\gamma^2
\quad  , \quad\gamma\ll 1\cr
\frac{4}{\pi\gamma}\ln \gamma \quad , \quad \gamma\gg 1
\end{cases}
\label{coscase}
\end{eqnarray}
where ${\bf K}(x)$ and ${\bf E}(x)$ are the complete elliptic integral functions \cite{ww}.

The expression \eqref{wkbprob} is a remarkable result. It looks nonperturbative, but in fact it interpolates between two very different physical regimes: a nonperturbative limit when the inhomogeneity parameter $\gamma\ll 1$, and a perturbative limit when $\gamma \gg 1$:
\begin{eqnarray}
P\sim
\begin{cases}\exp\left[-\pi\,\frac{m^2c^3}{e \mathcal E\hbar}\right]
\ , \hskip .3cm
\gamma\ll 1\quad {\rm (nonperturbative,\,\, ``tunneling")}\cr\cr \left(\frac{e \mathcal E}{\omega
mc}\right)^{4mc^2/\hbar
\omega}~\ ,\hskip .3cm \gamma \gg 1 \quad {\rm (perturbative,\,\, ``multiphoton")}
\end{cases}
\label{pnp}
\end{eqnarray}
When $\gamma\ll 1$ the time dependent field is slowly varying and we recover the leading nonperturbative exponential behavior of the Heisenberg-Euler result (\ref{imag}) for a constant electric field. On the other hand, when $\gamma\gg 1$ the logarithmic behavior of $g(\gamma)$ in \eqref{coscase} leads to the result in (\ref{pnp}), which has the perturbative form of the square of the gauge field strength, $(\frac{eA}{mc})^2$, raised to a power equal to the number of factors of the photon energy $\hbar \omega$ needed to make up the pair creation threshold energy $2mc^2$. Physically, this is the distinction between ``tunneling ionization'' and ``multiphoton ionization". From the point of view of resurgence, this transition is an example of a Stokes transition \cite{basar-dunne}, a change of dominant saddle configurations. 
\\

\begin{quote}
\centerline{Stokes transition: tunneling $\leftrightarrow$ multi-photon}
\end{quote}
These two regimes are well understood, both theoretically and experimentally, in the strong-field ionization of atoms and molecules, and there are several current proposals to analyze this transition for the QED vacuum \cite{meuren,luxe}.

{\color{Blue}
\begin{exercise}
\label{ex:3.4}
\noindent{\bf Resurgence in the Locally Constant Field Approximation (LCFA):}
Consider an electric field, in the $z$ direction, with a monochromatic inhomogeneity in time: $E(t)=\mathcal E \cos(\omega t)$. The LCFA is the zeroth order derivative expansion, in which we take the Heisenberg-Euler constant field result, replace the constant $E$ by the inhomogeneous $E$, and integrate.
\begin{enumerate}
\item
Compute the LCFA effective action, integrating the Heisenberg-Euler effective action over a period of the field, with the constant field replaced by its time-dependent form.

\item
Show that the coefficients of the LCFA weak field expansion grow factorially with perturbative order, and with subleading corrections of power-law and exponential form. Compute the first few power-law correction terms.

\item
Show that the power-law corrections in the previous part are related to the fluctuations about the instanton factors for the imaginary part of the effective action.

\end{enumerate}
\end{exercise}
}

\subsubsection{Borel Summation of the Derivative Expansion}
\label{sec:sech}

When the background field is inhomogeneous, the large mass expansion of the effective Lagrangian becomes an expansion in powers of the fields and also their spacetime derivatives.  We have already seen that for constant fields the Heisenberg-Euler weak field expansion is factorially divergent. The same is generically true for the derivative expansion. This is more difficult to quantify, because at higher orders in the large mass expansion more and more combinations of derivatives can occur.

However, for an inhomogeneous electric field characterized by {\it just one inhomogeneity parameter} (such as the monochromatic field discussed in the previous section), the large mass expansion is a double-series: an expansion in the field and in its inhomogeneity scale. Each of the two series is factorially divergent, so interesting new effects arise in Borel summation.
This can be analyzed in great detail for a particular kind of inhomogeneity, for which the Dirac  
and Klein-Gordon equations are exactly solvable. Consider a {\it single-bump}  field, such as
a time-dependent electric field of the form\footnote{By analytic continuation we can also obtain results for the fields: $E(x)={\mathcal E} \, {\rm sech}^2(k x)$ or a magnetic field $
B(x)={\mathcal B} \, {\rm sech}^2(k x)$. It also provides a smooth regularization of the constant field case.}
\begin{eqnarray}
E(t)&=&E \, {\rm sech}^2(\omega t) 
\label{eq:bump}
\end{eqnarray}
The effective action can be expressed exactly as an integral over certain momentum parameters \cite{nikishov}. In fact, these expressions can be further reduced to a single Borel integral \cite{cangemi}, which then permits analysis of the singularity structure and the large-order growth properties of the double-series.
Assuming that the electron mass sets the dominant scale (an excellent approximation for current lasers), the asymptotic expansion of the exact result yields the derivative expansion \cite{hall}
\begin{eqnarray}
  S\sim -\frac{m^4}{8\pi^{3/2}\omega }
\sum_{j}^\infty \sum_{k}^\infty
\frac{(-1)^{j+k}}{(m/\omega)^{2j}}
\left(\frac{2eE}{m^2}\right)^{2k}\frac{\Gamma(2k +
j)\Gamma(2k+ j -2) {\mathcal B}_{2k+2j}}{ j!(2k)!\Gamma(2k+j+\frac{1}{2})}
\label{derivel}
\end{eqnarray}
Fixing the order $j$ of the derivative expansion, the remaining sum over $k$ is divergent and nonalternating, and hence has a nonperturbative imaginary part. 
For fixed order $j$ of the derivative expansion, the perturbative expansion coefficients are non-alternating and grow factorially fast as the order $k\to\infty$:
\begin{eqnarray}  
c_k^{(j)}=\frac{(-1)^{j+k}\Gamma(2k+j)
\Gamma(2k+j+2){\mathcal B}_{2k+2j+2}} {\Gamma (2k+3)\Gamma
(2k+j+\frac{5}{2})}\sim 
  2 \,
\frac{\Gamma(2k+3j-\frac{1}{2})}{(2\pi)^{2j+2k+2}}
\label{dc}
\end{eqnarray}
Borel summation produces a leading nonperturbative imaginary part 
\begin{eqnarray}
{\rm Im} S^{(j)} \sim  \frac{1}{j!}\,
\left(\frac{\pi m^4 \omega^2}{4e^3 E^3}\right)^{j}
\exp\left[-\frac{m^2\pi}{eE}\right]
\label{jsum}
\end{eqnarray}
The sum over the derivative expansion order $j$ exponentiates:
\begin{eqnarray}
{ \rm  Im} S\sim \exp\left [-\frac{m^2\pi}{eE}
\left(1-\frac{1}{4}\left(\frac{m \omega}{eE}\right)^2\right)\right]\,.
\label{kthenj}
\end{eqnarray}
On the other hand, if we first consider a fixed order $k$ of the perturbative expansion, the remaining $j$ sum is divergent and non-alternating at large order ($j\to\infty$):
\begin{eqnarray}  
c_j^{(k)}=(-1)^{j+k}\frac{\Gamma(j+2k)
\Gamma(j+2k-2){\mathcal B}_{2k+2j}} {\Gamma (j+1)\Gamma
(j+2k+\frac{1}{2})}\sim  2^{\frac{9}{2}-2k} 
\frac{\Gamma(2j+4k-\frac{5}{2})}{ (2\pi)^{2j+2k}}
\label{dce}
\end{eqnarray}
This produces the leading nonperturbative imaginary part:
\begin{eqnarray}
{ \rm Im} S^{(k)} \sim  \frac{1}{(2k)!}\left(\frac{2\pi e E}{\omega^2}\right)^{2k} \,
\exp[-2\pi m/\omega]
\label{ksum}
\end{eqnarray}
The remaining $k$ sum also exponentiates, yielding the leading behavior
\begin{eqnarray}
{ \rm Im} S\sim \exp\left [-\frac{2\pi m}{\omega}
\left(1-\frac{eE}{m\omega}\right)\right]\,.
\label{jthenk}
\end{eqnarray}
To reconcile the two different expressions (\ref{kthenj}) and (\ref{jthenk}) for the imaginary part of the effective action, we recall  the WKB analysis and the Keldysh inhomogeneity parameter, which in this case is the ratio of the two expansion parameters in the double sum (\ref{derivel}):
\begin{eqnarray}
\gamma=\frac{m\,\omega}{eE}=\frac{\omega/m}{eE/m^2}\ ,
\label{ggder}
\end{eqnarray}
For this particular time dependent electric field, $E(t)=E\, {\rm sech}^2(\omega t)$, the semiclassical analysis in section \ref{sec:wkb} yields the exponent $g(\gamma)$  in the leading WKB pair production rate $P\sim \exp[-\pi m^2 g(\gamma)/(eE)]$ in (\ref{wkbprob}):
\begin{eqnarray}
g(\gamma)&=& \frac{2}{\pi}\int_{-1}^1 \frac{\sqrt{1-u^2}}{1+\gamma^2 u^2}\, du \\
&=& \frac{2}{1+\sqrt{1+\gamma^2}} \sim \begin{cases} 1-\frac{\gamma^2}{4}+\dots \qquad ,\quad \gamma\ll 1 \cr
\frac{2}{\gamma}\left(1-\frac{1}{\gamma}+\dots\right) \,, \quad \gamma\gg 1\ .\end{cases}
\label{ggcompare}
\end{eqnarray}
The leading terms match those in (\ref{kthenj}) and (\ref{jthenk}), obtained from Borel summation of the perturbative and derivative expansions.
Physically, there is a competition between two exponential factors, $\exp\left [-\frac{m^2\pi}{eE}\right]$ and $\exp\left [-\frac{2\pi m}{\omega}\right]$, which means a competition between two different Borel singularities.
The dominant exponential is determined by the magnitude of the inhomogeneity parameter $\gamma$. Thus, this Borel summation of the derivative expansion  is  consistent with the WKB analysis of the imaginary part of the effective action. It also demonstrates clearly that the derivative expansion is an {\it asymptotic} expansion, rather than a {\it convergent} expansion, because if it were convergent there would be no imaginary parts generated from the derivative expansion.
 
\subsubsection{Resurgence of the Derivative Expansion} 
\label{sec:derivative-resurgence}

We can go beyond these leading order results. Consider the {\it spatially inhomogeneous} magnetic field $B(x)=B\, {\rm sech}^2\left(x/ \lambda\right)$. This is also a soluble case, permitting precision tests of resummations and analytic continuations. The time-dependent field of the previous section is obtained by: $B^2\mapsto -E^2$, $\lambda^2\mapsto -\tau^2=-\frac{1}{\omega^2}$, with Keldysh inhomogeneity parameters
\begin{equation}
  \gamma=\frac{m}{eB\lambda}\mapsto\frac{m}{eE\tau}
  \label{eq:res-keldysh}
\end{equation}
The exact  spectrum leads to an exact Borel integral \cite{cangemi}, from which one can generate the weak $B$ field expansion
\begin{eqnarray}
\frac{S(B, \lambda)}{L^2\lambda T}=\frac{m^4}{\pi^2} \sum_{n\geq 0} {\color{Red} a_n(\gamma)}\left(\frac{B}{m^2}\right)^{2n+4}
\label{eq:inhomog-weak}
\end{eqnarray}
The coefficients $a_n(\gamma)$ are now polynomials in the inhomogeneity parameter $\gamma$, expressed in terms of a hypergeometric function \cite{harris}.

The exact Borel integral representation identifies three sequences of Borel singularities, which are integer multiples of the following Borel singularities
\begin{eqnarray}
t_1= 2i/(\sqrt{1+\gamma^2}+1) \quad; \quad t_2=- 2i/(\sqrt{1+\gamma^2}-1)
 \quad; \quad t_3 = 2i/\gamma
 \label{eq:he-borel-poles}
 \end{eqnarray}
and which depend on the inhomogeneity parameter. In the constant field limit ($\gamma\to 0$), both $t_2$ and $t_3$ recede to infinity, while $t_1\to i$,  the constant field result.
 
 These Borel singularities can also be identified in the large order behavior of the weak field coefficients $a_n(\gamma)$. For example, the leading large order behavior is:
\begin{eqnarray}
a_n(\gamma)&\sim& (-1)^n \Gamma(2n+\tfrac{3}{2})\frac{3\sqrt{2\pi}}{4}{\color{blue}(1+\gamma^2)^{5/4}\left(\frac{\sqrt{1+\gamma^2}+1}{2}\right)^{2n+3/2}}
\nonumber\\
    &&\times\left[{\color{Red} 1}-{\color{Red}\frac{5}{4}\frac{(1-\tfrac{3}{4}\gamma^2)}{\sqrt{1+\gamma^2}}}\frac{|t_1|}{(n+\tfrac{1}{4})}+{\color{Red}\frac{105}{32}\frac{(1+\tfrac{1}{4}\gamma^2)^2}{(1+\gamma^2)}}\frac{|t_1|^2}{(n+\tfrac{1}{4})(n-\tfrac{1}{4})}+\ldots\right]
\end{eqnarray}
Compare with the dominant nonperturbative imaginary part when we analytically continue from the spatially inhomogeneous magnetic field to the time-dependent electric field:
\begin{eqnarray}
\frac{{\rm Im} S(E,\tau)}{L^3\tau} &\sim& \frac{m^4}{8\pi^3}\left(\frac{E}{m^2}\right)^{5/2}{\color{blue}(1+\gamma^2)^{5/4}} \exp\left(-\frac{\pi m^2}{E}{\color{blue}\frac{2}{\sqrt{1+\gamma^2}+1}}\right) \nonumber\\
&&
\hskip -1.5cm \times\left[{\color{Red} 1}-{\color{Red}\frac{5}{4}\frac{(1-\tfrac{3}{4}\gamma^2)}{\sqrt{1+\gamma^2}}}\left(\frac{E}{\pi m^2}\right)
+{\color{Red} \frac{105}{32}\frac{(1+\tfrac{1}{2}\gamma^2+\tfrac{1}{16}\gamma^4)}{(1+\gamma^2)}}\left(\frac{E}{\pi m^2}\right)^2+\ldots \right]
\end{eqnarray}
The exponent is determined by the leading Borel singularity, $t_1$, and the Stokes constant and the subleading corrections determine the fluctuations around the leading nonperturbative imaginary part of the effective action. This is a nontrivial example of {\it parametric resurgence} in a nonperturbative QFT computation. In fact, the effects of the other two Borel singularities can also be deduced from {\it exponentially subleading} corrections to the large-order growth of the expansion coefficients $a_n(\gamma)$ \cite{harris}.

\subsection{Resurgence and Worldline Instantons}
\label{sec:wli}

Multi-dimensional WKB is analytically and computationally difficult, so other methods for computing nonperturbative effects are worth exploring for high intensity QFT. One such approach is the {\it worldline instanton} method \cite{wli1,wli2}, which takes advantage of Feynman's relativistic path integral representation of QED \cite{feynman,morette}. For notational simplicity we consider scalar QED. 
The log determinant of the Klein-Gordon operator can be expressed as a proper-time transform of a heat kernel trace:
\begin{eqnarray}
 \log\det(D_\mu^2+m^2) &=& -\int_0^\infty \frac{dT}{T} e^{-m^2 T} {\rm Tr}\left(e^{-D_\mu^2\, T}\right)
\label{eq:wli-path-integral}\\
&=&-\int_0^\infty \frac{dT}{T} e^{-m^2 T} \int \mathcal D x_\mu \exp\left[-\int_0^T d\tau\left(\frac{1}{2} \dot{x}_\mu^2+\dot{x}_\mu A_\mu(x)\right)\right]
\nonumber
\end{eqnarray}
In the last line the exponential of the Klein-Gordon operator is considered as a (Euclidean) evolution operator, whose trace is then expressed as a ``quantum mechanical'' path integral for 4-dimensional quantum mechanics, with proper time $T$ being the evolution ``time''. The paths $x_\mu(\tau)$ are parameterized by proper time, and are closed because of the trace. This proper-time evolution idea  identifies space-time paths that evolve forward in time as particles, and  space-time paths that evolve backward in time as antiparticles \cite{feynman,nambu}. The exponent is the action for a relativistic particle in the background of a classical electromagnetic field, with gauge field $A_\mu(x)$. 

Perturbative expansions of worldline representations have led to dramatic advances in computation of certain amplitudes in QFT \cite{bern,schubert}, and more recently for gravitational physics \cite{porto,plefka}. But one could also use saddle point methods. For example, the imaginary part of the effective action can be computed by a two-step saddle approximation \cite{wli1,wli2}. 
\begin{enumerate}

\item
A saddle solution of the path integral in \eqref{eq:wli-path-integral} is  a closed loop solution to the classical  equations of motion (a set of nonlinear coupled ODEs):
\begin{eqnarray}
\ddot{x}_\mu(\tau) =  F_{\mu\nu}(x(\tau))\, \dot{x}_\nu(\tau)
\label{eq:wli-saddle}
\end{eqnarray}
The leading semiclassical approximation to the path integral part of \eqref{eq:wli-path-integral} can be symbolically written as 
\begin{eqnarray}
\int \mathcal D x_\mu \exp\left[-\int_0^T d\tau\left(\frac{1}{2}\dot{x}_\mu^2+\dot{x}_\mu A_\mu(x)\right)\right]
\approx \frac{ \exp\left(-S[x_{\rm saddle}]\right)}{\sqrt{\det\frac{\delta^2 S}{\delta x^2}\big |_{x_{\rm saddle}(\tau)}}}
\end{eqnarray}
The Gaussian fluctuations lead to the determinant of a matrix of coupled {\it ordinary differential operators}, which can be evaluated efficiently using the Gelfand-Yaglom method \cite{coleman,dunne-qft-dets}.

\item
The classical action is in general a complicated function of the total proper time $T$, so there is a second saddle approximation for the $T$ integral in \eqref{eq:wli-path-integral}. The associated saddle point condition enforces
\begin{eqnarray}
\frac{\partial S_{\rm cl}}{\partial T}=-m^2
\label{eq:wli-second-saddle}
\end{eqnarray}
which is the  classical mechanics Legendre transform, identifying the variation of the action with respect to the period as the "energy".

\end{enumerate}
In \cite{wli1,wli2} it is shown that this worldline instanton approach produces the standard WKB results of Br\'ezin/Itzykson, and Marinov/Popov, but can also be extended to multi-dimensional inhomogeneous background fields.

\begin{figure}[h!]
\centerline{\includegraphics[scale=0.7]{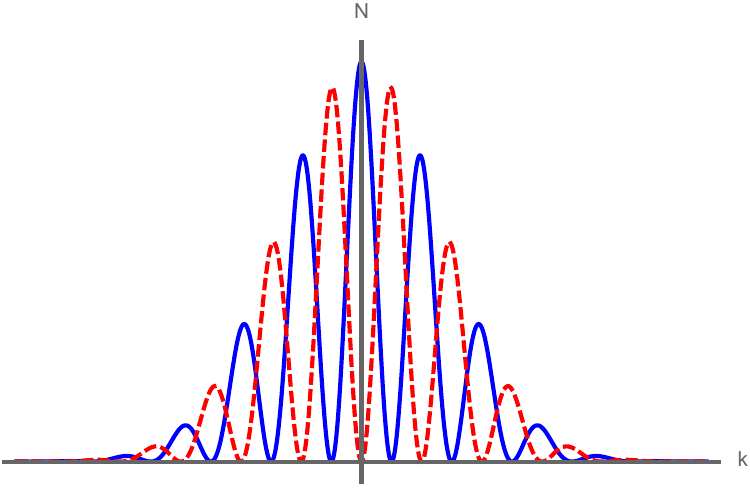} }
\caption{The form of the particle number $N$, as a function of longitudinal momentum $k$, in an electric field \eqref{eq:carrier}, with carrier phase $\varphi=\frac{\pi}{2}$. The blue (solid) curve is for spinor QED, and the red (dashed) curve is for scalar QED. See \cite{carrier,dumlu-stokes,ramsey}. }
\label{fig:carrier}
\end{figure}

These methods led to the realization that when a time-dependent background field is such that the electric field changes sign, quantum interference effects become important, and these require the consideration of {\it complex saddle} solutions  \cite{carrier,dumlu-stokes,ramsey}. For e.g., for a time-dependent oscillatory electric field with a Gaussian envelope
\begin{eqnarray}
E(t)=E\, \cos(\omega t+\varphi) \exp\left(-t^2/\tau^2\right)
\label{eq:carrier}
\end{eqnarray}
it is well-known in intense field atomic and molecular physics that the (nonperturbative) ionization spectrum is extremely sensitive to the carrier phase offset parameter $\varphi$ \cite{brabec}. This occurs also for the particle production analog, as can be seen in Fig. \ref{fig:carrier},
which shows the longitudinal momentum spectrum of produced particles from such a field with carrier phase $\varphi=\pi/2$. The difference between spinor and scalar QED emphasizes the significance of quantum interference.
The important implication can be represented qualitatively as:
\\

 \centerline{quantum interference\qquad  $\qquad \longleftrightarrow \qquad$ complex worldline saddles\\~}
 \bigskip
 
This shifts the focus to  Lefschetz thimbles for the Feynman worldline representation of the QED effective action in \eqref{eq:wli-path-integral}, to extract deeper nonperturbative information. This is an active direction of research, with many interesting recent results.

There is also an active parallel story of worldline methods for {\it gravitational} background fields, which is a highly productive method for studying black hole physics \cite{porto,plefka}.
{\color{Blue} 
\begin{exercise}
\label{ex:3.6}
{\bf Worldline Instantons in the Heisenberg-Euler Effective Action} Consider the classical Euclidean equations of motion for scalar QED in a (generally inhomogeneous) background electromagnetic field $F_{\mu\nu}(x)$:
$$\ddot{x}_\mu = F_{\mu\nu}(x)\, \dot{x}_\nu$$
where the dots refers to derivatives with respect to the proper-time, and $x_\mu(\tau)$ is the 4 dim spacetime coordinate.
\begin{enumerate}
\item
Show that for any solution,  $\dot{x}_{\mu}^2$ is a constant of motion.

\item
Show that the closed trajectory (with period $T$) for a constant $E$ field is a circle, and evaluate the classical action \cite{affleck}. Interpret this result physically.

\item
For the monochromatic electric field, $E(t)=E\, \cos(\omega t)$, use the Euclidean gauge field $A_3(x_4)=-\frac{i E}{\omega}\, \sinh(\omega x_4)$, and find the periodic solutions to the worldline instanton equations \eqref{eq:wli-saddle}. 

\item
Evaluate the corresponding classical action to obtain the result \eqref{wkbprob}-\eqref{eq:mono}. See \cite{wli1,wli2}.

\end{enumerate}
\end{exercise}
}

\section{Resurgence and Improved Summation Methods}
\label{sec:sum}

Given that resurgence generically relates expansions about different points, this suggests it should be possible to take advantage of resurgent structure to develop better extrapolation and analytic continuation methods. These are relevant in many physical applications
where we do not have the luxury of an explicit set of differential equations (or difference or integral equations) to solve.

Another generic feature of expansions of physical observables is that they are often factorially divergent formal asymptotic expansions, which typically means that there may be some interesting nonperturbative physics encoded in the given perturbative information. Then one can ask: what are the best methods for {\it decoding} this nonperturbative information? A simple but useful observation is that in this situation it is significantly more accurate to first regularize the problem by mapping to the Borel plane, and then trying to extract information about the analytic structure of the Borel transform function, rather than trying to apply extrapolation and analytic continuation methods directly to the original (truncated) formal asymptotic series. This gain in precision can be quantified \cite{plb,costin-recipe}.

It is also worth dispelling a common misunderstanding about asymptotic expansions: one often hears or reads statements suggesting that it is a waste of time trying to compute more terms of an asymptotic expansion, ``because it eventually diverges''. This is completely incorrect. The accurate statement is that if you evaluate a \underline{\bf truncated} factorially divergent asymptotic expansion, then adding more terms to the expansion eventually breaks down. However, truncation is  the wrong thing to do with an asymptotic expansion. If, instead, you  apply Borel summation, and associated methods outlined below, then adding more terms to the expansion leads to significant gains in precision, rather than a loss of precision. Regrettably, many theoretical and mathematical physics books are unclear about this. 

This is best illustrated by considering the generic situation of "factorial-over-power" large order behavior of a perturbative expansion  \cite{leguillou}. Since this frequently occurs in physical applications, this provides useful intuition, and also precise estimates of the relative precision of certain methods. An important lesson from this exercise is that the way you analyze an asymptotic expansion can make a {\it dramatic} difference in precision \cite{plb,costin-recipe}.

Consider a series that appears to be asymptotic, 
\begin{eqnarray}
f(x)
\sim \sum_{n=0}^\infty \frac{c_n}{x^{n+1}}
\qquad, \quad x\to+\infty
\label{eq:series1}
\end{eqnarray}
with  leading large order behavior of the coefficients having factorial-over-power form, with subleading power-law corrections:
\begin{eqnarray}
c_n\sim {\mathcal S}\, \frac{\Gamma(a\, n+b)}{A^n}\left(1+\frac{c}{n}+\dots\right) +\dots \qquad, \quad n\to\infty
\label{eq:bwl}
\end{eqnarray}
In physical applications, the various parameters in this large order behavior  \eqref{eq:bwl}  have the following interpretation:
\begin{itemize}
\item
The parameter $a$ is typically an integer (or a simple rational), often $a=1$ or 2. The physical significance of the parameter $a$ is that it tells you what is the correct power of the expansion variable to use: this is referred to as the {\it \'Ecalle critical variable}. 

\item
The parameter $A$ gives the location of the leading Borel singularity, and so is directly associated with the exponent in nonperturbative terms. It is related to the "instanton action".
It may be complex, corresponding to interesting oscillations of the perturbative coefficients.

\item
The parameter $b$ is associated with the leading  fluctuation power about the leading exponential. In many physical  applications it is a simple rational number. It is also related to the nature of the leading Borel singularity.

\item
The Stokes constant ${\mathcal S}$ determines the {\it strength} of the leading Borel singularity.

\item
The parameter $c$ relates to the next term in a fluctuation series about the leading nonperturbative exponential term.

\item
The final $+\dots$ terms refer to possible subleading (more distant) Borel singularities.

\end{itemize}

If we are solving an explicit nonlinear differential (or difference or integral) equation (or coupled set thereof), then these various parameters can often be determined analytically. See Exercise \ref{ex:4.2}. But even if we only have a list of perturbative coefficients $c_n$, there are still systematic ways to probe  the "data" (i.e., the coefficients) to extract information about these important nonperturbative physical parameters.

A low-precision, but simple, approach is to use ratio tests. For example, if $c_{n+1}/c_n\sim n/A$ for $n\to\infty$, then we deduce that $a=1$. However,  if $c_{n+1}/c_n\sim n^2/A$ for $n\to\infty$, then we deduce that $a=2$. This suggests that the nonperturbative terms will involve the {\it square root} of the expansion variable. This is what happens for perturbative expansions of atomic energy levels in the Zeeman and Stark effects, where the perturbative expansion coefficients grow like $(2n)!$, rather than $n!$. This is because the expansions are in {\it even powers} of the background field $B$ or $E$, respectively. Yet we know from simple tunneling arguments that the ionization rate has leading exponential form $\exp\left(-\#/|E|\right)$, and not $\exp\left(-\#/E^2\right)$. The same argument applies for the Heisenberg-Euler effective action discussed in Lecture 3.

\subsection{Richardson acceleration}
\label{sec:richardson}

A useful tool in analyzing the large-order behavior of series coefficients is {\it Richardson acceleration}, which is the simplest example of a class of series-acceleration methods. This is easy to implement, and when it works it is extremely impressive \cite{carl-book}.
Suppose some series coefficients $a_n$ have large-order behavior 
\begin{eqnarray}
a_n\sim b_0+\frac{b_1}{n}+\frac{b_2}{n^2}+\frac{b_3}{n^3}+\dots \qquad, \quad n\to\infty
\label{eq:an}
\end{eqnarray}
The goal is to extract numerically the limit value $b_0$ (and then to extract the subleading $b_j$ coefficients).

Richardson acceleration involves systematically forming carefully designed linear combinations of the $a_n$ for which the approach to the $n\to\infty$ limit (namely $b_0$) is accelerated.
This is analogous to the basic idea behind {\it  improved actions} in lattice gauge theory.
Assuming the behavior \eqref{eq:an},  the following combination of neighboring coefficients tends to the leading limit, $b_0$, with a reduced error $O(1/n^2)$, rather than the original  $O(1/n)$ error:
\begin{eqnarray}
{\color{blue} (n+1)\, a_{n+1}-n\, a_n} \sim  b_0+  b_2\left(\frac{1}{n+1}-\frac{1}{n}\right) +O\left(\frac{1}{n^3}\right) \sim  b_0 -  \frac{b_2}{n^2} +O\left(\frac{1}{n^3}\right)
\label{eq:rich1}
\end{eqnarray}
Similarly, we can combine a certain combination of three successive terms, $a_n$, $a_{n+1}$ and $a_{n+2}$, to cancel also the $O(1/n^2)$ term:
\begin{eqnarray}
{\color{blue}  \frac{1}{2} n^2\, a_n-(n+1)^2\, a_{n+1}+\frac{1}{2} (n+2)^2\, a_{n+2}} &\sim&  b_0 +\frac{b_3}{n^3}  +O\left(\frac{1}{n^4}\right)+\dots
\label{eq:an4b}
\end{eqnarray}
The pattern continues. At each order, multiply  $a_{n+k}$  by $(n+k)^N$ and form the following weighted linear combination:
\begin{eqnarray}
A_n:=\sum_{k=0}^N \frac{(-1)^{k+N}\,(n+k)^N \, a_{n+k} }{k! (N-k)!}
\quad \Rightarrow\quad A_n\sim b_0+O\left(\frac{1}{n^{N+1}}\right)
\nonumber
\label{eq:sum}
\end{eqnarray}
$N$ is called the ``order'' of the Richardson acceleration. If the coefficients have asymptotic behavior of the form in \eqref{eq:an},  the accelerated convergence can be extremely rapid. In certain situations, one can obtain sufficient precision for the leading term $b_0$ that it may be possible to guess the exact value of $b_0$. Given a conjecture for $b_0$, one can subtract it from the $a_n$ and repeat the process. If the guess is wrong the next iteration of Richardson acceleration will fail spectacularly.  Exercises \ref{ex:4.1} and \ref{ex:4.2} have some examples to illustrate how it works.

In physical applications we sometimes encounter large order growth involving logarithmic behavior. For example, in QM spectral problems, the leading perturbative coefficients typically grow factorially fast, $c_n\sim n!$. However, in the same problem, the coefficients of the perturbative fluctuations about the first exponentially small nonperturbative term (the ``first instanton factor''),  for example the splitting of degenerate levels in a double-well problem due to tunneling, typically grow like $n!\, \ln(n)$ \cite{nln}. One can adapt Richardson acceleration to treat series with this kind of growth \cite{hopf}. There are other variations of Richardson acceleration \cite{caliceti}, but the basic ideas are quite similar.

{\color{Blue}
\begin{exercise}
\label{ex:4.1}
{\bf Large Order/Low Order Resurgence Relations: Airy function:}\\
 Given the explicit formula \eqref{eq:airy-cn} for $c_n^{(\pm)}$ in terms of gamma functions, it is straightforward to derive the large-order behavior \eqref{eq:cn-resurgent} analytically. But in the absence of an explicit formula this can still be done numerically. Try this for the Airy function:
\begin{enumerate}
\item
Generate the first 100 $c_n^{(+)}$ coefficients in \eqref{eq:airy-eq2}, and use ratio tests and Richardson acceleration to deduce {\it numerically} the leading large order asymptotics.

\item
Deduce the Stokes constant {\it numerically}: $\mathcal S=\frac{1}{2\pi}$

\item
Extract {\it numerically} the first two subleading power-law corrections to the large order growth,  in decreasing-factorial form \eqref{eq:cn-resurgent}. Compare with the low order coefficients.

\end{enumerate}
\end{exercise}
}
{\color{Blue}
\begin{exercise}
\label{ex:4.2}
{\bf Large Order Growth in Yukawa Theory:}\\
 The perturbative expansion, $C(x)\sim \sum_{n=1}^\infty c_n \, x^n$, determines the anomalous dimension in the Hopf algebraic renormalization of 4 dimensional Yukawa theory. The coefficients $c_n$ are positive integers, enumerating combinatorial objects known as "connected chord diagrams". This sequence is listed on the OEIS as \url{https://oeis.org/A000699}. 
\begin{enumerate}
\item Generate 100 terms using the recursion formula listed on the OEIS (\url{https://oeis.org/A000699}) and then analyze them using Richardson acceleration to show that
$$
c_n \sim \frac{2^{n+\frac{1}{2}}\, \Gamma\left(n+\frac{1}{2}\right)}{e\sqrt{2\pi}}\left(1-\frac{\frac{5}{2}}{2\left(n-\frac{1}{2}\right)} -\frac{\frac{43}{8}}{2^2\left(n-\frac{1}{2}\right)\left(n-\frac{3}{2}\right)}+ O\left(\frac{1}{n^3}\right)\right)
$$
\item $C(x)$ satisfies a nonlinear ODE:  $C(x)\left(1-2x\frac{d}{dx}\right)C(x)=x-C(x)$. Show that the first nonperturbative correction term $C_{\rm np}(x)$ satisfies a linear ODE $\frac{d}{dx} \ln\left(C(x) C_{\rm np}(x)/x\right) =\frac{1}{2x C(x)}$. Therefore express $C_{\rm np}(x)$ in terms of the perturbative solution $C(x)$.
\item Hence expand $C_{\rm np}(x)$ at small $x$ and compare with part 1.
\end{enumerate}
\end{exercise}
}

\subsection{Darboux Theorem}
\label{sec:darboux}

A complementary perspective on the large order growth of perturbative expansions is given by Darboux's theorem \cite{henrici}. It is familiar that the radius of convergence of a series, corresponding to the power law in the large order growth of the coefficients,  is given by the distance from the expansion point to the nearest singularity in the complex plane. It is perhaps less well known that the {\it subleading} terms in the large order growth of the coefficients contain explicit information about the {\it local expansion around that singularity}. Applying this fact to a Borel transform function enables the extraction of detailed local information about a given Borel singularity (i.e local information of nonperturbative physics) from the coefficients of the original formal asymptotic series. This is one simple manifestation of resurgent behavior.

Suppose  $f(t)$ has the following  nonanalytic behavior near a singularity $t_0$:
\begin{eqnarray}
f(t)\sim  \phi(t)\, \left(1-\frac{t}{t_0}\right)^{-\beta}+\psi(t) \qquad, \quad t\to t_0 
\label{eq:darboux1}
\end{eqnarray}
with $\phi(t)$,  $\psi(t)$  analytic at $t_0$. This is  generic behavior at an isolated singularity.

Darboux's theorem states that the large-order growth of the Taylor  coefficients of $f(t)$ expanded at the origin, $f(t)=\sum_n b_n t^n$, have large-order growth:
\begin{equation}
b_n\sim \frac{\begin{pmatrix}
n+\beta-1\\ n
\end{pmatrix}}{t_0^n} \hskip -5pt \left[ \phi(t_0)- 
\frac{(\beta-1)\, t_0\, \phi^\prime(t_0)}{(n+\beta-1)}
+
\frac{(\beta-1)(\beta-2)\, t_0^2\, \phi^{\prime\prime}(t_0)}{2! (n+\beta-1)(n+\beta-2)}\, -\dots \right]
\label{eq:darboux2}
\end{equation}
The leading power law encodes the {\it location} $t_0$ of the singularity; the binomial factor encodes the {\it exponent} $-\beta$ of the singularity; subsequent power-law corrections determine the Taylor coefficients of the analytic function $\phi(t)$ that multiplies the non-analytic behavior.
In certain circumstances we can also determine the Taylor expansion of the analytic function $\psi(t)$:  multiply $f(t)$ by $\left(1-\frac{t}{t_0}\right)^{+\beta}$,  re-expand, and apply the Darboux theorem again.

Applying Darboux's theorem in the Borel plane provides some understanding of the existence of large-order/low-order resurgence relations, since the Taylor expansion of the Borel transform at the origin is determined by the coefficients of the original formal asymptotic series (after dividing by the factorial growth), and therefore by Darboux these determine the local behavior in the vicinity of the nearest singular point, thereby giving the location of the leading nonperturbative contribution, as well as its Stokes constant and fluctuations. If the Borel transform has many singularities, then there is an intricate network of relations between the expansions about these points. 
{\color{Blue}
\begin{exercise}
\label{ex:4.3}
{\bf Darboux Theorem:}
\begin{enumerate}
\item
Differentiate  \eqref{eq:darboux1}-\eqref{eq:darboux2} with respect to $\beta$ to derive the large order growth of the expansion coefficients of $f(t)$ at the origin, when the singularity at $t_0$ is a {\it logarithmic} branch point:
\begin{equation}
b_n\sim \frac{1}{t_0^n}\cdot \frac{1}{n} \left[\phi(t_0) - \frac{t_0\, \phi^\prime(t_0)}{(n-1)} +\frac{t_0^2\, \phi^{\prime\prime}(t_0)}{(n-1)(n-2)} -\dots \right]
\quad, \quad n\to\infty
\label{eq:darboux-log}
\end{equation}

\item
 Investigate Darboux's theorem numerically for the hypergeometric function, which has a branch point at $t=1$
$$
~_2F_1\left(a, b, c; t\right) =\frac{\Gamma(c)}{\Gamma(a)\Gamma(b)} \sum_{n=0}^\infty  \frac{\Gamma(n+a)\Gamma(n+b)}{\Gamma(n+c) n!} t^n
$$
Experiment with various choices for the parameters $a, b, c$, 
comparing with the exact expansion of the hypergeometric function about $t=1$ (see \href{https://dlmf.nist.gov/15.8}{dlmf.15.8}).

\item
 Investigate Darboux's theorem numerically for the function $(1-t+t^2)^{-1/3}$, which has branch points at $t=\exp[\pm i \pi/3]$.
 \end{enumerate}
 \end{exercise}
}

\subsection{Pad\'e-Borel Method}
\label{sec:pb-method}

In many physical applications  we only know a {\it finite} number of coefficients $c_n$ of the asymptotic expansion (\ref{eq:series1}). There are several different approaches to analyze the singularity structure of the associated Borel transform. Pad\'e approximants are a key tool in analytically continuing the truncated Borel transform beyond its radius of convergence. They can be significantly improved by combining them with conformal and uniformizing maps.

Ratio tests provide a  simple initial diagnostic, but  only extract rough information. Furthermore, they are less useful with {\it complex singularities}, as it is difficult to resolve oscillation patterns  from a {\it limited amount} of perturbative information. A  more powerful approach is to take advantage of complex analysis and look directly for information about the singularities of the function.\footnote{Here we concentrate on the singularities of a Borel transform function, but of course these methods can also be applied directly to any function with a finite radius of convergence.}
At first sight this appears to be an impossible problem: given a finite number of terms in an expansion of a function, how can we find the function's singularities, because the truncated series is just a {\it  polynomial}?
Pad\'e approximants are a remarkable tool for extracting this information \cite{carl-book}.

To sum a formal asymptotic series, first go to the Borel plane, converting the factorially divergent formal series to a Borel transform function, which has a finite radius of convergence. The main message is:

\begin{quote}
{\color{Blue} {\it The key to finding a precise summation, and/or analytic continuation, of an asymptotic series is to find a precise analytic continuation of the Borel transform function, especially near its singularities.}}
\end{quote}

Pad\'e is a simple and efficient method to analytically continue a series beyond its radius of convergence. It is a {\it rational} approximation to a polynomial, mapping a polynomial of order $N$ to a ratio of two polynomials, of orders $L$ and $M$, such that $L+M=N$, and the expansion of the ratio agrees with the original polynomial to order $N$:
\begin{eqnarray}
F_{N}(t)=\sum_n^{N} c_n\, t^n \quad \longrightarrow \quad 
P_{[L, M]}\left\{F_{N}\right\}(t) =\frac{R_L(t)}{S_M(t)}=\sum_n^{N} c_n\, t^n + \mathcal O \left(t^{N+1}\right)
\label{eq:pade}
\end{eqnarray}
We trade the $N$ coefficients of $F_N(t)$ for $L$ coefficients of $R_L(t)$ and $M$ coefficients of $S_M(t)$. There are explicit determinant expressions for these new coefficients, and  efficient algorithms to make these conversions. They are built-in functions in Maple and Mathematica. Implementing Pad\'e is simple. Understanding its implications is more interesting and more subtle.

The basic idea is that while $F_N(t)$ has no singularities, its Pad\'e approximant has singularities at the $M$ zeros of the denominator $S_M(t)$. 
This raises some obvious questions:
\begin{enumerate}
\item 
What do the poles  of the Pad\'e approximant ``know" about the {\it actual singularities} of the function to which $F_N(t)$ is an approximation?

\item
By construction, Pad\'e approximants can only have pole singularities, but the actual function may have branch point singularities. How does this work?

\end{enumerate}

The answers, and  the underlying physics, are very interesting. Near-diagonal Pad\'e (i.e., $L\approx M \approx N/2$) is deeply related to the theory of orthogonal polynomials, because the polynomials $R_N(t)$ and  $S_N(t)$ satisfy the same 3-term recursion relation, and 3-term recursion relations are associated with orthogonal polynomials \cite{carl-book}. An important consequence  is that the 
errors of sequences of Pad\'e approximants can be expressed in terms of orthogonal  polynomials \cite{plb}.
Therefore, at large order we can estimate errors using the asymptotics of orthogonal polynomials, a beautiful and well-developed subject. Some examples are discussed in section \ref{sec:potential}. 

The practical outcome is the following: Pad\'e is clearly very good at locating {\it pole} singularities of a function. But many physical functions have {\it branch point} singularities. Pad\'e represents a {\it branch point} singularity of a function by an arc of interlacing poles and zeros that accumulate to the location (in the complex plane) of the branch point. The {\it density} of the Pad\'e poles and zeros along this arc contains information about the {\it nature} (i.e., exponent) of the branch point singularity. However, this is not a particularly efficient or accurate way to extract this information about the nature of the singularity. There are better ways, as described below.

{\color{Blue}
\begin{exercise}
\label{ex:4.4}
\noindent{\bf Exploring Singularities with Pad\'e Approximants}

\begin{enumerate}

\item Generate 50 terms of the Taylor expansion of $1/(1-t+t^2)^{1/3}$ around $t=0$, and plot the poles and zeros of its diagonal Pad\'e approximant. Identify the singularities of the function. 

\item Generate 50 terms of the Taylor expansion of $1/(1-t+t^2)^{1/2}$ around $t=0$, and plot the poles and zeros of its diagonal Pad\'e approximant.  Identify the singularities of the function.

\item Explain the difference between the results of these two examples.

\end{enumerate}
\end{exercise}

}

In physical applications we expect Borel singularities to be associated with nonperturbative physics. This suggests that these singularities are physical, so we might expect them to have some  structure. Hence we might expect the Borel plane for a {\it physical problem} to have some simplicity and symmetry. Often there is a {\it dominant} Borel singularity (or a {\it dominant pair} of Borel singularities) associated with the leading divergence of the associated asymptotic series. In such cases we can quantify the possible precision of an analytic continuation based on a finite amount of input data.

\subsection{Potential Theory and the Physical Interpretation of Pad\'e Approximants}
\label{sec:potential}

Pad\'e approximants have an interesting physical interpretation,  based on old work of Szeg\"o \cite{szego}, with more recent developments due to Stahl \cite{stahl}.  For this interpretation, it is useful (for the electrostatic analogy) to invert the Borel variable ($t\to 1/t$) to move the point of analyticity from $t=0$ to $t=\infty$. Consider the Pad\'e poles  as ``charges" in two dimensional electrostatics, where the potential depends on the {\it logarithm} of the distance. Pad\'e produces a rational approximation that can be decomposed as a partial fraction expansion, each term representing the force due to a charge at the pole location and with strength given by the residue. In the limit $N\to \infty$, the Pad\'e prescription is actually minimizing a certain functional, which has the following physical interpretation: place the charges in the complex plane in such a way that as $N\to\infty$ they form flexible ``wires" with the property that the end points of the wires are fixed at the actual singularities of the function, but the shape of the wires, and their junctions, are deformed such that the {\it electrical capacitance} is minimized. 

Since Pad\'e approximants are  {\it rational approximations}, they can only converge where the function $B(t)$ is single-valued. Suppose $\mathcal{D}$ is a region of single-valuedness of  $B(t)$, with boundary   $\partial \mathcal{D}$. This boundary is effectively the set of cuts that Pad\'e creates. Consider $\partial \mathcal{D}$ to be an electrical conductor, with a unit charge, and normalize the electrostatic potential $V(x,y)=V(t)$, $t=x+iy$ (constant along a conductor) to vanish on $\partial \mathcal{D}$. Then the electrostatic capacitance of $\partial \mathcal{D}$ is cap$(\partial \mathcal{D})= 1/V(\infty)$. As $N\to\infty$, Pad\'e minimizes cap$(\partial \mathcal{D})$ by deforming the shape of $\partial \mathcal{D}$, while keeping the singularity locations fixed.
Furthermore, the equilibrium measure  on $\partial \mathcal{D}$ is the equilibrium density of charges on $\partial \mathcal{D}$ \cite{szego,stahl}.

This means that Pad\'e solves a 2d electrostatics problem. But recall that 2 dimensional electrostatics is solved by constructing analytic functions satisfying specified  boundary conditions, which essentially boils down to finding suitable conformal maps. So Pad\'e effectively generates its own conformal map: the cuts are deformable wires connecting fixed singularities, in such a way that the capacitance is minimized.

\subsection{One Branch Point Borel Singularity}
\label{sec:one-borel}

In many physical applications we encounter a formal asymptotic series with  leading large order behavior of the coefficients having factorial-over-power form \cite{leguillou}
\begin{eqnarray}
f(x)
\sim \sum_{n=0}^\infty \frac{c_n}{x^{n+1}}
\quad, \quad x\to+\infty
\quad; \quad 
c_n\sim {\mathcal S}\, \frac{\Gamma(a\, n+b)}{A^n} \quad, \quad n\to \infty
\label{eq:series2}
\end{eqnarray}
Here the parameters $a$, $A$, $\mathcal S$, and $b$ are independent of $x$. In treating such a problem, a natural approach is to apply Borel summation, leading to a Borel-Laplace integral representation of the function $f(x)$ whose asymptotic series is in (\ref{eq:series2}):
\begin{eqnarray}
f(x)=\int_0^\infty dt\, e^{-t\, x} \mathcal B(t)
\label{eq:laplace}
\end{eqnarray}
This maps the problem of extrapolating $f(x)$ in the complex $x$ plane to the problem of understanding the analytic structure of the Borel transform $\mathcal{B}(t)$, particularly its singularities in the complex Borel $t$ plane.

For example, the following function (based on the incomplete gamma function: recall Exercise \ref{ex:1.5})
\begin{eqnarray}
F(x; \alpha) \equiv x^{-1-\alpha} \, e^x \, \Gamma(1+\alpha, x)
=\int_0^\infty dt\, e^{-t\, x} (1+t)^\alpha 
\label{eq:i-gamma}
\end{eqnarray}
has an asymptotic expansion, as $x\to+\infty$, of the form in (\ref{eq:series2}):
\begin{eqnarray}
F(x; \alpha)
\sim \frac{1}{\Gamma(-\alpha)} \sum_{n=0}^\infty (-1)^n  \frac{\Gamma(n-\alpha)}{x^{n+1}}
\qquad, \quad x\to+\infty
\label{eq:inc-gamma3}
\end{eqnarray}
$F(x; \alpha)$ has nontrivial analytic continuation properties in the complex $x$ plane which are encoded in the analytic structure of the Borel transform function $\mathcal B(t)=(1+t)^\alpha$ in the Borel complex $t$ plane.

\subsubsection{Pad\'e-Borel Transform for One Branch Point Singularity} 
\label{sec:pb}

Consider making an $[N, N]$ diagonal  Pad\'e approximant based on $2N$ terms of the expansion on the Borel transform function $\mathcal B(t)=(1+t)^\alpha$. This approximant can be expressed as a ratio of Jacobi polynomials $P_N^{(\alpha, \beta)}$ \cite{plb}:
\begin{eqnarray}
{\rm Pade}{\rm -}{\rm Borel:} \qquad {\mathcal P \mathcal B}_{[N,N]}(t; \alpha)=\frac{P_N^{(\alpha,- \alpha)}\left(1+\frac{2}{t}\right)}{P_N^{(-\alpha, \alpha)}\left(1+\frac{2}{t}\right)}
\label{eq:pbp}
\end{eqnarray}
The zeros of the denominator lie in the range $-1\leq 1+\frac{2}{t} \leq +1$, so in the Borel $t$ plane these lie along the negative $t$ axis in the region $t\in (-\infty, -1]$, forming a cut along the negative $t$ axis accumulating to the branch point at $t=-1$.
The Pad\'e-Borel transform  (\ref{eq:pbp})  is remarkably accurate away from the cut, even for modest $N$. The large $N$ asymptotics of the Jacobi polynomials \cite{dunster,elliott-jacobi} quantifies this statement:
\begin{eqnarray}
\frac{{\mathcal P \mathcal B}_{[N,N]}(t; \alpha)}{(1+t)^\alpha}
&\sim &
\frac{I_\alpha\left(
\hskip -2pt\left(N+\frac{1}{2}\right)\hskip -2pt\ln\left[\frac{\sqrt{1+t}+1}{\sqrt{1+t}-1}\right]\right)}{I_{-\alpha}\left(
\hskip -2pt\left(N+\frac{1}{2}\right)\hskip -2pt\ln\left[\frac{\sqrt{1+t}+1}{\sqrt{1+t}-1}\right]\right)}
\\
&\sim& 1- 2\sin\left(\pi \alpha \right)\left(\frac{\sqrt{1+t}-1}{\sqrt{1+t}+1}\right)^{2N+1} +\dots \qquad , \quad t\to 0^+
\\
& \hskip -6cm \sim & \hskip -3cm  \frac{\Gamma(1-\alpha)}{\Gamma(1+\alpha)}\frac{\Gamma(N+1+\alpha)}{\Gamma(N+1-\alpha)}\, \frac{1}{(1+t)^\alpha} \left(1+\frac{2\alpha N(N+1)}{(\alpha^2-1)}\frac{1}{t} +\dots\right) \quad , \quad t\to \infty
\label{eq:pbp-limit}
\end{eqnarray}
$I_\alpha$ is the modified Bessel function. 
For Borel extrapolation, the small $x$ behavior is controlled by the large $t$ behavior of the Borel transform. Eq. (\ref{eq:pbp-limit}) implies ${\mathcal P \mathcal B}_{[N,N]}(t; \alpha)\sim t^\alpha \left(\frac{N^{2}}{t}\right)^\alpha$ as $t\to +\infty$. This implies that the $2N$ term truncated large $x$ asymptotic expansion
can be extrapolated with a chosen fractional error down to a small $x_{\rm min}$ which scales with $N$ as  $x_{\rm min}\sim c/N^2$, for some constant $c$. See \cite{plb} for detailed error estimates.

{\bf Remark:} At high order $N$, Pad\'e can become unstable due to very large coefficients appearing in the polynomials in the numerator and denominator, leading to delicate cancellations. This numerical problem can often be alleviated by converting the Pad\'e approximant to a partial fraction expansion, or to a continued fraction expansion, both of which can be done with high precision.

\subsubsection{Pad\'e Conformal Borel  Transform for One Branch Point Singularity}
\label{sec:pcb}

While the Pad\'e Borel method is quite impressive, especially considering how simple it is to implement, it has two important limitations:
\begin{itemize}
\item For a function with a branch point singularity, Pad\'e generates an approximate ``branch cut'' as an arc of interlacing Pad\'e poles and zeros accumulating to the branch point. Therefore Pad\'e is not accurate near the branch cut.

\item
In physical applications, in nonlinear problems, we frequently encounter the situation where a Borel branch point singularity is {\it repeated} in integer multiples, and these higher singularities may be hidden amongst the poles that are supposed to represent the cut associated with the leading singularity.

\end{itemize}

\begin{figure}[h!]
\centerline{\includegraphics[scale=.8]{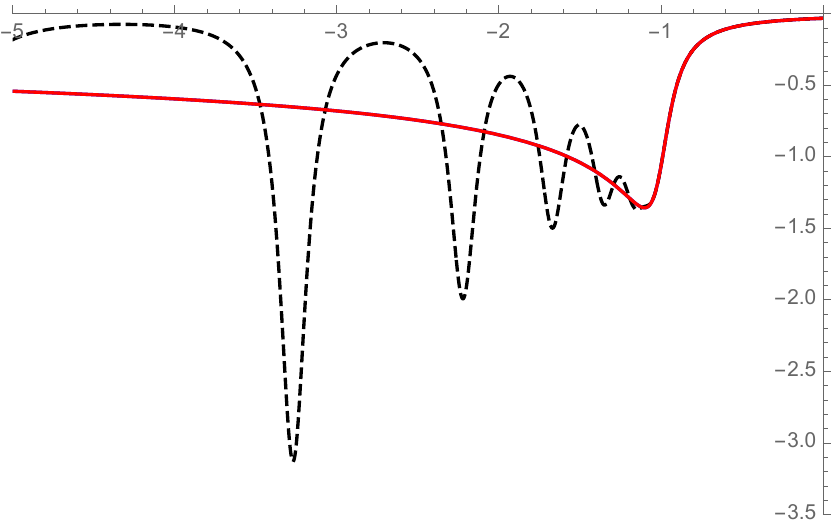}}
\caption{The Pad\'e and Pad\'e-Conformal approximations to the function $(1+t)^{-1/3}$, using as input 20 terms of the Taylor expansion about $t=0$. The imaginary part is plotted just above the cut ($t\to t+i/10$). The Pad\'e approximation (black, dashed) shows large peaks due to unphysical poles, while the Pad\'e-Conformal approximation (red, solid)  is indistinguishable from the actual function.}
\label{fig:pb-pcb}
\end{figure}

Fortunately there is a simple extra step which can resolve both these problems, and also produce a significant gain in precision. The improvement can be quite dramatic: see Figure \ref{fig:pb-pcb}. This is called the Pad\'e Conformal Borel method:
\begin{enumerate}
\item
Make a conformal map, $t=h(z)$, from the original Borel $t$ plane to the interior of the unit disk in the $z$ plane, mapping the origin to the origin.
\item
Re-expand about $z=0$ inside the unit disk,  to the same number of terms as the original truncated Borel expansion (this is provably optimal \cite{costin-uniform}), and {\it then} make a Pad\'e approximation in terms of the conformal variable $z$.
\item
Re-write this Pad\'e approximant in terms of the original Borel variable $t$ using the inverse conformal map.

\end{enumerate}

It is instructive to illustrate this procedure with the incomplete gamma function example from the previous section, and compare with the Pad\'e-Borel approximation. $B(t)=(1+t)^\alpha$ has a single branch point at $t=-1$, and a natural branch cut along the negative axis $t\in (-\infty, -1]$. The conformal map:
\begin{eqnarray}
t=\frac{4z}{(1-z)^2}
\quad \longleftrightarrow \quad 
z=\frac{\sqrt{1+t}-1}{\sqrt{1+t}+1}
\label{eq:cmap}
\end{eqnarray}
takes the singularity at $t=-1$ to $z=-1$, and $t=0$ to $z=0$, while $z=1$ corresponds to the point at infinity in the $t$ plane. The upper/lower edge of the cut $t\in (-\infty, -1]$ maps to upper/lower unit circle in the $z$ plane. For example, the points $t=-2\pm i\epsilon$ (as $\epsilon\to 0^+$) map to $z=\pm i$. Remarkably, once again we can express the exact diagonal Pad\'e approximant in terms of Jacobi polynomials \cite{plb}:
\begin{eqnarray}
{\rm Pade}{\rm -}{\rm Conformal}{\rm -}{\rm Borel:} \qquad {\mathcal P \mathcal C \mathcal B}_{[N,N]}(t; \alpha)=\frac{P_N^{(2\alpha,- 2\alpha)}\left(\frac{\sqrt{1+t}+1}{\sqrt{1+t}-1}\right)}{P_N^{(-2\alpha, 2\alpha)}\left(\frac{\sqrt{1+t}+1}{\sqrt{1+t}-1}\right)}
\label{eq:pcbp}
\end{eqnarray}
Observe that the argument of the Jacobi polynomials in \eqref{eq:pcbp} is now the inverse conformal variable! This has the consequence that the Pad\'e poles of \eqref{eq:pcbp} are now on the next sheet of the Riemann surface of the Borel transform. This makes the ${\mathcal P \mathcal C \mathcal B}$ approximation dramatically more accurate in the vicinity of the branch cut.  See Figure \ref{fig:pb-pcb}. The uniform asymptotics of the Jacobi polynomials \cite{dunster,elliott-jacobi} imply:
\begin{eqnarray}
\frac{{\mathcal P \mathcal C \mathcal B}_{[N,N]}(t; \alpha)}{(1+t)^\alpha}
& \sim &
\frac{I_{2\alpha}\left(
\hskip -2pt\left(N+\frac{1}{2}\right)\hskip -2pt\ln\left[\frac{(1+t)^{1/4}+1}{(1+t)^{1/4}-1}\right]\right)}
{I_{-2\alpha}\left(
\hskip -2pt\left(N+\frac{1}{2}\right)\hskip -2pt\ln\left[\frac{(1+t)^{1/4}+1}{(1+t)^{1/4}-1}\right]\right)}\\
&\hskip -2cm\sim & \hskip -1cm 1- 2\sin\left(2\pi \alpha \right)\left(\frac{(1+t)^{1/4}-1}{(1+t)^{1/4}+1}\right)^{2N+1} +\dots \qquad , \quad t\to 0^+
\\
&\hskip -6cm\sim & \hskip -3cm \frac{\Gamma(1-2\alpha)}{\Gamma(1+2\alpha)}\frac{\Gamma(N+1+2\alpha)}{\Gamma(N+1-2\alpha)}\, \frac{1}{(1+t)^\alpha} \left(1+\frac{4\alpha N(N+1)}{(4\alpha^2-1)}\frac{1}{\sqrt{t}} +\dots\right) 
 \, , \, t\to \infty
\label{eq:pcbp-limit}
\end{eqnarray}
An important difference compared to the Pad\'e Borel result (\ref{eq:pbp-limit}) is that the indices of the Jacobi polynomials have changed from $\pm \alpha$ to $\pm 2\alpha$. This enters the large $t$ asymptotics, which affects the extrapolation in the original variable from large $x$ down to small $x$. The net result is that the extrapolation limit $x_{\rm min}$ scales with $N$ as $x_{\rm min}\sim c/N^4$, which represents a significant improvement.

For example, compare the Pad\'e approximation with the  Pad\'e-Conformal approximation, each starting with a 20-term approximation to $\mathcal B(t)=(1+t)^{-1/3}$. In Figure \ref{fig:pb-pcb} we plot the actual function, and its Pad\'e and Pad\'e-Conformal approximations just above the cut $t\in (-\infty, -1]$. The Pad\'e approximation shows unphysical poles along the cut, while the Pad\'e-Conformal approximation is indistinguishable from the actual function.

{\color{Blue}
\begin{exercise}
\label{ex:4.5}
{\bf Pad\'e-Borel and Pad\'e-Conformal-Borel for the Airy Function}\\
The Airy function can be expressed as a Borel integral for $x>0$
\begin{eqnarray}
{\rm Ai}(x) =\frac{4}{3} x^{3/2} \, e^{-\frac{2}{3} x^{3/2}}\frac{1}{2\sqrt{\pi}\, x^{1/4}} \int_0^\infty dt\, e^{-\frac{4}{3} x^{3/2}\, t} \mathcal B(t)
\label{eq:airy-pb1}
\end{eqnarray}
where the Borel transform is the the hypergeometric function $\mathcal B(t)=~_2F_1\left(\frac{1}{6}, \frac{5}{6}, 1; -t\right)$.
\begin{enumerate}
\item
Make a diagonal $[5,5]$ Pad\'e approximant for  a $10$-term  truncated expansion of  $B(t)$, and plot the real and imaginary parts of this Pad\'e-Borel approximation to $B(t)$ along the complex Borel $t$ plane directions ${\rm arg}(t)=0$, ${\rm arg}(t)=\frac{\pi}{2}$, and ${\rm arg}(t)$ near $\pi$. 

\item
Repeat, with exactly the same input data, but using the Pad\'e-Conformal-Borel approximation to the Borel transform, and compare with the previous part.
\item
Use these results to compare plots of the (real and imaginary parts of the) Airy function ${\rm Ai}(x)$ in the complex $x$ plane, along the positive real axis, along the direction $\pi/3$, and along the  direction $2\pi/3$, based on the Pad\'e-Borel and Pad\'e-Conformal-Borel approximations. Plot the relative errors for each approximation.

\end{enumerate}
\end{exercise}
}

\noindent{\bf Remarks:} 
\begin{itemize}
\item
The ${\mathcal P \mathcal C \mathcal B}$ result (\ref{eq:pcbp}) is obtained from {\it exactly} the same input information ($2N$ terms of the truncated asymptotic series) as the ${\mathcal P \mathcal B}$ result (\ref{eq:pbp}), but the  ${\mathcal P \mathcal C \mathcal B}$ approximation is significantly more accurate.

\item
The increased precision is particularly dramatic near the Borel cut. See Figure \ref{fig:pb-pcb}. 

\item
The conformal map (\ref{eq:cmap}) does not require knowledge of the {\it nature} of the branch point, only  its {\it location} (rescaled here to be at $t=-1$).
However, square root branch points ($\alpha=\pm \frac{1}{2}$) are special, because the conformal map converts a square root branch point to a pole, and so the Pad\'e-Conformal-Borel transform is exact for all $N$. 
This extreme sensitivity underpins the {\it singularity elimination} method \cite{costin-uniform}, which can be  applied iteratively to obtain remarkably precise knowledge of the exact {\it location} and  {\it exponent} of a (Borel) singularity.

\item The use of conformal maps is natural because Pad\'e is fundamentally related to 2 dimensional electrostatics, which is solved by analytic functions (the real or imaginary part thereof). Recall section \ref{sec:potential}. So the only complication is imposing the boundary conditions, which means taking care of the boundaries or singularities, and this is achieved by suitable conformal maps.

\item An excellent catalogue of conformal maps is contained in \cite{kober}.

\item
In general there may be more than one Borel singularity. A common occurrence is the existence of two Borel singularities in the finite Borel plane and possibly a singularity at infinity. In this case the appropriate conformal map is known. Another case that appears in physical applications is a symmetric distribution of singularities, for example at $t_k=|t| e^{i k\, 2\pi/M}$ for $k=1, \dots , M$, for some integer $M$. In this case the appropriate conformal map is also known \cite{kober,costin-recipe}.

However, an empirical observation is that even when we use a simple conformal map based on just the {\it leading dominant singularity(ies)}, the increase in precision beyond the Pad\'e Borel approximation is significant \cite{p1,cs1}.

\item
An even more accurate method is to apply a {\it uniformizing map} rather than a conformal map. This is the essence of the  \underline{Optimality Theorem} \cite{costin-uniform}. Given information about the Riemann surface of the function (often known for resurgent functions), the optimal extrapolation procedure is to use a uniformizing map.

In applications with symmetry, the exact uniformizing map may be known, but even a uniformizing map based on leading singularities makes a dramatic difference. 
For a single singularity, at $\omega=+1$, the uniformizing map is the inverse elliptic nome function:
\begin{eqnarray}
\omega=\varphi(z) =16z -128z^2+704z^3+\dots  
\quad \longleftrightarrow\quad
z=\exp\left[-\pi\, \frac{\mathbb K (1-\omega)}{\mathbb K(\omega)}\right]
\label{eq:uniform}
\end{eqnarray}
For example consider the Airy Borel transform function $~_2F_1(\frac{1}{6}, \frac{5}{6}, 1; \omega)$. Compose with the map in \eqref{eq:uniform}, re-expand, make a Pad\'e approximation, and then map back to $\omega$ with the inverse uniformization map. The result is shown  in Figure \ref{fig:pub}, using just 10 terms of the expansion of $~_2F_1(\frac{1}{6}, \frac{5}{6}, 1; \omega)$. 
\begin{figure}[h!]
\centerline{\includegraphics[scale=0.6]{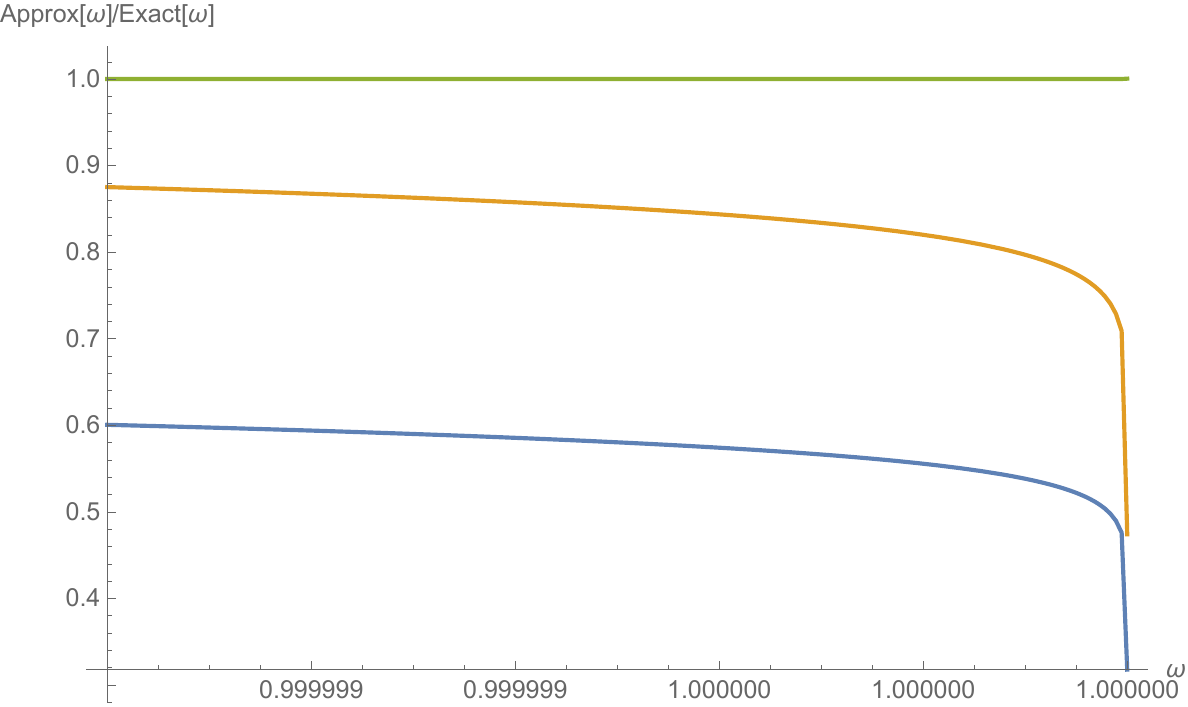}}
\caption{The ratio of the approximation to the exact function $~_2F_1(\frac{1}{6}, \frac{5}{6}, 1; \omega)$ is plotted very close to the singularity at $\omega=1$. The blue line (bottom) is the Pad\'e approximation; the orange curve (middle) is the Pad\'e conformal approximation;  the green curve (top) uses the uniformizing map in \eqref{eq:uniform}.}
\label{fig:pub}
\end{figure}
Note that each of these three approximation methods use exactly the same input information. The gain in precision from using the uniformizing map is remarkable. It is due to an inherent exponential distortion near a uniformized singularity \cite{costin-recipe}.
 \end{itemize}

\subsection{Resolving Hidden Singularities}
\label{sec:hidden}

Another advantage of making a conformal map, followed by  a Pad\'e approximation, is that a common physical situation is that a branch point singularity is repeated along the same line (e.g., in integer multiples of the leading singularity). But if Pad\'e places poles along this line in an attempt to represent a cut, then the {\it genuine}  more distant branch points will be "hidden" amongst those Pad\'e poles. But the conformal map resolves these genuine singularities, because they appear as accumulation points on the unit disk at points separated from the leading singularity. And the poles accumulate to the the genuine singularities from the {\it outside} of the unit disk. 

\begin{figure}[h!]
 \centering{\includegraphics[scale=0.55]{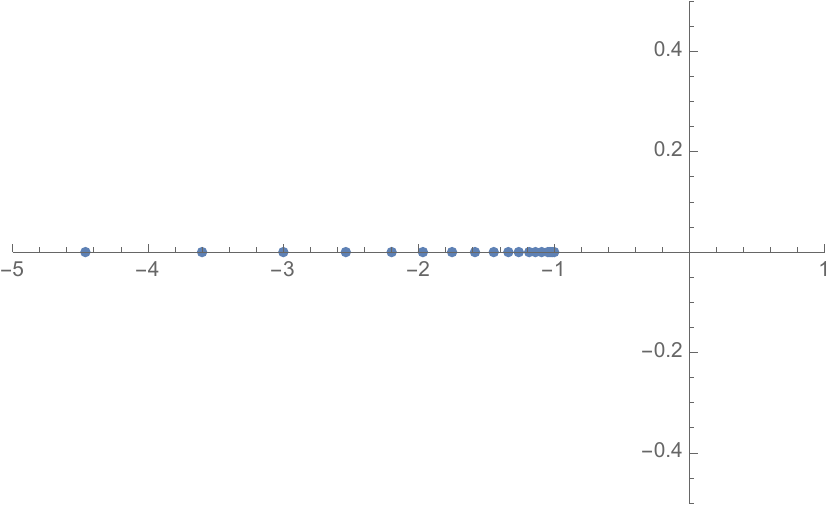}}
\caption{Pad\'e $t$-plane poles:  $N=50$. Suggests a  branch point at $t=-1$ and at $t=-\infty$.}
\label{fig:pade-p-poles}
\end{figure}

To illustrate the idea, consider the function
\begin{eqnarray}
B(t)=(1+t)^{-1/3}(2+t)^{-1/7}
\label{eq:bp}
\end{eqnarray}
which has a branch point at $t=-1$ and at $t=-2$. The dominant singularity is at $t=-1$. Expanding about $t=0$ to 50 terms, and making a diagonal Pad\'e approximant, we find Pad\'e poles accumulating to the singularity at $t=-1$. See Figure \ref{fig:pade-p-poles}. But it is not clear from this plot that there is also a singularity at $t=-2$. 

\begin{figure}[h!]
\centering{\includegraphics[scale=.4]{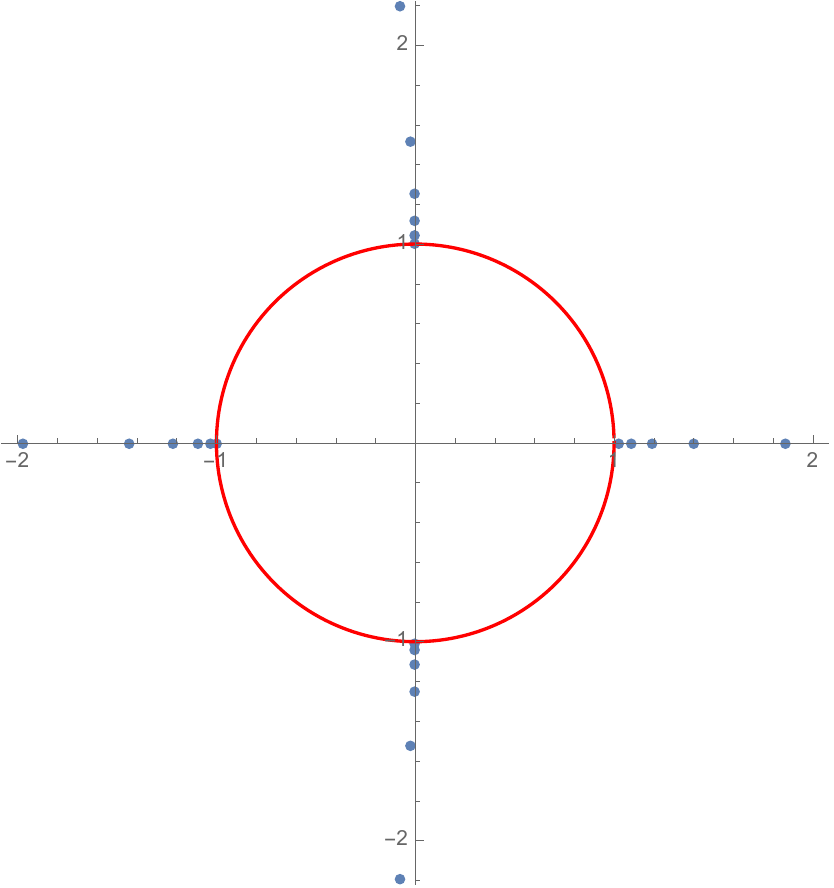}}
\caption{Pad\'e $z$-plane poles:  $N=50$. Suggests  branch points at $z=-1$, $z=-\infty$, and at $z=\pm i$.}
\label{fig:pade-z-poles}
\end{figure}

To resolve this ``hidden" singularity,   make a conformal map {\it based on the leading singularity at $t=-1$}:
\begin{eqnarray}
t=\frac{4z}{(1-z)^2} 
\qquad \longleftrightarrow\qquad
z=\frac{\sqrt{1+t}-1}{\sqrt{1+t}+1}
\label{eq:1cut}
\end{eqnarray}
Then re-expand about $z=0$ to order 50, and make a diagonal Pad\'e approximant in $z$. This produces Pad\'e poles accumulating to $z=-1$, which is the conformal map image of $t=-1$, and at $z=+1$, which is the conformal map image of $t=-\infty$, and also at $z=\pm i$, which are the conformal map images of $t=-2$ on either side of the cut. See Figure \ref{fig:pade-z-poles}.
Making the conformal map first, followed by a Pad\'e approximant in the conformal variable $z$,  effectively resolves the hidden singularity at $t=-2$.

This simple method can be applied in many examples where there are multiple collinear singularities. Indeed, in nonlinear problems, one often finds that a leading singularity is repeated an {\it infinite number} of times, for example in integer multiples of the leading singularity. Given enough coefficients, the Pad\'e Conformal method will resolve some number of these more distant genuine singularities which, without the conformal map, are hidden amongst the poles that Pad\'e places along the cut in an attempt to represent the leading branch cut. For physical applications, see \cite{harris,p1,cs1,cusp}.

{\color{Blue}
\begin{exercise}
\label{ex:4.6}
\noindent{\bf Resolving Hidden Singularities in Painlev\'e II:}\\
Consider the Borel transform of the perturbative $x\to-\infty$ solution to the Painlev\'e II equation \eqref{eq:p2}, as discussed in section \ref{sec:p2-neg}. 
\begin{enumerate}

\item Generate many coefficients of the Borel transform of the perturbative expansion \eqref{eq:p2minus}, and use a Pad\'e approximant to find the leading Borel singularities.

\item

Make a suitable conformal map to resolve any Borel singularities that you suspect may be hidden by the Pad\'e poles.

\item
Investigate what happens as you increase the number of terms in the expansion.

\end{enumerate}
\end{exercise}
}

\subsection{Continuing to Higher Sheets}
\label{sec:sheets}

The uniformizing map is even better than the conformal map, in terms of increased precision near the singularity(ies) and the cut(s) (recall Figure \ref{fig:pub}), and also because it may permit continuation from one Riemann sheet to another.  We illustrate with an example, the inversion of a cubic. 
Consider the "equation of state":
\begin{eqnarray}
 w=z+z^3
 \qquad \longrightarrow\qquad z=z(w)
\label{eq:wz}
\end{eqnarray}
This occurs for example in the mean field Ginzburg-Landau approach to the Ising singularity in a chiral random matrix theory of QCD \cite{basar-ly}. 
There are clearly 3 sheets: $z_1(w)$; $z_2(w)$ and $z_3(w)=-z_2(-w)$. The first  sheet has branch points at $w=\pm i\sqrt{\frac{4}{27}}$, while the second and third sheets each have just one branch point. On the first Riemann sheet the inversion has a Taylor series about the origin:
\begin{eqnarray}
z_1(w)=w-w^3+3w^5-12w^7+\dots
\label{eq:z1}
\end{eqnarray}
With the convenient rescaling, $\tilde{w}:=\sqrt{\frac{27}{4}}\, w$, the singularities are at $\tilde{w}=\pm i$, $\tilde{w}=i\, \infty$. The uniformization map for $\hat{\mathbb C}\setminus \{-i,i , i\,\infty\}$ is  in terms of the modular $\lambda$ function
\begin{eqnarray}
\tilde{w}(z)=i\left(-1+2\lambda\left(i\left(\frac{1+i \, z}{1-i \, z}\right)\right)\right)
\label{eq:lambda}
\end{eqnarray}
The optimal solution is to take the truncated expansion in \eqref{eq:z1} and re-expand in $z$ to the same order as the original expansion in $\tilde{w}$, after making the map in \eqref{eq:lambda}, and then make a Pad\'e approximation, followed by the inverse map.
This achieves several things. First, it gives a very high precision extrapolation on the first sheet. But, further, it produces a high precision extrapolation onto the second sheet \cite{basar-ly}.

\subsection{Singularity Elimination Method}
\label{sec: sing}

Another high precision method is "singularity elimination" \cite{costin-uniform}, which enables extreme sensitivity in the determination of the {\it location} and {\it exponent} of a singularity. The idea is simple. Suppose you have physical reason to expect a singularity at a location $t=t_0$, and with exponent $\beta_0$. Then one can make a fractional derivative (a linear transform) to adjust the exponent to be whatever you choose, without changing the location of the singularity. So choose to convert it to $\beta=\frac{1}{2}$. Then a conformal map based on this location will completely remove the singularity, resulting in an expansion that is analytic at that point. The singularity has been completely removed. Why is this useful? First, Pad\'e can numerically distinguish with extreme sensitivity  between an analytic point and a singular point, so you can confirm both the original conjectures (of the location and/or the exponent) with unprecedented precision. This can also be used to refine a conjecture to higher and higher precision.
For some recent physical applications, see \cite{cs1,cusp}.
Second, once the singularity has been removed, the expansion coefficients about this (now regular) point encode, using the linear transform,  the expansion coefficients  about the original singular point. This is useful for exploring large-order/low-order resurgent relations connecting the expansions about two different points.

\section{Acknowledgements}
I thank the organizers, and the participants, for an interesting and inspiring Summer School, and CERN for hosting this event.  I also gratefully acknowledge my collaborators M. \"Unsal, G. Ba\c sar, O. Costin and C. Schubert. This research is supported in part by the U.S. Department of Energy, Office of High Energy Physics, Award DE-SC0010339.

\end{document}